\documentclass[12pt]{article}
\usepackage[a4paper, total={6in, 9in}]{geometry}
\usepackage{graphicx}
\usepackage{amsfonts}
\usepackage{amsmath}
\usepackage{amssymb}
\usepackage{dsfont}
\usepackage{arydshln}
\usepackage{physics}
\usepackage{stmaryrd}
\usepackage{appendix}
\usepackage{hyperref}
\usepackage{simpler-wick}
\usepackage{subcaption}
\usepackage{authblk}

\numberwithin{equation}{section} 

\newcommand{\pf}{\text{pf}\,}

\begin{document}

\begin{titlepage}

\title{
\hfill\parbox{3cm}{ \normalsize YITP-25-167}\\   
\vspace{1cm} 
Real eigenvalue/vector distributions of random real 
antisymmetric tensors
}

\author[1]{Nicolas Delporte}
\author[1]{Giacomo La Scala}
\author[2, 3]{Naoki Sasakura}
\author[1]{Reiko Toriumi}

\affil[1]{\normalsize\it Okinawa Institute of Science and Technology Graduate University, 1919-1, Tancha, Onna, Kunigami District, Okinawa 904-0495, Japan. \hfill}
\affil[2]{\normalsize\it 
Yukawa Institute for Theoretical Physics, Kyoto University, Kitashirakawa, Sakyo-ku, Kyoto 606-8502, Japan.\hfill}
\affil[3]{\normalsize\it 
CGPQI, Yukawa Institute for Theoretical Physics, Kyoto University,
Kitashirakawa, Sakyo-ku, Kyoto 606-8502, Japan.
\authorcr
Emails: {\rm\url{nicolas.delporte@oist.jp}, \url{giacomo.lascala@oist.jp}, \url{sasakura@yukawa.kyoto-u.ac.jp}, \url{reiko.toriumi@oist.jp}.} \authorcr \hfill }

\date{}

\maketitle

\begin{abstract}
\noindent 
Real eigenpairs of a real antisymmetric tensor of order $p$ and dimension $N$ 
can be defined as pairs of a real eigenvalue and $p$ 
orthonormal $N$-dimensional real eigenvectors.
We compute the signed and the genuine distributions of such eigenvalues of 
Gaussian random real antisymmetric 
tensors by using a quantum field theoretical method. 
An analytic expression for finite $N$ is obtained for the signed
distribution and the analytic large-$N$ asymptotic forms for both.
We compute the edge of the distribution for large-$N$, one application of which 
is to give an upper bound (believed tight) of the injective norm
of the random real antisymmetric tensor. 
We find a large-$N$ universality across various tensor eigenvalue distributions: 
the large-$N$ asymptotic forms of the distributions
of the eigenvalues $z$ of the complex, complex symmetric, real symmetric, and real 
antisymmetric random tensors are all expressed by $e^{N\,B\, h_p(z_c^2/z^2)+o(N)}$,
where the function $h_p(\cdot)$ depends only on the order $p$, while $B$ and $z_c$ differ for each 
case, $NB$ being the total dimension of the eigenvectors and $z_c$ being
determined by the phase transition point of the quantum field theory.
\end{abstract}

\end{titlepage}

\tableofcontents

\section{Introduction}

Eigenvalues of matrices are classic very useful quantities, as they are invariants under change of basis of a vector space, significantly reducing the degrees of freedom of a matrix of dimension $N$ from $N \times N$ to $N$, still capturing the essence of a given matrix.
The asymptotic distributions of eigenvalues of random matrices (e.g., Wigner's semicircle law for Gaussian random matrices, Marchenko-Pastur law for rectangular random matrices) provide a powerful analytical tool 
in studying various subjects of science and mathematics \cite{akemann2011oxford}.
e.g., free probability \cite{voiculescu1997ams,nicabook,speichermingo}, matrix models (for quantum gravity for 2-dimensions and for QCD) \cite{FRANCESCO19951,anninos2020notes,tHooft:1973alw},
quantum chaos \cite{Camargo:2025zxr}, nuclear physics modeling the spectrum of a heavy atom nucleus \cite{c5bd8f0f-2576-3f83-a184-791e55682183}, neural network and deep learning \cite{louart2017randommatrixapproachneural}, etc.

Tensor eigenvalues/vectors (called eigenpairs) \cite{QI20051302,Lim2005SingularVA} 
are also important quantities which have various applications \cite{10.5555/3240740}.
However, the situation is much more intricate than for matrices, and the properties of eigenpairs of tensors are relatively poorly understood.
First of all, there are 
a multitude
of the definitions of eigenpairs depending on contexts \cite{QI20051302,Lim2005SingularVA,Cartwright_2013},
requiring more case-dependent studies.  
It is also known that computing tensor eigenpairs is NP-hard \cite{10.1145/2512329}. 
The number of eigenpairs of a tensor is proportional to the exponential of its dimension \cite{Cartwright_2013}. 
There is no obvious relation between eigenpairs  and decompositions of tensors, such as the 
tensor rank decomposition \cite{https://doi.org/10.1002/sapm192761164,CP}, which is analogous to the spectral decomposition of matrices. 

Considering the above intricacy, it would be more fruitful to study properties averaged over random tensors rather than those of any tensor. 
Random tensors generalise versatile and rich random matrix theories described above, and provide many interesting results to different areas of science and mathematics, e.g., quantum gravity for 3 and 4 dimensions \cite{ambjorn1991three,sasakura1991tensor,gross1992tensor,Gurau:2009tw,gurau2024quantum,Gurau:2011xp,Bonzom:2011zz}, spin glasses \cite{crisanti1992spherical}, quantum information (e.g., geometric measure of entanglement) \cite{Collins:2022akx,Dartois:2024zuc},
signal detection
\cite{Lahoche:2020txd,ouerfelli2021selectivemultiplepoweriteration},
stochastic equations \cite{Chandra:2023kpp}, free probability \cite{CollinsGurauLionni2024,Bonnin:2024lha}, holography \cite{Witten:2016iux,Gurau:2017xhf,klebanov2018tasi}, 
turbulence 
\cite{Dartois:2018kfy}, 
etc. 
As for eigenvalues of random tensors, the eigenvalue distribution of the real symmetric random tensor was computed in \cite{auffinger2013random}
as complexity of a spin glass model \cite{crisanti1992spherical}, a complex valued 
extension of the model was studied in \cite{PhysRevResearch.3.023064,Kent-Dobias_2022}, 
the number of eigenvalues of the random real tensor was studied in \cite{doi:10.1137/16M1089769},
the largest tensor eigenvalue was estimated by melonic dominance in \cite{Evnin:2020ddw}, and a generalisation
of Wigner's semi-circle law was proposed in \cite{Gurau:2020ehg}.

Systematic studies of eigenvalue/vector distributions of random tensors was initiated by one of the present authors 
by using a quantum field theoretical (QFT) method. The QFT method is powerful, precise, intuitive and general, having 
been sophisticated by the high-energy and the condensed matter communities for a long time.
The QFT method would be applicable to various kinds of random tensor problems, and, in fact, 
it has recently been successfully applied to a number of tensor eigenvalue/vector distributions  
\cite{Sasakura:2022zwc,Sasakura:2022iqd,Sasakura:2022axo,Kloos:2024hvy,Sasakura:2024awt,Delporte:2024izt,Majumder:2024ntn}.  
A noteworthy achievement was a new asymptotic upper bound (believed to be tight) of the injective norm of 
the complex random tensor \cite{Sasakura:2024awt,Dartois:2024zuc}, 
which was determined from the edge of the eigenvalue distribution of the complex random tensor \cite{Sasakura:2024awt}.  
This has importance in quantum information theory, since this determined the asymptotic
value of the geometric measure of quantum entanglement
\cite{https://doi.org/10.1111/j.1749-6632.1995.tb39008.x,H_Barnum_2001,PhysRevA.68.042307}
of random multipartite states for the first time. 

The purpose of the present paper is to apply the QFT method to the eigenvalue/vector distribution of the real antisymmetric random tensor.
An antisymmetric tensor can naturally be related to a multipartite state of fermions, having a motivation in quantum information theory.
We will first see that real eigenpairs of a real antisymmetric tensor of order $p$ 
and dimension $N$ can be defined by pairs of 
a real eigenvalue and $p$ orthonormal real $N$-dimensional eigenvectors. 
A new challenge will be to gauge-fix the non-Abelian SO($p$) gauge
symmetry of the eigenvector equations, which will be achieved by 
extending the gauge-fixing of an Abelian one 
in \cite{Sasakura:2024awt}.  We will compute the location of the edge of the distribution for large-$N$,
which determines the injective norm of the real antisymmetric random tensor, 
or equivalently the geometric measure of quantum entanglement of 
the random fermionic multipartite state.
We will find a universality of the large-$N$ asymptotic forms of the eigenvalue distributions, which holds across different kinds of random tensors.
This would suggest that various large-$N$ universalities of random matrices 
\cite{pastur} may be extended to random tensors. 

This paper is organised as follows.
In Section~\ref{sec:permutedeigendistri}, we define real eigenpairs of a real antisymmetric tensor as pairs of a real eigenvalue and 
$p$ orthonormal real eigenvectors. In fact, the eigenproblem is equivalent to only considering
$p$ {\it orthogonal} real eigenvectors by rescaling the eigenvalue to one.
We define the genuine and the signed distributions of the real eigenvectors of the real antisymmetric random tensor. 
In Section \ref{sec:partition-function}, we reformulate the signed distribution in a QFT partition function
by introducing fermions (Grassmann variables).  We introduce the gauge-fixing terms for the SO($p$) rotational gauge symmetry of the 
eigenvectors. Section \ref{sec:split} is devoted to the decomposition of the components of fields into the parallel and transverse
part against the eigenvectors. This decomposition simplifies later computations, because the random tensor decouples from the parallel part.  
In Section \ref{sec:computeintegrals} we integrate over all the fields to obtain the analytic expression of the signed distribution with 
the gauge-fixing. 
In Section \ref{sec:signeddistri}, the gauge-fixing is smeared away to obtain a gauge-free expression of the signed distribution. 
We also perform the integration over the rotational directions of the eigenvectors, and obtain the signed distribution expressed as a
function of the norm of the eigenvectors. This is equivalent to the signed distribution 
of the eigenvalues after rescaling.  
We compare the result with the Monte Carlo simulations. 
In Section~\ref{sec:signedlargeN} we take the large $N$ limit of the signed distribution using the technique of the Schwinger-Dyson equation. 
In Section \ref{sec:genuine}, we present the genuine distribution of the real eigenvectors 
of random antisymmetric real tensors also by a QFT formulation now with bosons and fermions,
which exhibit a supersymmetry. We use the Schwinger-Dyson method to compute the asymptotic
genuine distribution for large $N$.
The edge of the distribution is computed, which corresponds to the largest eigenvalue
and also to the injective norm of the real antisymmetric random tensor. 
In Section \ref{sec:universality} 
we finally present an important observation that there is a universality 
concerning the asymptotic forms of 
the eigenvalue distributions of random Gaussian tensors 
regardless of their properties (e.g., symmetry, real/complex, etc.).
We end with some concluding remarks in Section \ref{sec:concl}.
We present supplemental materials in the Appendix section.

\section{The Permuted Eigenvector Distribution}
\label{sec:permutedeigendistri}
\subsection{The Eigenvalue Problem}
\label{sec:eigenprob}
In order to introduce the problem of eigenvalues of antisymmetric tensors, let us temporarily restrict to an order 3 tensor. Let $T \in \otimes^3 \mathbb{R}^N$ be a real tensor, totally antisymmetric in the exchange of any two indices in the following sense:
\begin{equation}
    T_{abc} = -T_{bac} = T_{bca} = -T_{cba} = T_{cab} = -T_{acb}
    \,.
\end{equation}

The problem that we now consider is a generalisation of the eigenvalue problem, usually defined for matrices, inspired by the approach used in \cite{10.5555/3240740}. First, we introduce a family of distinct real vectors with two indices: a vector one in the $\mathbb{R}^N$ space we are considering our tensor over, and the second indicating a space we will call ``flavour". These vectors take the form $w^{i}_{a} \in \mathbb{R}$, with $i \in \{1,2,3\}$ being the flavour indices and $a \in \{1,\dots,N\}$ the real space indices. We consider the following system of coupled equations, obtained by permuting the contractions of this tensor with each vector. Repeated indices are summed over, employing Einstein summation notation:
\begin{equation} \label{eigenproblem-lambda-system}
    \begin{cases}
        T_{abc} w^{2}_{b} w^{3}_{c} = z^{1} w^{1}_{a}\\
        T_{abc} w^{3}_{b} w^{1}_{c} = z^{2} w^{2}_{a}\\
        T_{abc} w^{1}_{b} w^{2}_{c} = z^{3} w^{3}_{a}
        \,.
    \end{cases}
\end{equation}
This system defines what we will call a ``permuted eigenvector problem", and it is solved by a tuple $(z^{1},z^{2},z^{3},w^{1},w^{2},w^{3})$, with $z^{1},z^{2},z^{3} \in \mathbb{R}$, where we take all three vectors to be normalised to $|w^{i}| = 1$ \footnote{This can always be imposed by absorbing the norm of a non-normalised vector into redefinitions of $z^{1}, z^{2}, z^{3}$, so is without loss of generality.}. By contracting each vector of different flavour with itself and permuting tensor indices, we can show that all the $z^{i}$ are equal to each other:
\begin{align}
    &T_{abc} w^{1}_{a} w^{2}_{b} w^{3}_{c} = z^{1} w^{1}_{a} w^{1}_{a} = z^{2} w^{2}_{a} w^{2}_{a} = z^{3} w^{3}_{a} w^{3}_{a}\\
    \Rightarrow &z^{1} |w^{1}|^2 = z^{2} |w^{2}|^2 = z^{3} |w^{3}|^2\\
    \Rightarrow &z := z^{1} = z^{2} = z^{3}
    \,. \label{eq:eigenvalues-all-equal}
\end{align}
Therefore, we call $z \in \mathbb{R}$ one eigenvalue of the system, associated with the triple of vectors $(w^{1}, w^{2}, w^{3})$, which for convenience we will call an ``eigentuple". By tensor antisymmetry, the vectors in this tuple are all orthonormal:
\begin{gather}
    z w^{1}_{a} w^{2}_a = T_{abc} w^2_{a} w^{2}_{b} w^{3}_{c} = -T_{bac} w^2_{b} w^{2}_{a} w^{3}_{c} = 0, \nonumber \\
    z w^{2}_{a} w^{3}_a = \dots = 0, \quad z w^{3}_{a} w^{1}_a = \dots = 0 \\
    \Rightarrow w^{i}_{a} w^{j}_{a} = \delta^{ij} 
    \,.
    \label{eq:original-diagonal-property}
\end{gather}
We therefore see that the solution to our permuted eigenvalue problem can be written as $(z, w^{1}, w^{2}, w^{3})$, with all vectors in the eigentuple being orthonormal to one another. In this paper, we will consider only the case where $z \neq 0$. In the homogenous case of $z = 0$, the system has continuous degeneracies and the methods by which these eigentuples are studied cannot be applied. We also exclude trivial solutions where either $w^{i} = 0$. Differently from the case of matrix eigenvalues, the different number of vectors contracted appearing on the two sides of the equation system \eqref{eigenproblem-lambda-system} allows us to absorb the eigenvalue on the right into the norm of the vectors, by redefining them:
\begin{align}
    &w^{i}_{a} =: z v^{i}_{a}\\
    \Rightarrow & |z|^2 T_{abc} v^{2}_{b} v^{3}_{c} = |z|^2 v^{1}_{a} \Rightarrow T_{abc} v^{2}_{b} v^{3}_{c} = v^{1}_{a} \\
    \Rightarrow & |z|^2 = \frac{1}{|v^{i}|^2} \label{eq:norm-to-eigenval}
    \,.
\end{align}
When returning to the eigenvalue $z$, we can always take its positive value. Using this, we can rewrite the system of \eqref{eigenproblem-lambda-system} in the following single expression:
\begin{equation}
    \frac{1}{2} \epsilon^{ijk} T_{abc} v^{j}_{b} v^{k}_{c} = v^{i}_{a}, \quad \forall i \in \{1,2,3\}, a \in \{1, \dots, N\}
    \,,
\end{equation}
where $\epsilon^{ijk}$ denotes the Levi-Civita symbol.

From this equation, we can generalise the permuted eigenvalue problem to any $p \ge 3$. Let $T \in \otimes^{p} \mathbb{R}^{N}$ be an order $p$ totally antisymmetric tensor, and take the unnormalised real vectors $v^{i}_{a}$ to have flavours $i \in \{1, \dots, p\}$ and $a \in \{1, \dots , N\}$, then we define the eigenvalue problem as the simultaneous solution of:
\begin{equation} \label{eq:eigenproblem-condensed-form}
    \frac{1}{(p-1)!} \epsilon^{i i_{2} \dots i_{p}} T_{a a_{2} \dots a_{p}} v^{i_{2}}_{a_{2}} \dots v^{i_{p}}_{a_{p}} = v^{i}_{a}, \quad \forall i \in \{1,\dots,p\}, a \in \{1, \dots, N\} \,.
\end{equation}
In a similar fashion to equation \eqref{eq:norm-to-eigenval}, we then relate the unnormalised vectors to the eigenvalue of the normalised vectors by:
\begin{equation}
     z = \frac{1}{|v^{i}|^{p-2}} \,.
    \label{eq:lamvecrel}
\end{equation}
Note that, for general $p$,
one can always take positive eigenvalues by flipping the overall sign of $w^1_a$.
These constitute the definitions for general $p$ which we can now study in the context of a randomly distributed tensor $T$.

\subsection{Randomising the Tensor and Obtaining the Density}
\label{sec:randomtensor}
The permuted eigenvalue problem \eqref{eq:eigenproblem-condensed-form} has been formulated for a fixed tensor $T\in \otimes^{p} \mathbb{R}^N$, which we now promote to being random. We take the tensor components to be i.i.d. centered Gaussian variables with variance $(2\alpha p!)^{-1}$ and a constant $\alpha >0$. We define the measure:
\begin{gather}
    \int \mathcal{D}T\, e^{- \alpha \sum_{a_{1} \dots a_{p}}T^{2}_{a_{1} \dots a_{p}}}\,  \bullet := \frac{1}{\mathcal{Z}_{T}} \int \prod_{a_{1} < \dots < a_{p}} dT_{a_{1} \dots a_{p}}\, e^{- \alpha \sum_{a_{1} \dots a_{p}}T^{2}_{a_{1} \dots a_{p}}}\, \bullet \,, \nonumber\\
    \mathcal{Z}_{T} := \int \prod_{a_{1} < \dots < a_{p}} dT_{a_{1} \dots a_{p}}\, e^{- \alpha \sum_{a_{1} \dots a_{p}}T^{2}_{a_{1} \dots a_{p}}}
    \label{eq:tensor-gaussian-measure}
    \,.
\end{gather}
As a comparison to other matrix-valued probability distributions, this measure can be considered an adaptation of the Gaussian Orthogonal Ensemble (GOE) to tensors, with a further modification for the antisymmetry of the indices.

Given a fixed tensor, we define the following variables, which enforce that equation \eqref{eq:eigenproblem-condensed-form} is solved when they are set to zero:
\begin{equation} \label{eq:fia-def}
    f^{i}_{a} := v^{i}_{a} - \frac{1}{(p-1)!} \epsilon^{i i_{2} \dots i_{p}} T_{a a_{2} \dots a_{p}} v^{i_{2}}_{a_{2}} \dots v^{i_{p}}_{a_{p}}
    \,.
\end{equation}
Denote all eigentuples that solve \eqref{eq:eigenproblem-condensed-form}, indexed by $I$, as $\{v\}_{I}(T) = (v^{1},\dots, v^{p})_{I}(T)$. Then, using equation \eqref{eq:fia-def} and a multi-index determinant, we have the following identity:
\begin{equation} \label{eq:solution-delta-identity}
    \prod_{i,a} \delta(f^{i}_{a}) = \sum_{I} \Bigg|\det\left(\frac{\partial f^{i}_{a}}{\partial v^{j}_{b}}\big(\{v\}_{I}(T)\big)\right)\Bigg|^{-1} \delta(\{v\} - \{v\}_{I}(T)) \,.
\end{equation}
For a fixed tensor $T$, the eigenvector tuple density can be simply expressed in terms of $\delta$-functions at these $\{v\}_{I}(T)$ and simplified with \eqref{eq:solution-delta-identity}:
\begin{equation} \label{eq:fixed-tensor-density}
    \rho(\{v\},T) = \sum_{I} \delta(\{v\} - \{v\}_{I}(T)) = \Bigg|\det\left(\frac{\partial f^{i}_{a}}{\partial v^{j}_{b}}\right)\Bigg| \prod_{i,a} \delta(f^{i}_{a})
    \,.
\end{equation}
In the above, the determinant is with respect to the multi-index defined by the pair $(i,a)$, and we have used the transformation identity for multi-dimensional delta functions to shift zeros from solution vectors to the $f^{i}_{a}$'s. Also, it is worth noting that this density is not normalised as a probability density but integrates to the number of eigentuples solving the problem. We can now obtain the tensor-averaged distribution of eigentuples by averaging over equation \eqref{eq:fixed-tensor-density} with \eqref{eq:tensor-gaussian-measure} as:
\begin{equation} \label{eq:genuine-distr}
    \rho(\{v\}) ``:=" \Bigg\langle \Bigg|\det\left(\frac{\partial f^{i}_{a}}{\partial v^{j}_{b}}\right)\Bigg| 
    \;
    \prod_{i,a} \delta(f^{i}_{a}) \Bigg\rangle_{T}
    \,.
\end{equation}
We call this expression the \emph{genuine} distribution of eigentuples. We now introduce a modification which is easier for analysis using the partition function approach of Section \ref{sec:partition-function}, by removing the absolute value from the determinant:
\begin{equation} \label{eq:signed-unfixed-distr}
    \widetilde{\rho}(\{v\}) ``:=" \Bigg\langle \det\left(\frac{\partial f^{i}_{a}}{\partial v^{j}_{b}}\right) \prod_{i,a} \delta(f^{i}_{a}) \Bigg\rangle_{T}
    \,.
\end{equation}
We will call the latter the \emph{signed} distribution of eigenvectors. While it is not the proper distribution we seek, we can still extract from it important quantities of the genuine case, as discussed in \cite{Kloos:2024hvy}, most importantly the location of its ``edge" representing the largest eigenvalue. It also leads to exact computations at finite $N$ and finite $p$. In Section \ref{sec:gaugefreedom}, we will see that in reality, not all $f^{i}_{a}$ are truly independent and the above distributions are not well-defined (this is the reason for the double-quotes in our definitions). Both distributions will require a finite-$N$ modification introducing a gauge-fixing, the subject of Section \ref{sec:gauge-fixing-procedure}.

\section{The Signed Partition Function} \label{sec:partition-function}
\subsection{Introducing the Fields}
\label{sec:introfields}
Having expressed the tensor-averaged genuine and signed distributions of eigenvectors $\rho(v)$ and $\widetilde{\rho}(v)$ in terms of determinants and delta-functions of the $f^{i}_{a}$, we can adopt the quantum-field-theoretic perspective \cite{Sasakura:2022zwc,Sasakura:2022iqd,Sasakura:2022axo,Kloos:2024hvy,Sasakura:2024awt,Delporte:2024izt,Majumder:2024ntn} on the problem by rewriting the distributions in terms of integrals over bosonic and fermionic fields. Our goal here is to obtain a partition function with an associated action that we can manipulate. For the moment, we only consider the signed distribution $\widetilde{\rho}(\{v\})$ of \eqref{eq:signed-unfixed-distr}. To introduce the integrals, we rewrite the multi-index determinant in terms of Grassmann anticommuting fermionic variables $\bar{\psi}^{i}_{a}, \psi^{i}_{a}$ and we rewrite the delta functions in terms of bosonic real (Lagrange multiplier) variables $\lambda^{i}_{a}$ (see \eqref{eq:delta-function-convention} and \eqref{eq:fermi-measure-convention} in Appendix \ref{sec:appendix-conventions} for the conventions used), where again we have $i \in \{1,\dots,p\}, a \in \{1, \dots, N\}$. A final outer integral is given by the tensor averaging of \eqref{eq:tensor-gaussian-measure}. The expression that we obtain is:
\begin{equation} \label{eq:signed-unfixed-part-funct}
    \widetilde{\rho}(\{v\}) = \int \mathcal{D}T\, \mathcal{D}\bar{\psi}\, \mathcal{D}\psi\, \mathcal{D}\lambda\, e^{\mathcal{S}}
    \,,
\end{equation}
where we define and interpret the exponent as the following action:
\begin{equation} \label{eq:signed-unfixed-action}
    \mathcal{S} := -\alpha T^2 + \bar{\psi}^{i} \cdot \frac{\partial f^{i}}{\partial v^{j}} \cdot \psi^{j} + \mathbf{i} \lambda^{i} \cdot f^{i}
    \,,
\end{equation}
and $\mathbf{i}$ denotes the imaginary unit.
In the above we have represented contractions in the $a_{1}, \dots,a_{p}$ index space as $\lambda^{i} \cdot f^{i} = \lambda^{i}_{a} f^{i}_{a}$ or by squaring the tensor over all indices. Having an explicit expression for $f^{i}_{a}$ in \eqref{eq:fia-def}, we can compute the Jacobian derivative that appears in the above \eqref{eq:signed-unfixed-action}:
\begin{equation} \label{partition-f-derivative}
    \frac{\partial f^{i}_{a}}{\partial v^{j}_{b}} = \delta^{ij} \delta_{ab} - \frac{1}{(p-2)!} \epsilon^{ij i_{3} \dots i_{p}} T_{ab a_{3} \dots a_{p}} v^{i_{3}}_{a_{3}} \dots v^{i_{p}}_{a_{p}}
    \,.
\end{equation}
Thus, we can expand the partition function action for the signed distribution of eigentuples in the following explicit form:
\begin{align}
    \mathcal{S} = &-\alpha T_{a_{1} \dots a_{p}} T_{a_{1} \dots a_{p}} + \bar{\psi}^{i}_{a} \psi^{i}_{a} + \mathbf{i}\lambda^{i}_{a} v^{i}_{a} \nonumber\\
    &- \frac{1}{(p-2)!} \epsilon^{ij i_{3} \dots i_{p}} T_{ab a_{3} \dots a_{p}} \bar{\psi}^{i}_{a} \psi^{j}_{b} v^{i_{3}}_{a_{3}} \dots v^{i_{p}}_{a_{p}} - \frac{\mathbf{i}}{(p-1)!} \epsilon^{i i_{2} \dots i_{p}} T_{a a_{2} \dots a_{p}} \lambda^{i}_{a} v^{i_{2}}_{a_{2}} \dots v^{i_{p}}_{a_{p}}
    \,.
\end{align}
We will refer to separate parts of this action and the integrals over their components in the partition function as sectors.

\subsection{Gauge Freedom}
\label{sec:gaugefreedom}
The formulation \eqref{eq:eigenproblem-condensed-form} of the eigenvalue equation has some gauge freedom, given by $\text{SO}(p)$ transformations of the eigenvector tuple in the flavour index space. To see this, consider a transformation in the matrix $\text{GL}(p, \mathbb{R})$ representation of the group, transforming the vectors as:
\begin{equation}
    v^{i}_{a} \rightarrow M^{ii'} v^{i'}_{a}, \quad M \in \text{SO}(p) \label{eq:free-sop-transf}
    \,.
\end{equation}
Then, the left-hand side of equation \eqref{eq:eigenproblem-condensed-form} undergoes the transformation:
\begin{align}
    \frac{1}{(p-1)!} \epsilon^{i i_{2} \dots i_{p}} &T_{a a_{2} \dots a_{p}} v^{i_{2}}_{a_{2}} \dots v^{i_{p}}_{a_{p}} \nonumber\\
    &\rightarrow \frac{1}{(p-1)!} \epsilon^{i i_{2} \dots i_{p}} M^{i_{2} i'_{2}} \dots M^{i_{p} i'_{p}} T_{a a_{2} \dots a_{p}} v^{i'_{2}}_{a_{2}} \dots v^{i'_{p}}_{a_{p}} \nonumber\\
    &= \frac{1}{(p-1)!} \epsilon^{i' i'_{2} \dots i'_{p}} (M^{-1})^{i'i} \underbrace{\det(M)}_{=1} T_{a a_{2} \dots a_{p}} v^{i'_{2}}_{a_{2}} \dots v^{i'_{p}}_{a_{p}} \nonumber\\
    &= M^{ii'} \Big(\frac{1}{(p-1)!} \epsilon^{i' i'_{2} \dots i'_{p}} T_{a a_{2} \dots a_{p}} v^{i'_{2}}_{a_{2}} \dots v^{i'_{p}}_{a_{p}}\Big)
    \,.
\end{align}
Therefore, the eigenvalue equation \eqref{eq:eigenproblem-condensed-form} is covariant with respect to the transformation $M$ and the distributions of \eqref{eq:genuine-distr}, \eqref{eq:signed-unfixed-distr} are in reality not well-defined. By considering the generators of the Lie algebra of $\text{SO}(p)$, that is $p \times p$ antisymmetric matrices, and picking a basis for this space, we see that this gauge freedom is $\frac{1}{2}p(p-1)$-dimensional. Therefore, we can fix a gauge by specifying as many conditions on the vectors $v^{i}_{a}$, requiring that these can always be achieved by means of $\text{SO}(p)$ transformations in the flavour index space. Hinting at having one gauge choice for each antisymmetric generator of the symmetry group, we define and write these general gauge-fixing variables as:
\begin{equation} \label{eq:gen-tight-gauge-conditions}
    G^{i,i'} \overset{!}{=} 0, \quad i < i' \,.
\end{equation}
We also introduce the similar variables:
\begin{equation}
    G^{ii'} := \begin{cases}
        G^{i,i'}, &i < i'\\
        0, &i \ge i'
    \end{cases}\,,
\end{equation}
such that we allow both indices $i,i'$ to run unrestricted between $1, \dots, p$ for notational convenience.

Now, we wish to choose the $G^{i,i'}$ explicitly. A gauge choice that fixes the continuous $\text{SO}(p)$ symmetry which we are always able to make (and will justify below) is the following:
\begin{equation} \label{eq:new-cont-gauge-fix}
    v^{i_{1}}_{1} \overset{!}{=} 0, \text{ for } i_{1} > 1, \quad v^{i_{2}}_{2} \overset{!}{=} 0, \text{ for } i_{2} > 2, \quad \dots \quad v^{p}_{p-1} \overset{!}{=} 0
    \,.
\end{equation}
This indeed fixes $\frac{1}{2}p(p-1)$ continuous parameters. 
However, the gauge conditions \eqref{eq:eigenproblem-condensed-form} do not completely fix the SO($p$) symmetry: one can also see that SO($p$) transformation can flip 
the signs of any of the $v^{i}_{i}\, (i=1,2,\cdots,p-1$).  
Therefore, to remove this discrete gauge symmetry, we choose to focus on the sector of solutions such that:
\begin{equation} \label{eq:new-disc-gauge-fix}
    v^{i}_{i} > 0, \quad \text{for } i \in \{1, \dots, p-1\} \,.
\end{equation}

The gauge conditions \eqref{eq:new-cont-gauge-fix} and \eqref{eq:new-disc-gauge-fix} above can be justified as follows. First, consider the flavour-space ``vector" $v^{i}_{1}$, indexed by $i \in \{1, \dots, p\}$. Through the means of an $\text{SO}(p)$ transformation as in \eqref{eq:free-sop-transf}, we can freely choose to align this vector to the first ``axis" in its components. Specifically, we can ask that:
\begin{equation}
    v^{2}_{1} \overset{!}{=} 0\, \quad \dots \quad v^{p}_{1} \overset{!}{=} 0
    \,.
\end{equation}
Nevertheless, $v^{i}_{1}$ can now either be aligned to the positive or negative first axis, so we additionally
apply an SO($p$) transformation so that:
\begin{equation}
    v^{1}_{1} > 0
    \,.
\end{equation}
With this alignment to be preserved, we now have a residual and available $\text{SO}(p-1)$ transformation to fix. So, we consider the second vector $v^{i}_{2}$, in its residual $i \ge 2$ components and ask it, in an exactly analogous way, to satisfy:
\begin{equation}
    v^{3}_{2} \overset{!}{=} 0\,, \quad \dots \quad v^{p}_{2} \overset{!}{=} 0 \,, \quad v^{2}_{2} > 0
    \,.
\end{equation}
This further pushes us down to an $\text{SO}(p-2)$ symmetry. We can continue in this manner until we fix 
the $\text{SO}(2)$ symmetry. 

Therefore, to conclude, the gauge-fixing variables that we choose to set to zero as in \eqref{eq:gen-tight-gauge-conditions} are:
\begin{equation}
    G^{i,i'} := \beta v^{i'}_{i}\ (i<i'), \label{eq:chosen-gauge-conditions}
\end{equation}
where we have introduced an arbitrary fixed real parameter $\beta > 0$ in such a way that there is no effect on how the vector components are fixed (i.e. $G^{i,i'} = 0 \iff v^{i'}_{i} = 0$). This parameter is convenient to probe what happens to certain sectors of the action as $\beta \to 0$, where we recover non-gauge-fixed results. Since the parameter is arbitrary, we expect it to drop out from all meaningful quantities of our problem. Furthermore, we will take condition \eqref{eq:new-disc-gauge-fix} to implicitly restrict the sector of our distribution that we examine (see end of following Section \ref{sec:gauge-fixing-procedure}).

\subsection{Gauge Fixing Procedure} 
\label{sec:gauge-fixing-procedure}
The gauge-fixing procedure that we describe here is independent of the choice of gauge that we pick to study our problem. Therefore, we will explain it taking the $G^{i,i'}$ of \eqref{eq:gen-tight-gauge-conditions} as general, without restricting to a specific gauge choice like \eqref{eq:chosen-gauge-conditions}. First, consider the following contraction:
\begin{align}
    f^{i}_{a} v^{i'}_{a} &= \Big(v^{i}_{a} - \frac{1}{(p-1)!} \epsilon^{i i_{2} \dots i_{p}} T_{a a_{2} \dots a_{p}} v^{i_{2}}_{a_{2}} \dots v^{i_{p}}_{a_{p}} \Big) v^{i'}_{a} \nonumber\\
    &= v^{i}_{a} v^{i'}_{a} - \frac{1}{(p-1)!} \epsilon^{i i_{2} \dots i_{p}} \epsilon^{i' i_{2} \dots i_{p}} T_{a a_{2} \dots a_{p}} v^{1}_{a} v^{2}_{a_{p}} \dots v^{p}_{a_{p}} \nonumber\\
    &= v^{i}_{a} v^{i'}_{a} - \delta^{ii'} T_{a_{1} \dots a_{p}} v^{1}_{a_{1}} \dots v^{p}_{a_{p}}
    \,.
\end{align}
In particular, for $i \neq i'$, we have that:
\begin{equation} \label{eq:fv-off-diag-metric}
    f^{i}_{a} v^{i'}_{a} = v^{i}_{a} v^{i'}_{a}
    \,.
\end{equation}
For any choice of $i$ and $i'$, we have the following result, that expresses the redundancy in the non-gauge fixed system:
\begin{equation} \label{eq:fv-contraction-symmetric}
    f^{i}_{a}v^{i'}_{a} - f^{i'}_{a}v^{i}_{a} = 0, \quad \forall i,i' \in \{1, \dots, p\}
    \,.
\end{equation}
We use the above result to motivate the following modification to the system we are solving for, in order to incorporate the fixing conditions. Let $G^{i,i'}$ be arbitrary but attainable conditions to be satisfied and define the following:
\begin{align} \label{gauge-f-tildes}
    \widetilde{f}^{i}_{a} &:= f^{i}_{a} + \sum_{i' > i} v^{i'}_{a} G^{i,i'} = f^{i}_{a} + v^{i'}_{a} G^{ii'}\nonumber\\
    &= v^{i}_{a} - \frac{1}{(p-1)!} \epsilon^{i i_{2} \dots i_{p}} T_{a a_{2} \dots a_{p}} v^{i_{2}}_{a_{2}} \dots v^{i_{p}}_{a_{p}} + \sum_{i' > i} v^{i'}_{a} G^{i,i'}
    \,.
\end{align}
In order to show that this modification of the $f^{i}_{a}$ variables is a valid gauge-fixing procedure, we need to show both that $\tilde{f}^{i}_{a} = 0$ implies $G^{i,i'} = 0$ and therefore that $\tilde{f}^{i}_{a} = 0$ also implies $f^{i}_{a} = 0$. The way we do this is by first showing that the vectors $v^{i}_a$ of different flavour that solve $\tilde{f}^{i}_{a} = 0$, are all orthogonal. Then we show that this orthogonality also implies that all such vectors have the same norm. Finally, we show that this fixes all gauge-fixing conditions to be satisfied. We proceed.

To show orthogonality, we examine the contractions between the $\widetilde{f}$'s and the $v$'s. Assume $\widetilde{f}^{i}_{a}=0$ and take $i < p$ in the following:
\begin{gather}
    0=\widetilde{f}^{p}_{a} v^{i}_{a} = (f^{p}_{a} + \underbrace{\sum_{i'>p}}_{=0} v^{i'}_{a} G^{p,i'}) v^{i}_{a} = f^{p}_{a} v^{i}_{a} = v^{p}_{a} v^{i}_{a} \\
    \Rightarrow v^{p} \cdot v^{i} = 0 \quad \forall i < p \label{eq:vp-perp-vi}
    \,,
\end{gather}
where we have used \eqref{eq:fv-off-diag-metric}. Therefore, the vector $v^{p}$ is shown to be orthogonal to all other vectors. Using this fact, we can show the same for $v^{p-1}$, now taking $i < {p-1}$:
\begin{gather}
    0 = \widetilde{f}^{p-1}_{a} v^{i}_{a} = (f^{p-1}_{a} + \sum_{i' > p-1} v^{i'}_{a} G^{i, i'}) v^{i}_{a} = \underbrace{f^{p-1}_{a} v^{i}_{a}}_{v^{p-1} \cdot v^{i}} + \underbrace{v^{p}_{a} v^{i}_{a}}_{=0} G^{p-1, p} = v^{p-1}_{a} v^{i}_{a} \\
    \Rightarrow v^{p-1} \cdot v^{i} = 0 \quad \forall i < p-1
    \,,
\end{gather}
where we used  \eqref{eq:vp-perp-vi}. Therefore, we have $v^{p-1} \perp v^{i}, i \neq p-1$. Repeating the procedure done on $v^{p}$ and $v^{p-1}$ iteratively for lower $i$ and using the orthogonality derived from higher $i$ in the gauge-fixing sum condition over $i' > i$, we obtain that all vectors in the same eigentuple of the gauge-fixed system are orthogonal to each other:
\begin{equation} \label{gauge-orthogonality}
    v^{i} \cdot v^{j} = 0, i \neq j
    \,.
\end{equation}
Now, we perform the same kinds of contractions diagonally, for $i=i'$, to show that, again, all norms are equal. In the following, the repetition of the $i$ index is \textit{not} contracted over. We assume that $\tilde{f}^{i}_{a} = 0$ is already fixed, then for arbitrary $i$:
\begin{align}
    0 = \underbrace{\widetilde{f}^{i}_{a} v^{i}_{a}}_{\text{fixed } i} &= f^{i}_{a} v^{i}_{a} + \sum_{i' > i} v^{i'}_{a} v^{i}_{a} G^{i,i'} \nonumber\\
    &= v^{i}_{a} v^{i}_{a} - \delta^{ii} T_{a_{1} \dots a_{p}} v^{1}_{a_{1}} \dots v^{p}_{a_{p}} + \sum_{i' > i} \underbrace{v^{i'}_{a} v^{i}_{a}}_{= 0} G^{i,i'} \nonumber\\
    &= |v^{i}|^{2} - T_{a_{1} \dots a_{p}} v^{1}_{a_{1}} \dots v^{p}_{a_{p}}
    \,.
\end{align}
As this holds for any choice of $i$:
\begin{align} \label{gauge-same-norm}
    T_{a_{1} \dots a_{p}} v^{1}_{a_{1}} \dots v^{p}_{a_{p}} &= |v^{1}|^{2} = |v^{2}|^{2} = \dots = |v^{p}|^{2} \nonumber\\
    &=: \nu^2, \quad \nu > 0
    \,.
\end{align}
Hence, by combining equations \eqref{gauge-orthogonality} and \eqref{gauge-same-norm} we have shown that the eigenvector contractions in the $a_{1},\dots,a_{p}$ indices can be expressed as:
\begin{equation} \label{gauge-v-metric}
    v^{i}_{a} v^{j}_{a} = \nu^2 \delta^{ij}
    \,.
\end{equation}
Finally, we are in a position to use \eqref{eq:fv-contraction-symmetric} and \eqref{gauge-v-metric} to show that by imposing $\tilde{f}^{i}_{a} = 0$, we obtain that necessarily $G^{i,i'} = 0$. We start by taking $i<i'$ without loss of generality, for which we have that $\tilde{f}^{i}_{a}=0$ and $\tilde{f}^{i'}_{a}=0$, then:
\begin{align}
    0 = \widetilde{f}^{i}_{a} v^{i'}_{a} - \widetilde{f}^{i'}_{a} v^{i}_{a} &= \underbrace{f^{i}_{a} v^{i'}_{a} - f^{i'}_{a} v^{i}_{a}}_{=0} + \sum_{i'' > i} \underbrace{v^{i''}_{a} v^{i'}_{a}}_{= \nu^2 \delta^{i''i'}} G^{i,i''} - \sum_{i'' > i'} \underbrace{v^{i''}_{a} v^{i}_{a}}_{= 0 \text{ as } i<i'<i''} G^{i',i''} \nonumber\\
    &= \sum_{i'' > i} \nu^2 \delta^{i''i'} G^{i,i''} \nonumber\\
    &= \nu^2 G^{i,i'} \Rightarrow G^{i,i'} = 0
    \,.
\end{align}
Trivially, inserting all $G^{i,i'} = 0$ into equation \eqref{gauge-f-tildes} also shows that $f^{i}_{a} = 0$. Therefore, we see that the gauge is fixed successfully.

With the gauge-fixed system of equations, we must modify the signed distribution \eqref{eq:signed-unfixed-distr} to contain the additional terms studied above. We perform the following replacements:
\begin{align}
    \prod_{i,a} \delta(f^{i}_{a}) \quad &\longrightarrow \quad \prod_{i,a} \delta(\widetilde{f}^{i}_{a}) \label{eq:delta-f-to-ftilde}\\
    \Bigg|\det\left(\frac{\partial f^{i}_{a}}{\partial v^{j}_{b}}\right)\Bigg| \quad &\longrightarrow \quad \Bigg|\det\left(\frac{\partial \widetilde{f}^{i}_{a}}{\partial v^{j}_{b}}\right)\Bigg| 
    \,.
    \label{eq:det-f-to-ftilde}
\end{align}
For complete correctness, there should be a further modification imposing \eqref{eq:new-disc-gauge-fix}. This could be done by adding Heaviside functions to the distribution, but as the effect is just of reducing to one sector in its argument we will make this restriction choice implicit. Hence, the gauge-fixed signed density is reintroduced as a modification of \eqref{eq:signed-unfixed-distr} using the equations \eqref{eq:delta-f-to-ftilde}, \eqref{eq:det-f-to-ftilde} above:
\begin{equation} \label{eq:gauge-fixed-signed-density}
    \widetilde{\rho}(\{v\}) := \Bigg\langle \det\left(\frac{\partial \widetilde{f}^{i}_{a}}{\partial v^{j}_{b}}\right) \prod_{i,a} \delta(\widetilde{f}^{i}_{a}) \Bigg\rangle_{T}
    \,.
\end{equation}
A further observation to be made is that, given the $\delta$-function sector, we are able to use any result derived starting from the assumption $\widetilde{f}^{i}_{a} = 0$ inside the determinant sector. In particular, we can take $f^{i}_{a} = 0$ and $G^{i.i'} = 0$, from our reasoning above (the explicit justification in this case is covered in Appendix \ref{sec:appendix-justify-ftilde-f} regardless). With the modification to \eqref{eq:gauge-fixed-signed-density}, our gauge-fixed signed distribution requires a recomputation of the derivatives that appear in the fermionic determinant term:
\begin{align}
    \frac{\partial \widetilde{f}^{i}_{a}}{\partial v^{j}_{b}} &= \frac{\partial f^{i}_{a}}{\partial v^{j}_{b}} + \frac{\partial}{\partial v^{j}_{b}} \left(v^{i'}_{a} G^{ii'}\right) = \frac{\partial f^{i}_{a}}{\partial v^{j}_{b}} + \delta^{i'j} \delta_{ab} \underbrace{G^{ii'}}_{=0} + v^{i'}_{a} \frac{\partial G^{ii'}}{\partial v^{j}_{b}} \nonumber\\
    &= \frac{\partial f^{i}_{a}}{\partial v^{j}_{b}} + v^{i'}_{a} \frac{\partial G^{ii'}}{\partial v^{j}_{b}} \label{eq:new-explicit-dftilde}
    \,,
\end{align}
where we take $G^{ii'} = 0$ from the $\delta$-function sector. Using this simplification, we obtain the expression for the gauge-fixed signed distribution expressed as a partition function from \eqref{eq:gauge-fixed-signed-density}, in exactly the same way as for the ill-defined \eqref{eq:signed-unfixed-part-funct}:
\begin{align}
    \widetilde{\rho} (\{v\}) &= \int \mathcal{D}T\, \mathcal{D}\bar{\psi}\, \mathcal{D}\psi\, \mathcal{D}\lambda\, \exp \Big(-\alpha T^2 + \bar{\psi}^{i} \cdot \frac{\partial \widetilde{f}^{i}}{\partial v^{j}} \cdot \psi^{j} + \mathbf{i} \lambda^{i} \cdot \widetilde{f}^{i}\Big) \nonumber\\
    &=: \int \mathcal{D}T\, \mathcal{D}\bar{\psi}\, \mathcal{D}\psi\, \mathcal{D}\lambda\, e^{\widetilde{\mathcal{S}}} \label{eq:rho-stilde-rel}
    \,,
\end{align}
where we can write the gauge-fixed action $\tilde{\mathcal{S}}$ by defining the gauge-fixing contributions:
\begin{gather}
    \mathcal{S}_{G, \psi} := \bar{\psi}^{i}_{a} v^{i'}_{a} \frac{\partial G^{ii'}}{\partial v^{j}_{b}} \psi^{j}_{b}, \quad \mathcal{S}_{G, \lambda} := \mathbf{i} \lambda^{i}_{a} v^{i'}_{a} G^{ii'} \label{eq:action-gauge-fixing-terms} \\
    \mathcal{S}_{G} := \mathcal{S}_{G, \psi} + \mathcal{S}_{G, \lambda}
    \,,
\end{gather}
and summing them to the original action $\mathcal{S}$ of \eqref{eq:signed-unfixed-action}:
\begin{equation} \label{eq:raw-gauge-fixing-action}
    \widetilde{\mathcal{S}} := \mathcal{S} + \mathcal{S}_{G}
    \,.
\end{equation}

\section{Parallel-Transverse Splits}
\label{sec:split}
In multiple parts of the partition function action, we have direct index contractions between the variables of integration $\bar{\psi}, \psi, \lambda$ and the vectors in the eigentuple $v^{i}$. It turns out that important simplifications to the calculations can be made if variables are split between parts that live in the subspace spanned by these  vectors and parts in its orthogonal complement. We start by the following simplifying definition of a metric/Gramian-like matrix\footnote{We aim to use the vectors in the eigentuple as a basis, therefore there is a correspondence between the metric on this space and the Gramian of the vectors.} in the eigenvector tuple subspace:
\begin{equation} \label{eq:partrans-metric}
    g^{ij} := v^{i} \cdot v^{j}
    \,.
\end{equation}
Then, we define the following parallel and transverse projectors:
\begin{align}
    I^{\parallel}_{ab} 
    &
    = v^{i}_{a} (g^{-1})^{ij} v^{j}_{b} \label{eq:par-projector}
    \\
    I^{\perp}_{ab} &
    = \delta_{ab} - I^{\parallel}_{ab} \label{eq:trans-projector}
    \,.
\end{align}
By using the definition in \eqref{eq:partrans-metric}, the usual projector identities can be shown to be satisfied: $(I^{\parallel})^2 = I^{\parallel}$, $(I^{\perp})^2 = I^{\perp}$, $I^{\parallel}I^{\perp} = 0$ and $I^{\parallel} + I^{\perp} = \mathds{1}$. In particular, we also have that $I^{\perp} v^{i} = 0$. We note that, while we do eventually obtain $\delta$-functions that set $g^{ij} = \nu^2 \delta^{ij}$ as a consequence of gauge-fixing terms (in \eqref{eq:int-lambda-gforms}), we do not rely on this diagonal form of $g$ yet.

\subsection{$\bar{\psi}, \psi$ Decompositions} \label{sec:psi-bar-psi-decomp}
First, we look at the parallel-transverse decompositions of the pair of fermions $\bar{\psi}$ and $\psi$. As their parallel parts live in the subspace spanned by the vectors in the eigentuple\footnote{Strictly speaking, they are in the span when we allow for Grassmann coefficients, so we actually define this parallel-transverse division based entirely on applying the projectors.}, we can express them through newly defined coefficients $\bar{\psi}^{ij}$ and $\psi^{ij}$ that express this linear combination of $v^{i}$'s.
\begin{align}
    \bar{\psi}^{i}_{a} &= I^{\parallel}_{ab} \bar{\psi}^{i}_{b} + I^{\perp}_{ab} \bar{\psi}^{i}_{b} =: \bar{\psi}^{\parallel i}_{a} + \bar{\psi}^{\perp i}_{a} =: \bar{\psi}^{ij} v^{j}_{a} + \bar{\psi}^{\perp i}_{a} \label{eq:psi-bar-par-trans-decomp}\\
    \psi^{i}_{a} &= I^{\parallel}_{ab} \psi^{i}_{b} + I^{\perp}_{ab} \psi^{i}_{b} =: \psi^{\parallel i}_{a} + \psi^{\perp i}_{a} =: \psi^{ij} v^{j}_{a} + \psi^{\perp i}_{a} 
    \,.
    \label{eq:psi-par-trans-decomp}
\end{align}
We compute how this decomposition affects the contributions to the fermion determinant term that involves derivatives of $f^{i}_{a}$, the first in \eqref{eq:new-explicit-dftilde}, by expanding:
\begin{equation} \label{eq:psi-df-splits}
    \bar{\psi}^{i} \cdot \frac{\partial f^{i}}{\partial v^{j}} \cdot \psi^{j} = \bar{\psi}^{\parallel i} \cdot \frac{\partial f^{i}}{\partial v^{j}} \cdot \psi^{\parallel j} + \bar{\psi}^{\parallel i} \cdot \frac{\partial f^{i}}{\partial v^{j}} \cdot \psi^{\perp j} + \bar{\psi}^{\perp i} \cdot \frac{\partial f^{i}}{\partial v^{j}} \cdot \psi^{\parallel j} + \bar{\psi}^{\perp i} \cdot \frac{\partial f^{i}}{\partial v^{j}} \cdot \psi^{\perp j}
    \,.
\end{equation}
First, consider the following computation, which derives two simple identities assuming that $f^{i}_{a} = 0$:
\begin{align}
    f^{i}_{a} = 0 \Rightarrow v^{i}_{a} &= \frac{1}{(p-1)!} \epsilon^{i i'_{2} \dots i'_{p}} T_{a a_{2} \dots a_{p}} v^{i'_{2}}_{a_{2}} \dots v^{i'_{p}}_{a_{p}} \\
    \Rightarrow \epsilon^{i i_{2} \dots i_{p}} v^{i}_{a} &= \frac{1}{(p-1)!} \epsilon^{i i_{2} \dots i_{p}} \epsilon^{i i'_{2} \dots i'_{p}} T_{a a_{2} \dots a_{p}} v^{i'_{2}}_{a_{2}} \dots v^{i'_{p}}_{a_{p}} \nonumber \\
    &= \frac{1}{(p-1)!} \delta^{i_{2} \dots i_{p}}_{i'_{2} \dots i'_{p}} T_{a a_{2} \dots a_{p}} v^{i'_{2}}_{a_{2}} \dots v^{i'_{p}}_{a_{p}} \nonumber\\
    &= T_{a a_{2} \dots a_{p}} v^{i_{2}}_{a_{2}} \dots v^{i_{p}}_{a_{p}} \label{eq:partrans-v-epsilon-identity}\\
    \Rightarrow \epsilon^{i'j'k' \dots i_{p}} g^{kk'} &= T_{a_{1} a_{2} a_{3} \dots a_{p}} v^{i'}_{a_{1}} v^{j'}_{a_{2}} v^{k}_{a_{3}} \dots v^{i_{p}}_{a_{p}},
\label{eq:gtotformula}
\end{align}
where $\delta^{i_{1} \dots i_{p}}_{i'_{1} \dots i'_{p}}$ is defined in Appendix \ref{sec:appendix-conventions}. Since by Section \ref{sec:gauge-fixing-procedure} we have $\widetilde{f}^{i}_{a} = 0$ if and only if $f^{i}_{a} = 0, G^{i,i'} = 0$, we can use these simplifications because the fermionic sector multiplies the $\delta$-function sector. Now, we can both show that the cross terms of \eqref{eq:psi-df-splits} vanish and that the dependence on the tensor $T$ fully drops out of the strictly parallel part. For the first of the cross terms, using equation \eqref{partition-f-derivative} and the identity \eqref{eq:partrans-v-epsilon-identity} we show:
\begin{align}
    \bar{\psi}^{\parallel i}_{a} \frac{\partial f^{i}_{a}}{\partial v^{j}_{b}} \psi^{\perp j}_{b} &= \bar{\psi}^{ii'} v^{i'}_{a}\big(\delta^{ij} \delta_{ab} - \frac{1}{(p-2)!} \epsilon^{ij i_{3} \dots i_{p}} T_{ab a_{3} \dots a_{p}} v^{i_{3}}_{a_{3}} \dots v^{i_{p}}_{a_{p}}\big) \psi^{\perp j}_{b} \nonumber \\
    &= \bar{\psi}^{ii'} \underbrace{v^{i'}_{a} \psi^{\perp i}_{a}}_{= 0} - \frac{1}{(p-2)!} \bar{\psi}^{ii'} \epsilon^{i j i_{3} \dots i_{p}} T_{ab a_{3} \dots a_{p}} v^{i'}_{a} v^{i_{3}}_{a_{3}} \dots v^{i_{p}}_{a_{p}} \psi^{\perp j}_{b}  \nonumber\\
    &= \frac{1}{(p-2)!} \bar{\psi}^{ii'} \epsilon^{ij i_{3} \dots i_{p}} T_{baa_{3} \dots a_{p}} v^{i'}_{a} v^{i_{3}}_{a_{3}} \dots v^{i_{p}}_{a_{p}} \psi^{\perp j}_{b} \nonumber \\
    &= \frac{1}{(p-2)!} \bar{\psi}^{ii'} \epsilon^{ij i_{3} \dots i_{p}} \epsilon^{li' i_{3} \dots i_{p}} \underbrace{v^{l}_{b} \psi^{\perp j}_{b}}_{=0} = 0
    \,.
    \label{eq:psi-df-par-trans}
\end{align}
An almost identical computation results in the vanishing of the other cross term. On the other hand, concentrating on the parallel-parallel part:
\begin{align}
    \bar{\psi}^{\parallel i}_{a} \frac{\partial f^{i}_{a}}{\partial v^{j}_{b}} \psi^{\parallel j}_{b} &= \bar{\psi}^{ii'} v^{i'}_{a} \big(\delta^{ij} \delta_{ab} - \frac{1}{(p-2)!} \epsilon^{ij i_{3} \dots i_{p}} T_{ab a_{3} \dots a_{p}} v^{i_{3}}_{a_{3}} \dots v^{i_{p}}_{a_{p}} \big) \psi^{jj'} v^{j'}_{b} \nonumber \\
    &= \bar{\psi}^{ij} \psi^{ik} g^{jk} - \frac{1}{(p-2)!} \bar{\psi}^{ii'} \psi^{jj'} \epsilon^{ij i_{3} \dots i_{p}} T_{ab a_{3} \dots a_{p}} v^{i'}_{a} v^{j'}_{b} v^{i_{3}}_{a_{3}} \dots v^{i_{p}}_{a_{p}} \nonumber \\
    &= \bar{\psi}^{ij} \psi^{ik} g^{jk} - \frac{1}{(p-2)!} \bar{\psi}^{ii'} \psi^{jj'} \epsilon^{ijk i_{4} \dots i_{p}} \epsilon^{i'j'k' i_{4} \dots i_{p}} g^{kk'} \nonumber \\
    &= \bar{\psi}^{ij} \psi^{ik} g^{jk} - \frac{1}{p - 2} \bar{\psi}^{ii'} \psi^{jj'} g^{kk'} \delta^{ijk}_{i'j'k'}
    \,,
    \label{eq:psi-df-par-par}
\end{align}
where \eqref{eq:gtotformula} has been used. Note that the $T$ dependence has disappeared in the last expression. Collecting equations \eqref{eq:psi-df-splits}, \eqref{eq:psi-df-par-trans} and \eqref{eq:psi-df-par-par}:
\begin{equation}\label{eq:par-trans-split-psi-df}
    \bar{\psi}^{i} \cdot \frac{\partial f^{i}}{\partial v^{j}} \cdot \psi^{j} = \underbrace{\bar{\psi}^{ij} \psi^{ik} g^{jk} - \frac{1}{p - 2} \bar{\psi}^{ii'} \psi^{jj'} g^{kk'} \delta^{ijk}_{i'j'k'}}_{T-\text{independent}} + \underbrace{\bar{\psi}^{\perp i} \cdot \frac{\partial f^{i}}{\partial v^{j}} \cdot \psi^{\perp j}}_{T-\text{dependent}}
    \,.
\end{equation}
We can now also apply the parallel-transverse splitting to the fermionic gauge-fixing terms of \eqref{eq:action-gauge-fixing-terms}:
\begin{align}
    \mathcal{S}_{G, \psi} &= \bar{\psi}^{i}_{a} v^{i'}_{a} \frac{\partial G^{ii'}}{\partial v^{j}_{b}} \psi^{j}_{b} \nonumber\\
    &= \bar{\psi}^{ii''} g^{i''i'}\frac{\partial G^{ii'}}{\partial v^{j}_{b}} \psi^{j}_{b} = \bar{\psi}^{ii''} g^{i''i'} \frac{\partial G^{ii'}}{\partial v^{j}_{b}} (\psi^{jk} v^{k}_{b} + \psi^{\perp j}_{b}) \label{eq:psi-partrans-gauge-splitting}
    \,.
\end{align}
Equation \eqref{eq:psi-partrans-gauge-splitting} includes cross terms between the parallel and transverse components of the fermions. Nevertheless, these will drop out when performing integration over parallel sectors (seen in the computation in \eqref{eq:psi-par-integral-computation}) and only the parallel-parallel terms will remain.

\subsection{$\lambda$ Decompositions}
In an analogous manner to \eqref{eq:psi-bar-par-trans-decomp} and \eqref{eq:psi-par-trans-decomp}, we can decompose the $\lambda$ variables into parallel and transverse parts with respect to eigentuple-vector projections, once again also introducing the new parallel coefficients $\lambda^{ij}$:
\begin{equation}
    \lambda^{i}_{a} = I^{\parallel}_{ab} \lambda^{i}_{b} + I^{\perp}_{ab} \lambda^{i}_{b} = \lambda^{\parallel i}_{a} + \lambda^{\perp i}_{a} = \lambda^{ij} v^{j}_{a} + \lambda^{\perp i}_{a}
    \,.
\end{equation}
Differently from the $\bar{\psi},\psi$ sector, we cannot use this kind of decomposition to simplify what components of $\lambda$ couple to the tensor $T$ because, differently from Section \ref{sec:psi-bar-psi-decomp}, we now are in the $\delta(\tilde{f})$ function sector. Therefore, we use the decomposition only to simplify the gauge-fixing terms in the action of \eqref{eq:action-gauge-fixing-terms}:
\begin{equation}
    \mathcal{S}_{G, \lambda} = \mathbf{i} \lambda^{i}_{a} v^{i'}_{a} G^{ii'} = \mathbf{i} \lambda^{ii''} g^{i''i'} G^{ii'} \label{eq:lambda-partrans-gauge-splitting}
    \,.
\end{equation}
So, we see that only the parallel components of the different flavours of $\lambda$ appear in their contribution to the gauge-fixing sector.

\section{Computing the Integrals}
\label{sec:computeintegrals}
Collecting all the contributions to the gauge-fixed partition function action coming from equations \eqref{eq:raw-gauge-fixing-action}, \eqref{eq:par-trans-split-psi-df}, \eqref{eq:psi-partrans-gauge-splitting}, \eqref{eq:lambda-partrans-gauge-splitting}, the total action appearing in \eqref{eq:rho-stilde-rel} is:
\begin{align}
    \widetilde{\mathcal{S}} = &-\alpha T_{a_{1} \dots a_{p}} T_{a_{1} \dots a_{p}} + \bar{\psi}^{\perp i}_{a} \psi^{\perp i}_{a} + \mathbf{i}\lambda^{i}_{a} v^{i}_{a} \nonumber\\
    &- \epsilon^{i_{1} i_{2} i_{3} \dots i_{p}} T_{a_{1} a_{2} a_{3} \dots a_{p}} \Big(\frac{1}{(p-2)!}\bar{\psi}^{\perp i_{1}}_{a_{1}} \psi^{\perp i_{2}}_{a_{2}} v^{i_{3}}_{a_{3}} \dots v^{i_{p}}_{a_{p}} + \frac{\mathbf{i}}{(p-1)!} \lambda^{i_{1}}_{a_{1}} v^{i_{2}}_{a_{2}} v^{i_{3}}_{a_{3}} \dots v^{i_{p}}_{a_{p}} \Big) \nonumber\\
    &+ \bar{\psi}^{ij} \psi^{ik} g^{jk} - \frac{1}{p - 2} \bar{\psi}^{ii'} \psi^{jj'} g^{kk'} \delta^{ijk}_{i'j'k'} \nonumber\\
    &+ \bar{\psi}^{ij} g^{jk} \frac{\partial G^{ik}}{\partial v^{i'}_{b}} v^{j'}_{b}\psi^{i'j'} + \bar{\psi}^{ij} g^{jk} \frac{\partial G^{ik}}{\partial v^{l}_{b}} \psi^{\perp l}_{b} + \mathbf{i} \lambda^{ij} g^{jk} G^{ik} \label{eq:int-full-action}
    \,. 
\end{align}
Again, we have not decomposed the $\lambda$ variables that are contracted with the tensor $T$ into parallel and transverse components, as this does not allow an immediate simplification before tensor averaging is performed.

Before we move to the following sections, we introduce some notation for the intermediate effective actions obtained after integrating sectors of the partition function \eqref{eq:rho-stilde-rel}, that aims to illustrate what fields have been integrated out already and what sub-terms we consider. Consider the following:
\begin{equation}
    \widetilde{\mathcal{S}}^{\;\setminus\{\bullet\}}_{\star}
    \,.
\end{equation}
Then, by $\setminus\{\bullet\}$ we mean that the fields $\bullet$ have been integrated out of the action and by $\star$ we mean that we only collect terms dependent on the fields $\star$. For example $\widetilde{\mathcal{S}}^{\;\setminus \{T\}}_{\lambda}$ represents all the $\lambda$-dependent contributions to the effective action obtained following integration over the tensor $T$.

\subsection{Integration Over $T$}
First, we isolate all tensor-dependent contributions of equation \eqref{eq:int-full-action}:
\begin{align}
    \widetilde{\mathcal{S}}_{T} = &- \alpha T_{abc} T_{abc} \nonumber\\
    &- \epsilon^{i_{1} i_{2} i_{3} \dots i_{p}} T_{a_{1} a_{2} a_{3} \dots a_{p}} \Big(\frac{1}{(p-2)!}\bar{\psi}^{\perp i_{1}}_{a_{1}} \psi^{\perp i_{2}}_{a_{2}} v^{i_{3}}_{a_{3}} \dots v^{i_{p}}_{a_{p}} + \frac{\mathbf{i}}{(p-1)!} \lambda^{i_{1}}_{a_{1}} v^{i_{2}}_{a_{2}} v^{i_{3}}_{a_{3}} \dots v^{i_{p}}_{a_{p}} \Big)
    \,.
    \label{eq:tensoraction}
\end{align}
From the partition function perspective, the integral which we wish to compute is the innermost, over $\mathcal{D}T$, in the following:
\begin{equation}
    \widetilde{\rho}(\{v\}) = \int \mathcal{D}\bar{\psi}\, \mathcal{D}\psi\, \mathcal{D}\lambda\, \underbrace{e^{\widetilde{\mathcal{S}} - \widetilde{\mathcal{S}}_{T}}}_{T\text{-indep.}} \int \mathcal{D}T\, e^{\widetilde{\mathcal{S}}_{T}}
    \,.
\end{equation}
Since $\tilde{\mathcal{S}}_{T}$ is quadratic in the tensor, we can completely perform the integral over $T$ using Gaussian integration. Also, we notice that the gauge-fixing terms never couple to the tensor and so do not appear in this integral. Define for convenience:
\begin{equation}
    P_{a_{1} a_{2} a_{2} \dots a_{p}} := \frac{1}{\alpha} \epsilon^{i_{1} i_{2} i_{3} \dots i_{p}} \Big(\frac{1}{(p-2)!}\bar{\psi}^{\perp i_{1}}_{a_{1}} \psi^{\perp i_{2}}_{a_{2}} v^{i_{3}}_{a_{3}} \dots v^{i_{p}}_{a_{p}} + \frac{\mathbf{i}}{(p-1)!} \lambda^{i_{1}}_{a_{1}} v^{i_{2}}_{a_{2}} v^{i_{3}}_{a_{3}} \dots v^{i_{p}}_{a_{p}}\Big)
    \,.
\end{equation}
Then we look at the inner integral (and make summation over the $a_{1}, \dots,a_{p}$ indices explicit, for computational clarity) to compute:
\begin{align}
    \int \mathcal{D}T\, e^{\widetilde{\mathcal{S}}_{T}} &= \int \mathcal{D}T\, \exp \Big(- \alpha \sum_{a_{1} \dots a_{p}}\big( T^2_{a_{1} \dots a_{p}} + T_{a_{1} \dots a_{p}} P_{a_{1} \dots a_{p}} \big)\Big) \nonumber \\
    &= \int \mathcal{D}T\, \exp \Big(- \alpha p! \sum_{a_{1} < \dots < a_{p}} \big(\big( T_{a_{1} \dots a_{p}} + \frac{1}{2} \llbracket P_{a_{1} \dots a_{p}} \rrbracket \big)^2 - \frac{1}{4} \llbracket P_{a_{1} \dots a_{p}} \rrbracket^2 \big)\Big) \nonumber \\
    &= \exp ( \frac{1}{4} \alpha \sum_{a_{1} \dots a_{p}} \llbracket P_{a_{1} \dots a_{p}} \rrbracket^2)
    \,.
    \label{eq:integrated-tensor}
\end{align}
In the above, we have used the notation of $\llbracket \bullet \rrbracket$ to represent full antisymmetrisation on the $a_{1}, \dots, a_{p}$ indices induced by the antisymmetry of the tensor (see Appendix \ref{sec:appendix-conventions}). We expand the quadratic $P_{a_{1} \dots a_{p}}$ term in \eqref{eq:integrated-tensor}, where we notice that cross terms between the $\lambda$ and $\psi^{\perp}$ always involve a contraction between a transverse fermion and a vector, irrespective of the antisymmetrisation, and therefore vanish:
\begin{align}
    \frac{1}{4} \alpha \sum_{a_{1} \dots a_{p}} &\llbracket P_{a_{1} \dots a_{p}} \rrbracket^2 \nonumber\\
    = \frac{1}{4 \alpha} \sum_{a_{1} \dots a_{p}} \Big(&\frac{1}{(p-2)!^2} \epsilon^{i_{1} i_{2} i_{3} \dots i_{p}} \epsilon^{i'_{1} i'_{2} i'_{3} \dots i'_{p}} \llbracket \bar{\psi}^{\perp i_{1}}_{a_{1}} \psi^{\perp i_{2}}_{a_{2}} v^{i_{3}}_{a_{3}} \dots v^{i_{p}}_{a_{p}} \rrbracket \llbracket \bar{\psi}^{\perp i'_{1}}_{a_{1}} \psi^{\perp i'_{2}}_{a_{2}} v^{i'_{3}}_{a_{3}} \dots v^{i'_{p}}_{a_{p}} \rrbracket \nonumber\\
    - &\frac{1}{(p-1)!^2} \epsilon^{i_{1} i_{2} i_{3} \dots i_{p}} \epsilon^{i'_{1} i'_{2} i'_{3} \dots i'_{p}} \llbracket \lambda^{i_{1}}_{a_{1}} v^{i_{2}}_{a_{2}} v^{i_{3}}_{a_{3}} \dots v^{i_{p}}_{a_{p}} \rrbracket \llbracket \lambda^{i'_{1}}_{a_{1}} v^{i'_{2}}_{a_{2}} v^{i'_{3}}_{a_{3}} \dots v^{i'_{p}}_{a_{p}} \rrbracket \nonumber\\
    + &2 \frac{\mathbf{i}}{(p-1)!} \frac{1}{(p-2)!} \epsilon^{i_{1} i_{2} i_{3} \dots i_{p}} \epsilon^{i'_{1} i'_{2} i'_{3} \dots i'_{p}} \underbrace{\llbracket \bar{\psi}^{\perp i_{1}}_{a_{1}} \psi^{\perp i_{2}}_{a_{2}} v^{i_{3}}_{a_{3}} \dots v^{i_{p}}_{a_{p}} \rrbracket \llbracket \lambda^{i'_{1}}_{a_{1}} v^{i'_{2}}_{a_{2}} v^{i'_{3}}_{a_{3}} \dots v^{i'_{p}}_{a_{p}} \rrbracket}_{=0} \Big) \label{eq:tensor-averaging-with-par-perp}
    \,.
\end{align}
Therefore, after a careful computation involving squaring the antisymmetrised terms, we finally obtain the tensor averaged contribution to our first effective action $\tilde{\mathcal{S}}^{\setminus \{T\}}$:
\begin{align}
    \widetilde{\mathcal{S}}_{T} \overset{\int \mathcal{D}T}{\rightarrow}
    & \frac{1}{4\alpha p!(p-2)!} \epsilon^{i_{1} i_{2} i_{3} \dots i_{p}} \epsilon^{i'_{1} i'_{2} i'_{3} \dots i'_{p}} \big(- (\bar{\psi}^{\perp i_{1}} \cdot \bar{\psi}^{\perp i'_{1}})(\psi^{\perp i_{2}} \cdot \psi^{\perp i'_{2}}) \nonumber\\
    &\qquad\qquad\qquad\quad + (\bar{\psi}^{\perp i_{1}} \cdot \psi^{\perp i'_{2}})(\bar{\psi}^{\perp i'_{1}} \cdot \psi^{\perp i_{2}})\big) g^{i_{3} i'_{3}} \dots g^{i_{p} i'_{p}} \nonumber\\
    - &\frac{1}{4\alpha p! (p-1)!} \epsilon^{i_{1} i_{2} i_{3} \dots i_{p}} \epsilon^{i'_{1} i'_{2} i'_{3} \dots i'_{p}} \big((\lambda^{i_{1}} \cdot \lambda^{i'_{1}}) g^{i_{2} i'_{2}} g^{i_{3} i'_{3}} \dots g^{i_{p} i'_{p}} \nonumber\\
    &\qquad\qquad\qquad\quad + (p-1) (\lambda^{i_{1}} \cdot v^{i'_{1}}) (v^{i_{2}} \cdot \lambda^{i'_{2}}) g^{i_{3} i'_{3}} \dots g^{i_{p} i'_{p}}\big) \nonumber \\
    =:& \ \widetilde{\mathcal{S}}^{\setminus \{T\}} - (\widetilde{\mathcal{S}} - \widetilde{\mathcal{S}}_{T}) \label{eq:int-tensor-full-expr}
    \,.
\end{align}

\subsection{Integration Over $\lambda$}
\label{sec:lamint}
Following tensor averaging, we recollect all terms of the action that contain any $\lambda$ dependence from \eqref{eq:int-tensor-full-expr} and the $T$-independent remaining part of \eqref{eq:int-full-action} and proceed to rewrite these in terms of parallel and transverse components:
\begin{align}
    \widetilde{\mathcal{S}}^{\;\setminus \{T\}}_{\lambda} &= \mathbf{i} \lambda^{i} \cdot v^{i} - \frac{1}{4\alpha p! (p-1)!} \epsilon^{i_{1} i_{2} i_{3} \dots i_{p}} \epsilon^{i'_{1} i'_{2} i'_{3} \dots i'_{p}} \big((\lambda^{i_{1}} \cdot \lambda^{i'_{1}}) g^{i_{2} i'_{2}} g^{i_{3} i'_{3}} \dots g^{i_{p} i'_{p}} \nonumber\\
    &\quad + (p-1) (\lambda^{i_{1}} \cdot v^{i'_{1}}) (v^{i_{2}} \cdot \lambda^{i'_{2}}) g^{i_{3} i'_{3}} \dots g^{i_{p} i'_{p}}\big) + \mathbf{i} \lambda^{ij} g^{jk} G^{ik} \nonumber\\
    &= \widetilde{\mathcal{S}}^{\;\setminus \{T\}}_{\lambda^{\parallel}} + \widetilde{\mathcal{S}}^{\;\setminus \{T\}}_{\lambda^{\perp}}
    \,
\end{align}
where we separate fully into:
\begin{align}
    \widetilde{\mathcal{S}}^{\;\setminus \{T\}}_{\lambda^{\parallel}} &:= \mathbf{i} \lambda^{ij} g^{ij} - \frac{1}{4\alpha p! (p-1)!} \epsilon^{i_{1} i_{2} i_{3} \dots i_{p}} \epsilon^{i'_{1} i'_{2} i'_{3} \dots i'_{p}} \big(\lambda^{i_{1} j_{1}} \lambda^{i'_{1} j'_{1}} g^{j_{1} j'_{1}} g^{i_{2} i'_{2}} g^{i_{3} i'_{3}} \dots g^{i_{p} i'_{p}} \nonumber\\
    &\quad + (p-1) \lambda^{i_{1} j_{1}} g^{j_{1} i'_{1}} g^{i_{2} j'_{2}} \lambda^{i'_{2} j'_{2}} g^{i_{3} i'_{3}} \dots g^{i_{p} i'_{p}}\big) + \mathbf{i} \lambda^{ij} g^{jk} G^{ik}\,, \label{eq:lambda-par-action} \\
    \widetilde{\mathcal{S}}^{\;\setminus \{T\}}_{\lambda^{\perp}} &:= - \frac{1}{4\alpha p! (p-1)!} \epsilon^{i_{1} i_{2} i_{3} \dots i_{p}} \epsilon^{i'_{1} i'_{2} i'_{3} \dots i'_{p}} (\lambda^{\perp i_{1}} \cdot \lambda^{\perp i'_{1}}) g^{i_{2} i'_{2}} g^{i_{3} i'_{3}} \dots g^{i_{p} i'_{p}}
    \label{eq:lambda-trans-action}
    \,.
\end{align}
From the signed distribution point of view, we now want to compute the innermost integral over $\mathcal{D}\lambda$ in:
\begin{equation}
    \widetilde{\rho}(\{v\}) = \int \mathcal{D}\bar{\psi}\, \mathcal{D}\psi\, \underbrace{e^{\widetilde{\mathcal{S}}^{\setminus \{T\}} - \widetilde{\mathcal{S}}^{\setminus \{T\}}_{\lambda}}}_{\lambda \text{-indep.}} \int \mathcal{D}\lambda\, e^{\widetilde{\mathcal{S}}^{\;\setminus \{T\}}_{\lambda}}
    \,.
\end{equation}
The separation between parallel and transverse components allows us to carry out the $\lambda$-integration over these components separately (see explanation of the following integration measure and notation in Appendix \ref{sec:appendix-conventions-partrans}):
\begin{align}
    \int \mathcal{D}\lambda\, e^{\widetilde{\mathcal{S}}^{\;\setminus \{T\}}_{\lambda}} &= \int \mathcal{D}\lambda^{\parallel}\, e^{\widetilde{\mathcal{S}}^{\;\setminus \{T\}}_{\lambda^{\parallel}}} \int \mathcal{D}\lambda^{\perp}\, e^{\tilde{\mathcal{S}}^{\;\setminus \{T\}}_{\lambda^{\perp}}} \nonumber \\
    &= |\det g|^{\frac{p}{2}} \int \mathcal{D}[\lambda^{\parallel}]\, e^{\widetilde{\mathcal{S}}^{\;\setminus \{T\}}_{\lambda^{\parallel}}} \int \mathcal{D}\lambda^{\perp}\, e^{\widetilde{\mathcal{S}}^{\;\setminus \{T\}}_{\lambda^{\perp}}}
    \,.
\end{align}
The above expression for $\widetilde{\mathcal{S}}_{\lambda^{\parallel}}$ can be simplified by using the epsilon form of the matrix determinant, namely as in \eqref{eq:eps-det-2m}, after which $\epsilon$-symbols may be contracted. Using this, our final expression for the parallel $\lambda$-sector can be shown to be equal to:
\begin{equation} 
\label{eq:lambda-par-action-simpl}
    \widetilde{\mathcal{S}}^{\;\setminus \{T\}}_{\lambda^{\parallel}} = \mathbf{i} \lambda^{ij} g^{ij} - \frac{1}{4\alpha} \frac{1}{p!} (\det g) \lambda^{ii} \lambda^{jj} + \mathbf{i} \lambda^{ij} g^{jk} G^{ik}
    \,.
\end{equation}
We can treat this action by splitting the various summations into diagonal and off-diagonal contributions (following first and second lines respectively) in the $\lambda$ variables:
\begin{align}
    \widetilde{\mathcal{S}}^{\;\setminus \{T\}}_{\lambda^{\parallel}} = &\sum_{i} \mathbf{i} \lambda^{ii} \left(g^{ii} + g^{ik} G^{ik} \right) - \sum_{i,j} \frac{1}{4 \alpha} \frac{1}{p!} (\det g) \lambda^{ii} \lambda^{jj} \label{eq:lambda-par-diag-action}\\
    & +\sum_{i \neq j} \mathbf{i} \lambda^{ij} \left(g^{ij} + G^{ik} g^{kj}\right) \label{eq:lambda-par-off-diag-action}
    \,.
\end{align}
Like so, we can isolate on the off-diagonal integral of \eqref{eq:lambda-par-off-diag-action} and further split up the summation between the upper and lower triangles:
\begin{align}
    &\frac{1}{(2\pi)^{p(p-1)}} \int (\prod_{i \neq j} d\lambda^{ij})\, \exp \left( \sum_{i \neq j}\mathbf{i} \lambda^{ij} \left(g^{ij} + G^{ik} g^{kj}\right) \right) \nonumber \\
    \begin{split}
        &= \frac{1}{(2\pi)^{p(p-1)}}\int (\prod_{i < j} d\lambda^{ij})\, \exp \left( \sum_{i < j}\mathbf{i} \lambda^{ij} \left(g^{ij} + G^{ik} g^{kj}\right) \right)\\ &\qquad \times\int (\prod_{i > j} d\lambda^{ij})\, \exp \left( \sum_{i > j}\mathbf{i} \lambda^{ij} \left(g^{ij} + G^{ik} g^{kj}\right) \right)\,.
    \end{split}
\end{align}
If we restrict to the lower triangle, we can see that it integrates to $\delta$-functions zeroing the off-diagonal components of $g$:
\begin{align}
    &\frac{1}{(2\pi)^{\frac{1}{2} p(p-1)}} \int (\prod_{i > j} d\lambda^{ij})\, \exp \left( \sum_{i > j}\mathbf{i}  
    \lambda^{ij} \left(g^{ij} + G^{ik} g^{kj}\right) \right) \nonumber \\
    \begin{split}
        &= \frac{1}{(2\pi)^{\frac{1}{2} p(p-1)}} \int (\prod_{i > j} d\lambda^{ij})\, \exp \Bigg( \sum_{j < p} \mathbf{i} \lambda^{pj} \left(g^{pj} + \underbrace{G^{pk}}_{=0} g^{kj}\right) \nonumber \\
        &\hspace{2.5cm} + \sum_{j < p-1} \mathbf{i} \lambda^{p-1\,j} \left(g^{p-1\,j} + G^{p-1\,k} g^{kj}\right) + \dots \Bigg)  \nonumber
    \end{split}
    \\
    \begin{split}
        &= \prod_{j < p} \delta(g^{pj}) \times \frac{1}{(2\pi)^{ \frac{1}{2} (p-1)(p-2)}} \nonumber\\
        &\hspace{1.8cm}\times\int (\prod_{p-1 \geq  i > j} d\lambda^{ij})\, \exp \left( \sum_{j < p-1} \mathbf{i} \lambda^{p-1\,j} \left(g^{p-1\,j} + G^{p-1\,p} \underbrace{g^{pj}}_{= 0}\right) + \dots \right)  \nonumber
    \end{split}\\
    &= \prod_{i > j} \delta(g^{ij}) \label{eq:lambda-par-makes-g-diag}
    \,.
\end{align}
Therefore $g$ is diagonal. Under such $\delta$-functions, the upper triangle subsequently integrates to impose that the gauge-fixing variables must be equal to zero:
\begin{align}
    &\frac{1}{(2\pi)^{\frac{1}{2} p(p-1)}} \int (\prod_{i < j} d\lambda^{ij})\, \exp \left( \sum_{i < j}\mathbf{i} \lambda^{ij} \left(g^{ij} + G^{ik} g^{kj}\right) \right) \nonumber \\
    &= \frac{1}{(2\pi)^{\frac{1}{2} p(p-1)}} \int (\prod_{i < j} d\lambda^{ij})\, \exp \left( \sum_{i < j}\mathbf{i} \lambda^{ij} G^{ij} g^{jj} \right) \nonumber \\
    &= \prod_{i < j} \frac{1}{|g^{jj}|} \delta(G^{ij}) \label{eq:lambda-par-makes-gauge-deltas}
    \,.
\end{align}
In the final line, we use the fact that $g^{ii} = v^{i} \cdot v^{i} > 0$ for $v^{i} \neq 0$ to justify the transformation. With equation \eqref{eq:lambda-par-makes-gauge-deltas}, we can set the gauge-fixing variables to zero in \eqref{eq:lambda-par-diag-action} and, in view of computing the remaining part of the integral, we carry out the following transformation (with unit Jacobian):
\begin{equation}
    \lambda^{ii} = \begin{cases}
        \mu^{i} - \mu^{i+1}, &i<p \\
        \mu^{p}, &i = p
    \end{cases}  
    \,.
\end{equation}
Then we can compute the remaining diagonal-component integral as:
\begin{align}
    &\frac{1}{(2\pi)^{p}} \int (\prod_{i} d\lambda^{ii})\, \exp \Big(- \frac{1}{4\alpha} \frac{1}{p!} (\det g) \sum_{ij}\lambda^{ii} \lambda^{jj} + \mathbf{i}\sum_{i} \lambda^{ii} g^{ii}\Big) \nonumber \\
    = &\frac{1}{(2\pi)^{p}} \int (\prod_{i} d\mu^{i})\, \exp \Big(- \frac{1}{4\alpha} \frac{1}{p!} (\det g) (\mu^{1})^2 + \mathbf{i}\mu^{1}g^{11} \nonumber\\
    &\qquad\qquad\qquad\qquad\qquad\qquad+ \mathbf{i}\sum_{i=2}^{p} \mu^{i} (g^{ii} - g^{i-1\,i-1}) \Big) \nonumber \\
    = &\prod_{i=2}^{p} \delta(g^{ii} - g^{i-1\,i-1}) \int \frac{d\mu^{1}}{2\pi}\, \exp \Big(- \frac{1}{4\alpha} \frac{1}{p!} (\det g) (\mu^{1})^2 + \mathbf{i}\mu^{1}g^{11} \Big)
    \,,
\end{align}
where the $\delta$-functions ensure that all diagonal elements are equal, $g^{ii} = \nu^2, \forall i$, giving:
\begin{align}
    &= \prod_{i=2}^{p} \delta(g^{ii} - g^{i-1\,i-1}) \int \frac{d\mu^{1}}{2\pi}\, \exp \Big(- \frac{1}{4\alpha} \frac{1}{p!} (\nu^2)^{p} (\mu^{1})^2 + \mathbf{i}\nu^2 \mu^{1} \Big)\,, \nonumber \\
    &= \sqrt{\frac{\alpha p!}{\pi(\nu^2)^p}} e^{- \frac{\alpha p!}{(\nu^2)^{p-2}}} \prod_{i=2}^{p} \delta(g^{ii} - g^{i-1\,i-1}) \label{eq:g-remaining-deltas-factor}
    \,.
\end{align}
From the $\delta$-functions in equations \eqref{eq:lambda-par-makes-gauge-deltas} and \eqref{eq:g-remaining-deltas-factor}, we have shown explicitly that we can take the identities:
\begin{gather}
    g^{ij} = \nu^2 \delta_{ij} \label{eq:int-lambda-gforms}\\
    (g^{-1})^{ij} = \frac{1}{\nu^2} \delta^{ij}
    \,,
\end{gather}
to hold in all remaining sectors of the partition function. For example, we immediately deduce that, in \eqref{eq:lambda-par-makes-gauge-deltas}:
\begin{equation}
    \prod_{i < j} \frac{1}{|g^{jj}|} \delta(G^{ij}) = \frac{1}{(\nu^2)^{\frac{1}{2}p(p-1)}} \prod_{i < j} \delta(G^{i,j})
\end{equation}
Collecting all contributions from \eqref{eq:lambda-par-makes-g-diag}, \eqref{eq:lambda-par-makes-gauge-deltas} and \eqref{eq:g-remaining-deltas-factor} we conclude the parallel sector integration:
\begin{equation}
    \underbrace{|\det g|^{\frac{p}{2}}}_{= (\nu^2)^{\frac{1}{2}p^2}} \int \mathcal{D}[\lambda^{\parallel}] e^{\widetilde{\mathcal{S}}^{\;\setminus \{T\}}_{\lambda^{\parallel}}} = \sqrt{\frac{\alpha p!}{\pi}} e^{-\frac{\alpha p!}{(\nu^2)^{p-2}}} \prod_{i < j} \delta(g^{ij}) \prod_{i=2}^{p} \delta(g^{ii} - 
    g^{i-1 \, i-1}) \prod_{i < j} \delta(G^{i,j}) \label{eq:lambda-par-integrated}
    \,.
\end{equation}

Now we can shift to the computation of the integral involving the transverse components of the original $\lambda$ fields. Once again, here we use the epsilon form of the determinant and equation \eqref{eq:eps-det-2m} on the transverse action in \eqref{eq:lambda-trans-action}:
\begin{align}
    &\int \mathcal{D}\lambda^{\perp}\, e^{\widetilde{\mathcal{S}}^{\;\setminus \{T\}}_{\lambda^{\perp}}} \nonumber\\
    &= \int \mathcal{D}\lambda^{\perp}\, \exp \left(- \frac{1}{4\alpha p! (p-1)!} \epsilon^{i_{1} i_{2} i_{3} \dots i_{p}} \epsilon^{i'_{1} i'_{2} i'_{3} \dots i'_{p}} (\lambda^{\perp i_{1}} \cdot \lambda^{\perp i'_{1}}) g^{i_{2} i'_{2}} g^{i_{3} i'_{3}} \dots g^{i_{p} i'_{p}} \right) \nonumber \\
    &= \int \mathcal{D}\lambda^{\perp}\, \exp \left(- \frac{1}{4 \alpha p! (p-1)!}(\lambda^{\perp i_{1}} \cdot \lambda^{\perp i'_{1}}) \epsilon^{i_{1} i_{2} i_{3} \dots i_{p}} \epsilon^{i''_{1} i_{2} i_{3} \dots i_{p}} (\det g) (g^{-1})^{i'_{1} i''_{1}}\right)
    \nonumber \\
    &= \int \mathcal{D}\lambda^{\perp}\, \exp \left(- \frac{1}{4 \alpha p! (p-1)!}(\lambda^{\perp i_{1}} \cdot \lambda^{\perp i'_{1}}) (p-1)! \delta^{i_{1} i''_{1}} (\det g) (g^{-1})^{i'_{1} i''_{1}}\right)
    \nonumber \\
    &= \int \mathcal{D}\lambda^{\perp}\, \exp \left(- \frac{1}{2} \lambda^{\perp i}_{a} \Big( \frac{1}{2\alpha p!} (\det g) g^{-1} \Big)^{ii'} \lambda^{\perp i'}_{a}\right) \nonumber\\
    &= \frac{1}{(2\pi)^{p(N-p)}}\left(\frac{(4 \pi \alpha p!)^{p}}{(\det g)^{p-1}}\right)^{\frac{N - p}{2}} = \frac{1}{(2\pi)^{p(N-p)}} \left(\frac{(4 \pi \alpha p!)^{p}}{(\nu^2)^{p(p-1)}}\right)^{\frac{N - p}{2}}
    \,.
    \label{eq:lambda-trans-integrated}
\end{align}
Hence, we see that the computation of this transverse sector is much simpler than the parallel one, as it does not contain any gauge-fixing terms and simply amounts to a Gaussian integral.

In all of the calculations above, there has been no coupling between the $\lambda$-fields and the $\bar{\psi}, \psi$-fields. So, we highlight that integrating over $\lambda$ to obtain \eqref{eq:lambda-par-integrated} and \eqref{eq:lambda-trans-integrated} produced no additional terms in the $\bar{\psi}, \psi$ sector of the progressively integrated effective action. Using the results of the parallel and transverse integrals from these equations, we define $\tilde{\mathcal{S}}^{\setminus\{T,\lambda\}}$ such that:
\begin{equation} \label{eq:t-lambda-integrated-action}
    \widetilde{\rho}(v) = \int \mathcal{D}\bar{\psi}\, \mathcal{D}\psi\, e^{\widetilde{\mathcal{S}}^{\setminus \{T\}} - \widetilde{\mathcal{S}}^{\setminus \{T\}}_{\lambda}} \int \mathcal{D}\lambda^{\parallel}\, e^{\widetilde{\mathcal{S}}^{\setminus \{T\}}_{\lambda^{\parallel}}} \int \mathcal{D}\lambda^{\perp}\, e^{\widetilde{\mathcal{S}}^{\setminus \{T\}}_{\lambda^{\perp}}} =: \int \mathcal{D}\bar{\psi}\, \mathcal{D}\psi\, e^{\widetilde{\mathcal{S}}^{\setminus\{T,\lambda\}}}
    \,.
\end{equation}
Now, we can proceed to compute the $\bar{\psi}$ and $\psi$ integrations.

\subsection{Integration Over $\bar{\psi},\,\psi$}
\label{sec:psiintegrals}
First, we consider integrating the parallel sector, that is we consider computing the integrals over $\psi^{\parallel}$ and $\bar{\psi}^{\parallel}$. Schematically, we can represent these integrations' effect in the signed distribution as\footnote{Note that by the definition of our Grassmann measure in \ref{eq:fermi-measure-convention} we do not introduce any overall sign when permuting parallel and transverse parts.}:
\begin{equation} \label{eq:par-psis-rho-tilde-integrations}
    \widetilde{\rho}(\{v\}) = \int \mathcal{D}\bar{\psi}^{\perp}\, \mathcal{D}\psi^{\perp}\, \underbrace{e^{\widetilde{\mathcal{S}}^{\;\setminus\{T,\lambda\}} - \widetilde{\mathcal{S}}^{\;\setminus\{T,\lambda\}}_{\psi^{\parallel}}}}_{\bar{\psi}^{\parallel}, \psi^{\parallel} \text{-indep.}} \int \mathcal{D}\bar{\psi}^{\parallel}\, \mathcal{D}\psi^{\parallel}\, e^{\widetilde{\mathcal{S}}^{\;\setminus\{T,\lambda\}}_{\psi^{\parallel}}}
    \,,
\end{equation}
where we have the $T,\lambda$-integrated action $\widetilde{\mathcal{S}}^{\setminus\{T,\lambda\}}$ of equation \eqref{eq:t-lambda-integrated-action} and where we define the parallel fermion action by collecting all terms that depend on $\bar{\psi}^{\parallel}, \psi^{\parallel}$ in \eqref{eq:int-full-action}. For general $p$, this action reads as:
\begin{align}
    \widetilde{\mathcal{S}}^{\;\setminus\{T,\lambda\}}_{\psi^{\parallel}} &= \bar{\psi}^{ii'} \psi^{ij'} g^{i'j'} - \frac{1}{p-2} \bar{\psi}^{ii'} \psi^{jj'} g^{kk'} \delta^{ijk}_{i'j'k'} \nonumber\\
    &\quad + \bar{\psi}^{ij} g^{jk} \frac{\partial G^{ik}}{\partial v^{i'}_{b}} v^{j'}_{b}\psi^{i'j'} + \bar{\psi}^{ij} g^{jk} \frac{\partial G^{ik}}{\partial v^{l}_{b}} \psi^{\perp l}_{b}\,,
\end{align}
Using equation \eqref{eq:int-lambda-gforms}, we derive the expression:
\begin{equation}
    g^{kk'} \delta^{ijk}_{i'j'k'} = (p-2) \nu^2 (\delta^{ii'}\delta^{jj'} - \delta^{ij'}\delta^{ji'})
    \,.
\end{equation}
Substituting the expression into the action we obtain:
\begin{equation}
    \widetilde{\mathcal{S}}^{\;\setminus\{T,\lambda\}}_{\psi^{\parallel}} = \bar{\psi}^{ij} \widetilde{M}^{ij \ i'j'} \psi^{i'j'} + \bar{\psi}^{ij} X^{ij}_{\perp},
\end{equation}
where we have defined the following multi-indexed matrix:
\begin{equation}
    \widetilde{M}^{ij \ i'j'} := \nu^2 M^{ij \ i'j'} + \nu^2 \widetilde{G}^{ij \ i'j'}\,,
\end{equation}
with:
\begin{align}
    M^{ij \ i'j'} &:=\delta^{ii'}\delta^{jj'} - \delta^{ij}\delta^{i'j'} + \delta^{ij'}\delta^{ji'}\,, \\
    \widetilde{G}^{ij \ i'j'} &:= \frac{\partial G^{ij}}{\partial v^{i'}_{b}} v^{j'}_{b} \label{eq:big-gtilde-def}
    \,,
\end{align}
and the multi-indexed vector:
\begin{equation}
    X^{ij}_{\perp} := \nu^2 \frac{\partial G^{ij}}{\partial v^{k}_{b}} \psi^{\perp k}_{b} \label{eq:big-psi-transverse-p}
    \,.
\end{equation}
As we have shown in \eqref{eq:par-trans-split-psi-df}, the parallel-$\psi$ sector was unaffected by tensor averaging. In particular, this also means that no four-fermion terms are present in this sector and that we are able to fully evaluate the integral using Gaussian integration. In \eqref{eq:big-psi-transverse-p} we see that we have a dependence on transverse components of $\psi$, but the following calculation for the inner integral over $\mathcal{D}\bar{\psi}^{\parallel}$ and $\mathcal{D}\psi^{\parallel}$ in \eqref{eq:par-psis-rho-tilde-integrations} shows that such terms do not contribute:
\begin{align}
    \int \mathcal{D}\bar{\psi}^{\parallel}\, \mathcal{D}\psi^{\parallel}\, e^{\widetilde{\mathcal{S}}^{\;\setminus\{T,\lambda\}}_{\psi^{\parallel}}} &= |\det g|^{-\frac{p}{2}} |\det g|^{-\frac{p}{2}}\int \mathcal{D}[\bar{\psi}^{\parallel}]\, \mathcal{D}[\psi^{\parallel}]\, \exp \big(\bar{\psi}^{ij} \widetilde{M}^{ij \ i'j'} \psi^{i'j'} + \bar{\psi}^{ij} X^{ij}_{\perp}\big) \nonumber\\
    &= |\det g|^{-p} \int \mathcal{D}[\bar{\psi}^{\parallel}]\, \mathcal{D}[\psi^{\parallel}]\, \exp \big(\bar{\psi}^{ij} \widetilde{M}^{ij \ i'j'} (\psi^{i'j'} + (\widetilde{M}^{-1})^{i'j' \ i''j''} X^{i''j''}_{\perp})\big) \nonumber \\
    &= |\det g|^{-p} \int \mathcal{D}[\bar{\psi}^{\parallel}]\, \mathcal{D}[\psi^{\parallel}]\, \exp \big(\bar{\psi}^{ij} \widetilde{M}^{ij \ i'j'} \psi^{i'j'}\big) \nonumber\\
    &= |\det g|^{-p} \det \widetilde{M} =  (\nu^2)^{-p^2} \det \widetilde{M}
    \,.
    \label{eq:psi-par-integral-computation}
\end{align}
We see that the transverse components appearing in $X^{ij}_{\perp}$ are just absorbed by the Gaussian shift, which we can perform because $\bar{\psi}$ and $\psi$ are independent fermions. In order to compute $\det \widetilde{M}$ for general $p$, we start by looking at the symmetrisations that $M$ and $\widetilde{G}$ induce on the indices of the fermions that contract with them. Firstly:
\begin{align}
    \bar{\psi}^{ij} M^{ij \ i'j'} \psi^{i'j'} &= \bar{\psi}^{ij} (\psi^{ij} - \delta^{ij} \psi^{kk} + \psi^{ji}) \nonumber\\
    &= \bar{\psi}^{ij} (2 \psi^{(ij)} - \delta^{ij} \psi^{(kk)}) \nonumber\\
    &= 2 \bar{\psi}^{(ij)} \psi^{(ij)} - \bar{\psi}^{(ii)} \psi^{(jj)} \,,
\end{align}
so we see that $M$ contractions only involve the symmetric parts of the $\bar{\psi}, \psi$ coefficients. Then, we can note that:
\begin{equation} \label{eq:gtilde-zero-triangle}
    \widetilde{G}^{ij \ i'j'} = 0 \quad \text{if } i \ge j \,.
\end{equation}
Such that, for the other case of $i < j$, the above \eqref{eq:gtilde-zero-triangle} implies that:
\begin{align}
    \widetilde{G}^{(ij) \ i'j'} &= \frac{1}{2}(\widetilde{G}^{ij \ i'j'} + \underbrace{\widetilde{G}^{ji \ i'j'}}_{= 0}) = \frac{1}{2} \widetilde{G}^{ij \ i'j'} \,,\\
    \widetilde{G}^{[ij] \ i'j'} &= \frac{1}{2}(\widetilde{G}^{ij \ i'j'} - \underbrace{\widetilde{G}^{ji \ i'j'}}_{= 0}) = \frac{1}{2} \widetilde{G}^{ij \ i'j'} \,,
\end{align}
and therefore:
\begin{equation} \label{eq:gtilde-sym-antisym-eq}
    \Rightarrow \widetilde{G}^{(ij) \ i'j'} = \widetilde{G}^{[ij] \ i'j'} \quad \text{if } i< j \,.
\end{equation}
So, we can perform the following manipulations:
\begin{align}
    \bar{\psi}^{ij} \widetilde{G}^{ij \ i'j'} \psi^{i'j'} &= \bar{\psi}^{(ij)} \widetilde{G}^{(ij) \ i'j'} \psi^{i'j'} + \bar{\psi}^{[ij]} \widetilde{G}^{[ij] \ i'j'} \psi^{i'j'} \nonumber \\
    &= \sum_{i} \bar{\psi}^{(ii)} \widetilde{G}^{(ii) \ i'j'} \psi^{i'j'} + \sum_{i \neq j} \bar{\psi}^{(ij)} \widetilde{G}^{(ij) \ i'j'} \psi^{i'j'} + \sum_{i \neq j} \bar{\psi}^{[ij]} \widetilde{G}^{[ij] \ i'j'} \psi^{i'j'} \nonumber\\
    &= 2\sum_{i < j} \bar{\psi}^{(ij)} \widetilde{G}^{(ij) \ i'j'} \psi^{i'j'} + 2 \sum_{i < j} \bar{\psi}^{[ij]} \widetilde{G}^{[ij] \ i'j'} \psi^{i'j'} \nonumber\\
    &\rightarrow 2 \sum_{i < j} \bar{\psi}^{[ij]} \widetilde{G}^{[ij] i'j'} \psi^{i'j'} \nonumber\\
    &= \bar{\psi}^{[ij]} \widetilde{G}^{[ij] (i'j')} \psi^{(i'j')} + \bar{\psi}^{[ij]} \widetilde{G}^{[ij] [i'j']} \psi^{[i'j']} \,,
\end{align}
where (in the second-to-last step) we justify absorbing the symmetric $\bar{\psi}^{(ij)}$ terms by an appropriate linear shift of the integration measure on the antisymmetric $\bar{\psi}^{[ij]}$ components, which are independent of the $\psi^{[ij]}$, as well as \eqref{eq:gtilde-sym-antisym-eq}. Now, with these manipulations, we see that:
\begin{align}
    \bar{\psi}^{ij} \widetilde{M}^{ij \ i'j'} \psi^{i'j'} &= \nu^2 (2 \bar{\psi}^{(ij)} \psi^{(ij)} - \bar{\psi}^{(ii)} \psi^{(jj)}) \nonumber\\
    &\quad + \nu^2 \bar{\psi}^{[ij]} \widetilde{G}^{[ij] (i'j')} \psi^{(i'j')} + \nu^2 \bar{\psi}^{[ij]} \widetilde{G}^{[ij] [i'j']} \psi^{[i'j']} \nonumber \\
    &= \nu^2 (2 \bar{\psi}^{(ij)} \psi^{(ij)} - \bar{\psi}^{(ii)} \psi^{(jj)}) + \nu^2 \bar{\psi}^{[ij]} \widetilde{G}^{[ij] [i'j']} \psi^{[i'j']} + \underbrace{\dots}_{\text{irrelevant}}\,, \label{eq:mtilde-intermed-calc}
\end{align}
where we have separated out the symmetric-antisymmetric term that is irrelevant in the computation of the determinant of $\widetilde{M}$. To perform the integration, we transform the fermion measures to integrate over these symmetric-antisymmetric components:
\begin{equation} \label{eq:mtilde-psis-sym-antisym-meas}
    \mathcal{D}[\bar{\psi}^{\parallel}]\, \mathcal{D}[\psi^{\parallel}] = 2^{-p(p-1)} \mathcal{D}[\bar{\psi}^{(\parallel)}]\, \mathcal{D}[\bar{\psi}^{[\parallel]}]\, \mathcal{D}[\psi^{(\parallel)}]\, \mathcal{D}[\psi^{[\parallel]}]\,.
\end{equation}
With this, we can rewrite \eqref{eq:mtilde-intermed-calc} as:
\begin{align}
    \bar{\psi}^{ij} \widetilde{M}^{ij \ i'j'} \psi^{i'j'} &= 4\nu^2 \sum_{i<j} \bar{\psi}^{(ij)} \psi^{(ij)} + \nu^2 \left( 2\sum_{i} \bar{\psi}^{(ii)} \psi^{(ii)} - \sum_{i,j} \bar{\psi}^{(ii)} \psi^{(jj)} \right)  \nonumber \\
    &\quad + 4 \nu^2 \sum_{i < j} \sum_{i'<j'} \bar{\psi}^{[ij]} \tilde{G}^{[ij] [i'j']} \psi^{[i'j']} + \dots \nonumber \\
    &=: \sum_{i<j} \sum_{i'<j'}\bar{\psi}^{(ij)} \widetilde{M}^{ij \ i'j'}_{S} \psi^{(i'j')} + \sum_{i}\sum_{j} \bar{\psi}^{(ii)} \widetilde{M}^{ij}_{D} \psi^{(jj)} \nonumber \\
    &\quad + \sum_{i<j} \sum_{i'<j'} \bar{\psi}^{[ij]} \widetilde{M}^{ij \ i'j'}_{A} \psi^{[i'j']} + \dots \,.\label{eq:mtilde-ms-rewrite}
\end{align}
Therefore, in order to compute equation \eqref{eq:psi-par-integral-computation} using \eqref{eq:mtilde-psis-sym-antisym-meas} and \eqref{eq:mtilde-ms-rewrite}, we have that:
\begin{equation} \label{eq:mtilde-det-ms}
    \det \widetilde{M} = 2^{-p(p-1)} \det \widetilde{M}_{S}\, \det \widetilde{M}_{D} \, \det \widetilde{M}_{A} \,,
\end{equation}
where all multi-index determinants are computed along $i<j$, $i'<j'$. The first determinant is simply diagonal:
\begin{equation}
    \det \widetilde{M}_{S} = (4\nu^2)^{\frac{1}{2}p(p-1)}\,.
\end{equation}
The second determinant can be computed, for example by computing eigenvalues and respective multiplicities, to be:
\begin{equation} \label{eq:mtilde-det-md}
    \det \widetilde{M}_{D} = (\nu^2)^{p} \big(2 - p\big) 2^{p-1}
    \,.
\end{equation}
For the third determinant, we have:
\begin{align}
    \det \widetilde{M}_{A} &= (\nu^2)^{\frac{1}{2}p(p-1)} \det_{i<j, i'<j'} (4\widetilde{G}^{[ij][i'j']}) \nonumber\\
    &= (\nu^2)^{\frac{1}{2}p(p-1)} \det_{i<j, i'<j'} \Big(\widetilde{G}^{ij \ i'j'} - \widetilde{G}^{ij \ j'i'}\Big) \label{eq:mtilde-det-ma}
    \,.
\end{align}
Therefore, using the results in \eqref{eq:mtilde-det-ms}, \eqref{eq:mtilde-det-md} and \eqref{eq:mtilde-det-ma}, we have that:
\begin{align}
    \det \widetilde{M} &= 2^{-p(p-1)} (4\nu^2)^{\frac{1}{2}p(p-1)} (\nu^2)^{p} (2-p) 2^{p-1} (\nu^2)^{\frac{1}{2}p(p-1)} \Delta \nonumber \\
    &= - (p-2) 2^{p-1} (\nu^2)^{p^2}
    \Delta\,,
\end{align}
where we define:
\begin{equation}
    \Delta := \det_{i<j, i'<j'} \Big(\widetilde{G}^{ij \ i'j'} - \widetilde{G}^{ij \ j'i'}\Big)
    \,.
\end{equation}
Therefore:
\begin{equation}
    \Rightarrow \int \mathcal{D}\bar{\psi}^{\parallel}\, \mathcal{D}\psi^{\parallel}\, e^{\widetilde{\mathcal{S}}^{\;\setminus\{T,\lambda\}}_{\psi^{\parallel}}} = - (p-2) 2^{p-1} \Delta \label{eq:psi-par-integrated}
    \,.
\end{equation}
If we use the specific gauge-fixing condition of \eqref{eq:chosen-gauge-conditions}, we see that $\Delta \propto \beta^{\frac{1}{2}p(p-1)}$. Therefore, we briefly comment that without gauge-fixing (i.e. $\beta = 0$) our determinant integral would be equal to zero. Now, we define $\widetilde{\mathcal{S}}^{\;\setminus\{T,\lambda, \psi^{\parallel}\}}$ to incorporate parallel sector integrations for the $\bar{\psi}, \psi$-fields such that:
\begin{align}
    \widetilde{\rho}(\{v\}) 
    &=
    \int \mathcal{D}\bar{\psi}^{\perp}\, \mathcal{D}\psi^{\perp}\, e^{\widetilde{\mathcal{S}}^{\;\setminus\{T,\lambda\}} - \tilde{\mathcal{S}}^{\setminus\{T,\lambda\}}_{\psi^{\parallel}}} \int \mathcal{D}\bar{\psi}^{\parallel}\, \mathcal{D}\psi^{\parallel}\, e^{\widetilde{\mathcal{S}}^{\;\setminus\{T,\lambda\}}_{\psi^{\parallel}}} 
     \nonumber\\
    &=: 
    \int \mathcal{D}\bar{\psi}^{\perp}\, \mathcal{D}\psi^{\perp}\, e^{\widetilde{\mathcal{S}}^{\;\setminus\{T,\lambda, \psi^{\parallel}\}}}
    \,.
\end{align}

Therefore, the last remaining integration to compute to obtain the full partition function is the integral:
\begin{equation}
    \widetilde{\rho}(\{v\}) = \underbrace{e^{\widetilde{\mathcal{S}}^{\;\setminus\{T, \lambda, \psi^{\parallel}\}} - \widetilde{\mathcal{S}}^{\;\setminus\{T, \lambda, \psi^{\parallel}\}}_{\psi^{\perp}}}}_{\text{all const. w.r.t. } T,\lambda,\psi} \int \mathcal{D}\bar{\psi}^{\perp}\, \mathcal{D}\psi^{\perp}\, e^{\widetilde{\mathcal{S}}^{\;\setminus\{T, \lambda, \psi^{\parallel}\}}_{\psi^{\perp}}}
    \,.
\end{equation}
For $\widetilde{\mathcal{S}}^{\;\setminus\{T,\lambda, \psi^{\parallel}\}}_{\psi^{\perp}}$, we combine the remainder of the $\psi$-dependent part of \eqref{eq:int-full-action} after parallel-sector integration with equation \eqref{eq:int-tensor-full-expr}, that gives the quartic terms after $T$-integration, and use \eqref{eq:int-lambda-gforms} to contract $\epsilon$-symbols to obtain:
\begin{align}
    \widetilde{\mathcal{S}}^{\;\setminus\{T,\lambda, \psi^{\parallel}\}}_{\psi^{\perp}} = \bar{\psi}^{\perp i} \cdot \psi^{\perp i} + &\frac{1}{4\alpha p!(p-2)!} \epsilon^{i_{1} i_{2} i_{3} \dots i_{p}} \epsilon^{i'_{1} i'_{2} i'_{3} \dots i'_{p}} \big(- (\bar{\psi}^{\perp i_{1}} \cdot \bar{\psi}^{\perp i'_{1}})(\psi^{\perp i_{2}} \cdot \psi^{\perp i'_{2}}) \nonumber\\
    &\qquad\qquad\qquad\quad + (\bar{\psi}^{\perp i_{1}} \cdot \psi^{\perp i'_{2}})(\bar{\psi}^{\perp i'_{1}} \cdot \psi^{\perp i_{2}})\big) g^{i_{3} i'_{3}} \dots g^{i_{p} i'_{p}} \nonumber\\
    = \bar{\psi}^{\perp i} \cdot \psi^{\perp i} + &\frac{(\nu^2)^{p-2}}{4\alpha p!} \delta^{i_{1} i_{2}}_{i'_{1} i'_{2}} \big(- (\bar{\psi}^{\perp i_{1}} \cdot \bar{\psi}^{\perp i'_{1}})(\psi^{\perp i_{2}} \cdot \psi^{\perp i'_{2}}) \nonumber\\
    &\qquad\qquad\qquad\quad + (\bar{\psi}^{\perp i_{1}} \cdot \psi^{\perp i'_{2}})(\bar{\psi}^{\perp i'_{1}} \cdot \psi^{\perp i_{2}})\big) \nonumber\\
    = \bar{\psi}^{\perp i} \cdot \psi^{\perp i} - &\frac{(\nu^2)^{p-2}}{4\alpha p!} \big((\bar{\psi}^{\perp i} \cdot \bar{\psi}^{\perp j})(\psi^{\perp i} \cdot \psi^{\perp j}) + (\bar{\psi}^{\perp i} \cdot \psi^{\perp i})(\bar{\psi}^{\perp j} \cdot \psi^{\perp j}) \nonumber\\
    &\qquad\qquad\qquad\quad - (\bar{\psi}^{\perp i} \cdot \psi^{\perp j})(\bar{\psi}^{\perp i} \cdot \psi^{\perp j})\big) \label{eq:spsi-final-fourfermi}
    \,.
\end{align}
In the above, we have used anticommutativity of the Grassmann variables to eliminate the term including $\bar{\psi}^{\perp i} \cdot \bar{\psi}^{\perp i} = \bar{\psi}^{\perp i}_{a} \bar{\psi}^{\perp i}_{a} = -\bar{\psi}^{\perp i}_{a} \bar{\psi}^{\perp i}_{a} = 0$ and similarly $\psi^{\perp i} \cdot \psi^{\perp i} = 0$. The computation of the final partition function integral over this action $ \tilde{\mathcal{S}}^{\setminus\{T,\lambda, \psi^{\parallel}\}}_{\psi^{\perp}}$  requires using a different specific method, so is covered in isolation in the following Section \ref{sec:four-fermi}.

\subsection{Four-Fermi Theory Evaluation}\label{sec:four-fermi}
Equation \eqref{eq:spsi-final-fourfermi} shows that we obtain a four-fermion theory in the transverse sector of the $\psi$ fields after integrating out the tensor. This means that the final remaining integral of the partition function that we have to compute is:
\begin{equation} \label{eq:psi-trans-integral-def}
    \mathcal{Z}_{\perp} := \int \mathcal{D}\bar{\psi}^{\perp}\, \mathcal{D}\psi^{\perp}\,
    e^{\widetilde{\mathcal{S}}^{\;\setminus\{T,\lambda, \psi^{\parallel}\}}_{\psi^{\perp}}}
    \,.
\end{equation}
Once this integral is computed, we will have the final form of the signed distribution given schematically by the fully integrated action $\widetilde{\mathcal{S}}^{\setminus\{T, \lambda, \psi\}}$:
\begin{equation} \label{eq:full-integrated-partition-schematic}
    \widetilde{\rho}(\{v\}) = e^{\widetilde{\mathcal{S}}^{\;\setminus\{T, \lambda, \psi^{\parallel}\}} - \widetilde{\mathcal{S}}^{\;\setminus\{T, \lambda, \psi^{\parallel}\}}_{\psi^{\perp}}} \mathcal{Z}_{\perp} =: e^{\widetilde{\mathcal{S}}^{\;\setminus\{T, \lambda, \psi\}}}
    \,.
\end{equation}
In this section, we discuss how the integration can be evaluated, in principle in fully closed form when $N$ is finite. Define the following quadratic fermion action and associated partition function:
\begin{gather} \label{eq:fourfermi-squad}
    \widetilde{\mathcal{S}}_{\perp,2} := \bar{k}_{ij}(\bar{\psi}^{\perp i} \cdot \bar{\psi}^{\perp j}) + k_{ij} (\psi^{\perp i} \cdot \psi^{\perp j}) + \widetilde{k}_{ij} (\bar{\psi}^{\perp i} \cdot \psi^{\perp j}) \\
    \mathcal{Z}_{\perp, 2} := \int \mathcal{D}\bar{\psi}^{\perp}\, \mathcal{D}\psi^{\perp}\, e^{\widetilde{\mathcal{S}}_{\perp,2}}
    \,.
\end{gather}
Here, $\bar{k}, k, \tilde{k}$ are taken to be arbitrary-valued variables. Define the following differential operator:
\begin{equation} \label{fourfermi-op}
    \frac{\partial^{2}}{\partial k^{2}} := \sum_{ij} \left( \frac{\partial^{2}}{\partial \bar{k}_{ij} \partial k_{ij}} + \frac{\partial^{2}}{\partial \widetilde{k}_{ii} \partial \widetilde{k}_{jj}} - \frac{\partial^{2}}{\partial \widetilde{k}_{ij} \partial \widetilde{k}_{ij}} \right)
    \,.
\end{equation}
Now, using equation \eqref{eq:exponenting-differentiation} in the appendix, we recognise that the four-fermion action can be written by exponentiating the differential operator of \eqref{fourfermi-op}:
\begin{align}
    \mathcal{Z}_{\perp} &= e^{- \frac{(\nu^2)^{p-2}}{4\alpha p!} \frac{\partial^2}{\partial k^2}} \mathcal{Z}_{\perp, 2} \big|_{\bar{k} = k = 0, \tilde{k} = \mathds{1}} \nonumber \\
    &= e^{- \frac{(\nu^2)^{p-2}}{4\alpha p!} \frac{\partial^2}{\partial k^2}} \int \mathcal{D}\bar{\psi}^{\perp}\, \mathcal{D}\psi^{\perp}\, e^{\tilde{\mathcal{S}}_{\perp,2}} \big|_{\bar{k} = k = 0, \tilde{k} = \mathds{1}} 
    \,.
    \label{eq:exponential-derivative-full-formulation}
\end{align}
In this form, the quadratic partition function can be fully evaluated. While $\bar{k}, k$ are not restricted, we notice that their contractions with Grassmann anticommuting in variables in \eqref{eq:fourfermi-squad} only keep their antisymmetric components. Hence, we can manipulate the integral to yield a Pfaffian of a block matrix we call $K$:
\begin{align}
    \mathcal{Z}_{\perp, 2} &= \int \mathcal{D}\bar{\psi}^{\perp}\, \mathcal{D}\psi^{\perp}\, \exp \Big(
    \begin{pmatrix}
        \bar{\psi}^{\perp}_{a} & \psi^{\perp}_{a}
    \end{pmatrix}
    \underbrace{
    \begin{pmatrix}
        \frac{1}{2}(\bar{k} - \bar{k}^{T}) & \frac{1}{2} \tilde{k} \\
        -\frac{1}{2} \tilde{k}^{T} & \frac{1}{2}(k - k^{T})
    \end{pmatrix}
    }_{=: -K/2}
    \begin{pmatrix}
        \bar{\psi}^{\perp}_{a} \\
        \psi^{\perp}_{a}
    \end{pmatrix}
    \Big) \nonumber\\
    &= (\pf K)^{N-p}
    \,.
\end{align}
We notice that the Pfaffian is a finite polynomial, therefore the application of the exponential differential operator to any of its finite powers terminates even prior to setting $\bar{k} = k = 0, \tilde{k} = \mathds{1}$. This follows from the fact that both the quadratic action and the original four-fermi action are Grassmann integrals, and are therefore terminating series. We can express \eqref{eq:exponential-derivative-full-formulation} through Wick contractions, for which we define the following block matrix $A$, schematically acting on multi-indexed vectors of the form $\partial/\partial k = (\partial/\partial \bar{k_{ij}}, \partial/\partial k_{ij}, \partial/\partial \tilde{k}_{ij})$:
\begin{equation} \label{eq:a-matrix-def}
    e^{- \frac{(\nu^2)^{p-2}}{4\alpha p!} \frac{\partial^2}{\partial k^2}} =: e^{\frac{1}{2} \frac{\partial}{\partial k} \cdot A \cdot \frac{\partial}{\partial k}}\,.
\end{equation}
Then, the transverse fermion action can be computed as:
\begin{align}
    \mathcal{Z}_{\perp} &= e^{- \frac{(\nu^2)^{p-2}}{4\alpha p!} \frac{\partial^2}{\partial k^2}} (\pf K[\bar{k}, k, \tilde{k}])^{N-p} \Big|_{\bar{k} = k = 0, \tilde{k} = \mathds{1}} \nonumber \\
    &= \sum_{\wick{\c Q_{i} \c Q_{j}} = A_{ij}} (\pf K[- \bar{q}, - q, \mathds{1} - \tilde{q}])^{N-p} 
    \,,
    \label{eq:pf-wick-contractions}
\end{align}
where the notation below the sum represents summing over Wick contractions and we take $Q = (\bar{q}, q, \tilde{q})$ to be the multi-indexed vector containing dummy variables to Wick-contract. We briefly note that, had the integration variables been bosonic (i.e. non-Grassmann), this method might have not been justified and could have presented nonperturbative effects to be tackled with other techniques.

Once these contractions are computed, we have a closed form expression for $\mathcal{Z}_{\perp}$ that depends on a finite $N$ and on no fields, giving the end of the integration of all sectors in our partition function. For future reference, we have computed a few instances of this polynomial using Mathematica giving for example for $p=3$ with $\alpha=1$
(see App.~\ref{app:numsim}):
\begin{align}
    \mathcal{Z}_{\perp}(\nu^2, N) =\begin{cases}
        1\,, & N=4\,,\\
        1 - \frac{1}{2}\nu^2 + \frac{5}{48}\nu^4\,,& N=5\,,\\
        1 - \frac{3}{2}\nu^2 + \frac{13}{16} \nu^4 - \frac{6}{36} \nu^6 + \frac{5}{1152}\nu^8\,,& N=6\,,\\
        1 - 3\nu^2 + \frac{27}{8}\nu^4 - \frac{79}{48}\nu^6 + \frac{85}{256} \nu^8 - \frac{25}{1152} \nu^{10}+ \frac{5}{13824} \nu^{12}\,,& N=7
        \,.
    \end{cases}
\end{align}
For $p=4$, 
\begin{align}
    \mathcal{Z}_{\perp}(\nu^2, N) =\begin{cases}
        1\,, & N=5\,,\\
       1 - \frac{1}{4}\nu^4 + \frac{3}{128} \nu^8  - \frac{5} {9216}\nu^{12} + \frac{5}{589824} \nu^{16}\,,& N=6\,,\\
        1 - \frac{3}{4} \nu^4 + \frac{27}{128} \nu^8 - \frac{79}{3072} \nu^{12} + \frac{85}{65536} \nu^{16} - \frac{25}{1179648} \nu^{20} + \frac{5}{56623104} \nu^{24}\,,& N=7\ \,.
    \end{cases}
\end{align}

\section{The Signed Distribution}
\label{sec:signeddistri}
\subsection{Collecting the Partition Function}
Finally, we are able to collect all contributions to the partition function represented schematically in equation \eqref{eq:full-integrated-partition-schematic} and present them as the one single expression for our signed eigenvector distribution $\widetilde{\rho}$. First, we combine equations \eqref{eq:lambda-par-integrated}, \eqref{eq:lambda-trans-integrated}, \eqref{eq:psi-par-integrated} and \eqref{eq:psi-trans-integral-def} and we make the dependencies of $\Delta$ and $\mathcal{Z}_{\perp}$ explicit:
\begin{align}
    \widetilde{\rho}(\{v\}) &= - \frac{(p - 2) 2^{p-1} \Delta(\{v\})}{(2\pi)^{p(N-p)}} \left(\frac{(4 \pi \alpha p!)^{p}}{(\nu^2)^{p(p-1)}}\right)^{\frac{N - p}{2}} \sqrt{\frac{\alpha p!}{\pi}} e^{-\frac{\alpha p!}{(\nu^2)^{p-2}}}\ \mathcal{Z}_{\perp}(\nu^2,N)  \nonumber\\
    &\quad\times \prod_{i < j} \delta(g^{ij}) \prod_{i=2}^{p} \delta(g^{ii} - g^{i-1\,i-1}) \prod_{i < j} \delta(G^{i,j})
    \label{eq:unsimple-signed-distribution}
    \,,
\end{align}
which, simplifying and rescaling $\delta$-functions, becomes:
\begin{align}
    \widetilde{\rho}(\{v\}) &= - (p - 2) \Delta(\{v\}) \left(\frac{\alpha p!}{\pi}\right)^{\frac{1}{2}(p(N-p) + 1)} \nu^{-p(p-1)(N-p) - (p+1)(p-1)} e^{-\frac{\alpha p!}{(\nu^2)^{p-2}}} \nonumber\\
    &\quad\times \mathcal{Z}_{\perp}(\nu^2, N) \prod_{i < j} \delta(\hat{v}^{i} \cdot \hat{v}^{j}) \prod_{i=2}^{p} \delta(|v^{i}| - |v^{i-1}|) \prod_{i < j} \delta(G^{i,j})
     \label{eq:rhotildebare}
     \,,
\end{align}
where by setting $v^{i} = |v^{i}| \hat{v}^{i}$, we separate vector magnitude and direction. Above, in \eqref{eq:rhotildebare}, we keep using $\nu = |v^{1}| = |v^{2}| = |v^{3}|$ to highlight the parts of the action that only depend on the vector norm. Also, it is worth to point out that the $\beta$ parameter dependence in $\tilde{\rho}$ can be cancelled. In fact, the distribution should not depend on arbitrary gauge-fixing parameters and we expect this to happen.

\subsection{Gauge-Invariant Expression and Norm Distribution}
\label{sec:gaugeinv}
Now, we can use \eqref{eq:rhotildebare} to extract the density for the eigenvector norm, corresponding to the signed distribution of inverse eigenvalues of the random tensors. The strategy is to first obtain a gauge invariant expression for \eqref{eq:rhotildebare} using a DeWitt-Faddeev-Popov method (see e.g. \cite{Weinberg_1996}), then integrate the $pN$-dimensional density of the vector components to the $1$-dimensional density of their norm. Firstly, we define the gauge-invariant density by smearing the signed density over gauge orbits of $\text{SO}(p)$, integrating with respect to a Haar measure on the group:
\begin{equation} \label{eq:inv-signed-density-haar-def}
    \tilde{\rho}_{\text{inv}}(\{v\}) = \frac{1}{|\text{SO}(p)|} \int_{\text{SO}(p)} d\Omega\, \tilde{\rho}(\Omega \{v\})
    \,,
\end{equation}
where we are using the shorthand notation for the $\text{SO}(p)$ transformation:
\begin{equation}
    \Omega \{v\} := \Omega (v^{1}, \dots, v^{p}) := (\Omega^{1i}v^{i}, \dots, \Omega^{pk}v^{k}) \,.
\end{equation}
The translation invariance of the Haar measure then explicitly gives us that:
\begin{equation}
    \tilde{\rho}_{\text{inv}}(\Omega \{v\}) = \tilde{\rho}_{\text{inv}}(\{v\}), \quad \forall \Omega \in \text{SO}(p)
    \,.
\end{equation}

Let us make the dependence of the gauge-fixing variables $G^{i,i'}$ on the components of the vectors $v^{i}$ explicit, as well as collecting the multi-index $I = (i,i')$, $i<i'$. Then, we can split the density of \eqref{eq:rhotildebare} and define $\hat{\rho}$ by:
\begin{equation} \label{eq:gauge-fixes-rho-split}
    \widetilde{\rho}(\{v\}) =: \hat{\rho}(\{v\}) \prod_{I} \delta(G^{I}(\{v\}))
    \,.
\end{equation}
We begin with the following manipulation using properties of the Haar integral.
Through the smearing in \eqref{eq:inv-signed-density-haar-def}, we assign a non-zero density to all tuples $\{v\}$ that can be reached from a gauge-fixed solution to the permuted eigenvalue problem $\{v_{0}\}$ by an $\text{SO}(p)$ transformation, which we know will all be solutions of \eqref{eq:eigenproblem-condensed-form}. On the other hand, if $\{v\}$ solves \eqref{eq:eigenproblem-condensed-form} then, given that the gauge-fixing correctly removes continuous symmetries and we restrict to one of the sectors equivalent under the residual discrete symmetry, such a $\{v_{0}\}$ satisfying the gauge-fixing conditions and representing the gauge orbit is determined uniquely. Let $\Omega_{0} \in \text{SO}(p)$ be the ($\{v\}$-dependent) transformation relating the two. Then, on the support of $\tilde{\rho}_{\text{inv}}$:
\begin{align}
    \tilde{\rho}_{\text{inv}}(\{v\}) = \frac{1}{|\text{SO}(p)|} &\int d\Omega\, \tilde{\rho}(\Omega \{v\}) \nonumber \\
    \overset{\Omega \to \Omega \Omega_{0}}{=} \frac{1}{|\text{SO}(p)|} &\int d\Omega\, \hat{\rho}(\Omega \{v\}) \prod_{I} \delta(G^{I}(\Omega \{v\})) \nonumber \\
    = \frac{1}{|\text{SO}(p)|} &\int d\Omega\, \hat{\rho}(\Omega \{v_{0}\}) \prod_{I} \delta(G^{I}(\Omega \{v_{0}\})) \label{eq:rho-inv-mid}
    \,.
\end{align}
Now, the gauge-fixing delta functions can be manipulated by expanding the group transformation element $\Omega \in \text{SO}(p)$ in generators $T^{J}$ of its Lie algebra multiplied by coefficients $\theta_{J}$, as well as considering the limiting support of the delta-functions in the small neighbourhood around $v_{0}$. Concentrating on them:
\begin{align}
    \prod_{I} \delta(G^{I}(\Omega \{v_{0}\})) &= \prod_{I} \delta(G^{I}(e^{\theta_{J} T^{J}} \{v_{0}\}))  \nonumber\\
    &= \prod_{I} \delta(G^{I}(\{v_{0}\} + \theta_{J} T^{J} \{v_{0}\} + \mathcal{O}(\theta^{2}))) \nonumber \\
    &= \prod_{I} \delta\Big(\underbrace{G^{I}(\{v_{0}\})}_{= 0} + \frac{\partial G^{I}}{\partial v^{i}_{a}} \Big|_{\{v\} = \{v_{0}\}} (\theta_{J} T^{J} \{v_{0}\})^{i}_{a} + \mathcal{O}(\theta^{2})\Big) \nonumber \\
    &= \prod_{I} \delta\Big(\theta_{J} \frac{\partial G^{I}}{\partial v^{i}_{a}} \Big|_{\{v\} = \{v_{0}\}} (T^{J} \{v_{0}\})^{i}_{a} \Big) \nonumber \\
    &= \frac{1}{|\Delta_{0}(\{v_{0}\})|} \prod_{I} \delta(\theta_{I}) \label{eq:delta-g-lie-expand}
    \,,
\end{align}
with:
\begin{equation}
    \Delta_0(\{v_{0}\}) := \det \Big(\frac{\partial G^{I}}{\partial v} \Big|_{\{v\} = \{v_{0}\}} \cdot T^{J} \{v_{0}\}\Big) \,,
\end{equation}
where the determinant is computed in the multi-index sense with $I = (i,j)$ and $J = (i',j')$ (with $i < j, i' < j'$) and the dot product stands for $i, a$ index contractions. By using an explicit multi-indexed form of the generators of $\text{SO}(p)$ \footnote{The convention that we take here has an effect on our volume of $\text{SO}(p)$ in the later equation \eqref{eq:convention-sop-volume}.}, we can compute an element of the multi-index matrix of which the determinant is taken:
\begin{align}
    \frac{\partial G^{I}}{\partial v} \Big|_{\{v\} = \{v_{0}\}} \cdot T^{J} \{v_{0}\} &= \frac{\partial G^{i,j}}{\partial v^{k}_{a}} \Big|_{\{v\} = \{v_{0}\}} T^{i'j' k l} (v_{0})^{l}_{a} \nonumber \\
    &= \frac{\partial G^{i,j}}{\partial v^{k}_{a}} \Big|_{\{v\} = \{v_{0}\}} (v_{0})^{l}_{a} (\delta^{i'k} \delta^{j'l} - \delta^{i'l} \delta^{j'k}) \nonumber \\
    &= \frac{\partial G^{i,j}}{\partial v^{i'}_{a}} \Big|_{\{v\} = \{v_{0}\}} (v_{0})^{j'}_{a} - \frac{\partial G^{i,j}}{\partial 
    v^{j'}_{a}
    } \Big|_{\{v\} = \{v_{0}\}} (v_{0})^{i'}_{a} \nonumber\\
    &= \big(\widetilde{G}^{ij\, i'j'} - \widetilde{G}^{ij\, j'i'}\big) \big|_{\{v\} = \{v_{0}\}} \label{eq:recognise-gtilde-v0}
    \,,
\end{align}
where in the last line \eqref{eq:recognise-gtilde-v0} we recognise that this is the same as \eqref{eq:big-gtilde-def} evaluated at the representative in the $\text{SO}(p)$ orbit of $\{v\}$ that satisfies the gauge condition. Therefore:
\begin{equation}
    \Delta_{0}(\{v_{0}\}) = \Delta(\{v_{0}\}) \quad \Rightarrow \quad \pm |\Delta_{0}(\{v_{0}\})| = \Delta(\{v_{0}\})\,
    \label{eq:pmdeltaamb}
\end{equation}
where the adequate sign is to be determined based on the chosen $\{v\}$ sector between the ones related by the residual discrete symmetry. For the purposes of our results, it is safe for us to ignore such a global sign, as it is related to the specifics of residual gauge-fixing and can be compensated by similarly discarding it from numerical results (see Appendix \ref{app:numsim}). Using \eqref{eq:delta-g-lie-expand} in the gauge-invariant distribution of \eqref{eq:rho-inv-mid}, we obtain:
\begin{align}
    \tilde{\rho}_{\text{inv}}(\{v\}) &= \frac{1}{|\text{SO}(p)|_{\theta}} \int d\theta\, \hat{\rho}((\mathds{1} + \theta_{J} T^{J} + \dots) \{v_{0}\}) \frac{1}{\Delta(\{v_{0}\})} \prod_{I} \delta(\theta_{I}) \nonumber \\
    &= \frac{1}{|\text{SO}(p)|_{\theta}} \frac{1}{\Delta(\{v_{0}\})} \hat{\rho}(\{v_{0}\}) \label{eq:rho-doublehat-def1} \\
    &=: \frac{1}{|\text{SO}(p)|_{\theta}} \hat{\hat{\rho}}(\{v\}) \label{eq:rho-doublehat-def2}
    \,,
\end{align}
where the newly defined $\hat{\hat{\rho}}(\{v\}) $ depends only on gauge-invariant quantities and where we compute the volume of the Haar measure in our parametrisation, represented by the subscript $\theta$, by adapting the result in \cite{website:infn-haar-measures} to our chosen Lie-algebra generators:
\begin{equation} \label{eq:convention-sop-volume}
    |\text{SO}(p)|_{\theta} = \frac{2^{p-1} \pi^{\frac{1}{4}p(p+1)}}{\prod_{k = 1}^{p} \Gamma[\frac{k}{2}]}
    \,.
\end{equation}
Using \eqref{eq:rho-doublehat-def1} and \eqref{eq:rho-doublehat-def2} above together with \eqref{eq:gauge-fixes-rho-split}, we finally extract the gauge invariant signed distribution, which still takes $\{v\}$ as argument but is manifestly only dependent on the vector norms (all equal to $\nu$) and relative vector angles:
\begin{align}
    \tilde{\rho}_{\text{inv}}(\{v\}) &= - \frac{1}{|\text{SO}(p)|_{\theta}} (p - 2) \left(\frac{\alpha p!}{\pi}\right)^{\frac{1}{2}(p(N-p) + 1)} \nu^{-p(p-1)(N-p) - (p+1)(p-1)} e^{-\frac{\alpha p!}{(\nu^2)^{p-2}}} \nonumber\\
    &\quad\times \mathcal{Z}_{\perp}(\nu^2,N) \prod_{i=2}^{p} \delta(|v^{i}| - |v^{i-1}|) \prod_{i < j} \delta(\hat{v}^{i} \cdot \hat{v}^{j})
    \,.
\end{align}

For the norm distribution $\tilde{\rho}(\nu)$, consider an arbitrary region $D \subset \mathbb{R}^{+}$ given by the image of the region $\mathcal{D} \subset \mathbb{R}^{3N}$ under the operation of taking the norm of any vector in the tuple. Then we express the integral of the signed distribution in the region $D$ as:
\begin{align}
   \int_{D} d\nu\, \tilde{\rho}(\nu) &= \int_{\mathcal{D}} dv^{1}\, \dots\, dv^{p}\, \tilde{\rho}_{\text{inv}}(v^{1}, \dots, v^{p}) \nonumber \\
    &= \int_{\mathcal{D}} dv^{1}\, \dots\, dv^{p}\, \tilde{\rho}_{\text{inv}}(v^{1}, \dots, v^{p}) \underbrace{\int_{D} d\nu\, \delta(\nu - |v^{1}|)}_{=1} \nonumber \\
    &= \int_{D} d\nu\, \left(\int dv^{1}\, \dots\, dv^{p}\, \delta(\nu - |v^{1}|) \tilde{\rho}_{\text{inv}}(v^{1}, \dots, v^{p})\right)
    \,.
\end{align}
Given $\mathcal{D}, D$ dependent but arbitrary, we reach the expression for the norm distribution as:
\begin{equation}
    \tilde{\rho}(\nu) = \int dv^{1}\, \dots\, dv^{p}\, \delta(\nu - |v^{1}|) \tilde{\rho}_{\text{inv}}(v^{1}, \dots, v^{p})
    \,.
\end{equation}
To evaluate this integral, we can use generalised spherical coordinates. Define the coordinate system and corresponding Jacobians in $\mathbb{R}^{N}$ as follows:
\begin{gather}
    |v^{i}| \in [0,\infty), \quad \theta^{i}_{1}, \dots, \theta^{i}_{N-2} \in [0,\pi], \quad \theta^{i}_{N-1} \in [0, 2\pi) \\
    d\Theta^{i} = d\theta^{i}_{1}\, d\theta^{i}_{2}\, \dots\, d\theta^{i}_{N-1}\\
    |\det J(\theta^{i})| = (\sin^{N-2}\theta^{i}_{1})(\sin^{N-3}\theta^{i}_{2})\dots(\sin \theta^{i}_{N-2})
    \,.
\end{gather}
Then we can proceed to integrate:
\begin{align}
    \tilde{\rho}(\nu) &= \frac{1}{|\text{SO}(p)|_{\theta}} \int \Big(\prod_{i = 1}^{p} d|v^{i}|\, d\Theta^{i}\, |v^{i}|^{N-1}|\det J(\theta^{i})|\Big)\, \delta(\nu - |v^{1}|)\nonumber\\
    &\quad\quad\quad\quad\quad\quad\times \hat{\hat{\rho}}\big(\{(|v^{i}|, \theta^{i})_{i=1,\dots,p}\}\big) \nonumber\\
    &= - \frac{1}{|\text{SO}(p)|_{\theta}} (p - 2) \left(\frac{\alpha p!}{\pi}\right)^{\frac{1}{2}(p(N-p) + 1)} \nu^{-p(p-1)(N-p) - (p+1)(p-1) +p(N-1)} \nonumber\\
    &\quad\times e^{-\frac{\alpha p!}{\nu^{2(p-2)}}} \mathcal{Z}_{\perp}(\nu^2,N)\int \Big(\prod_{i = 1}^{p} d\Theta^{i}\, |\det J(\theta^{i})|\Big)\, \prod_{i<j} \delta(\hat{v}^{i} \cdot \hat{v}^{j}) 
    \,.
\end{align}

With the last integral amounting to unit hypersphere volumes (convention in Appendix \ref{sec:appendix-conventions}), we obtain the final finite $N$ expression of the signed norm distribution:
\begin{align}
\label{eq:analSignedRho}
    \tilde{\rho}(\nu) = &-\frac{\prod_{i = 1}^{p}S_{N-i}}{|\text{SO}(p)|_{\theta}} (p - 2) \left(\frac{\alpha p!}{\pi}\right)^{\frac{1}{2}(p(N-p) + 1)}\nonumber\\
    &\quad\times \nu^{-p(p-1)(N-p) - (p+1)(p-1) +p(N-1)}e^{-\frac{\alpha p!}{\nu^{2(p-2)}}} \mathcal{Z}_{\perp}(\nu^2,N)
    \,.
\end{align}

\begin{figure}[t!]
    \centering
    \begin{subfigure}[b]{0.5\textwidth}
        \centering
        \includegraphics[height=1.8in]{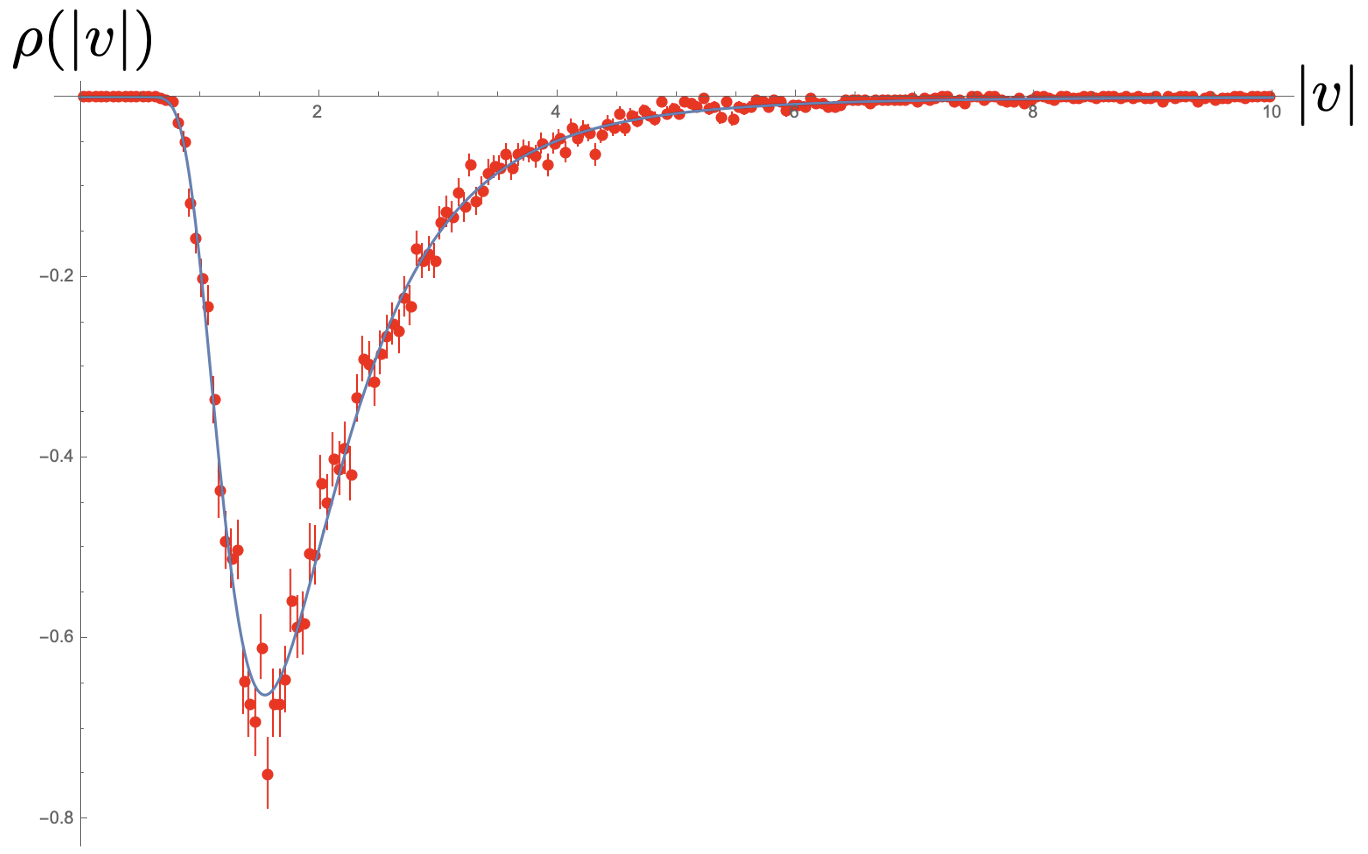}
        \caption{N=4}
    \end{subfigure}%
    ~ 
    \begin{subfigure}[b]{0.5\textwidth}
        \centering
        \includegraphics[height=1.8in]{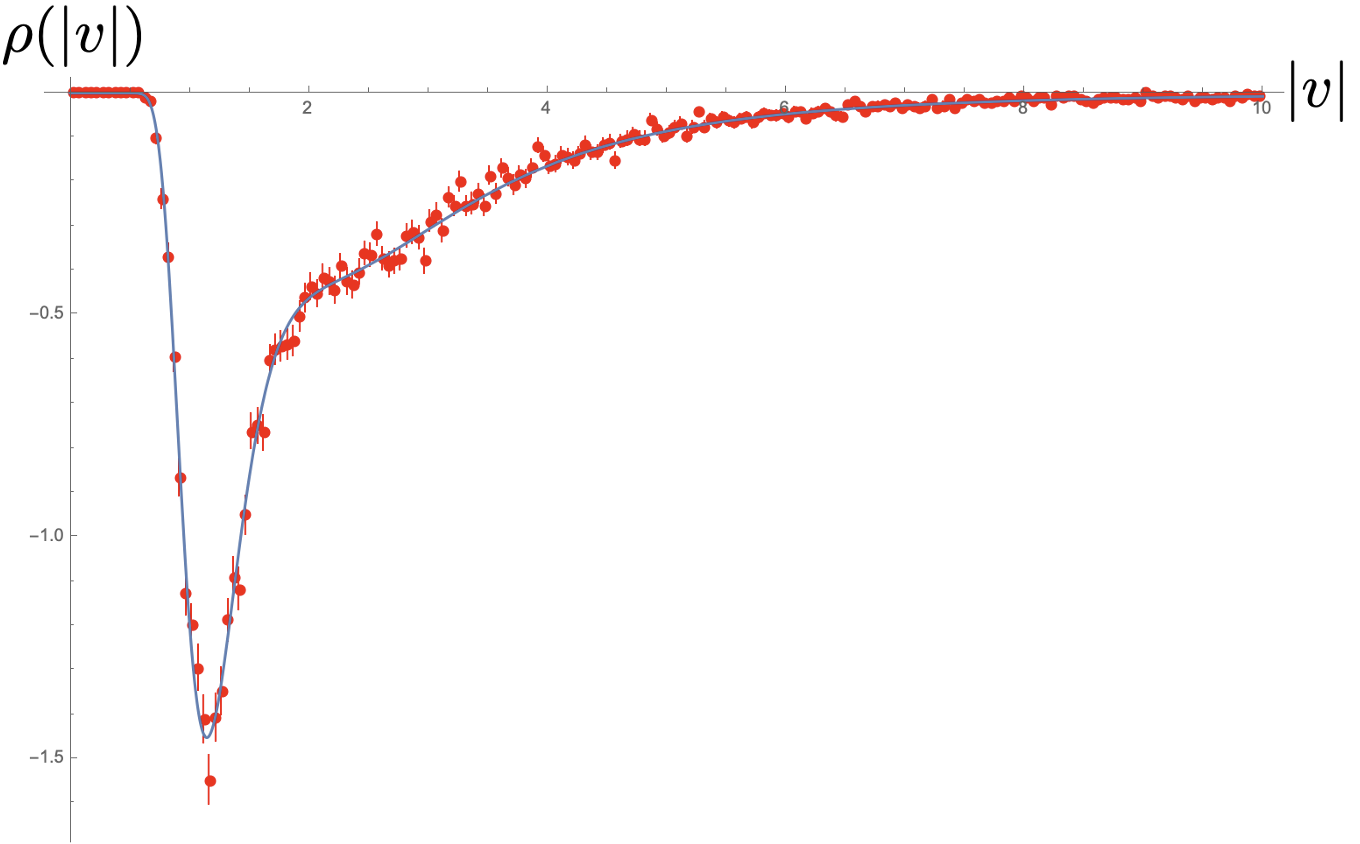}
        \caption{N=5}
    \end{subfigure}
        \centering
    \begin{subfigure}[b]{0.5\textwidth}
        \centering
        \includegraphics[height=1.8in]{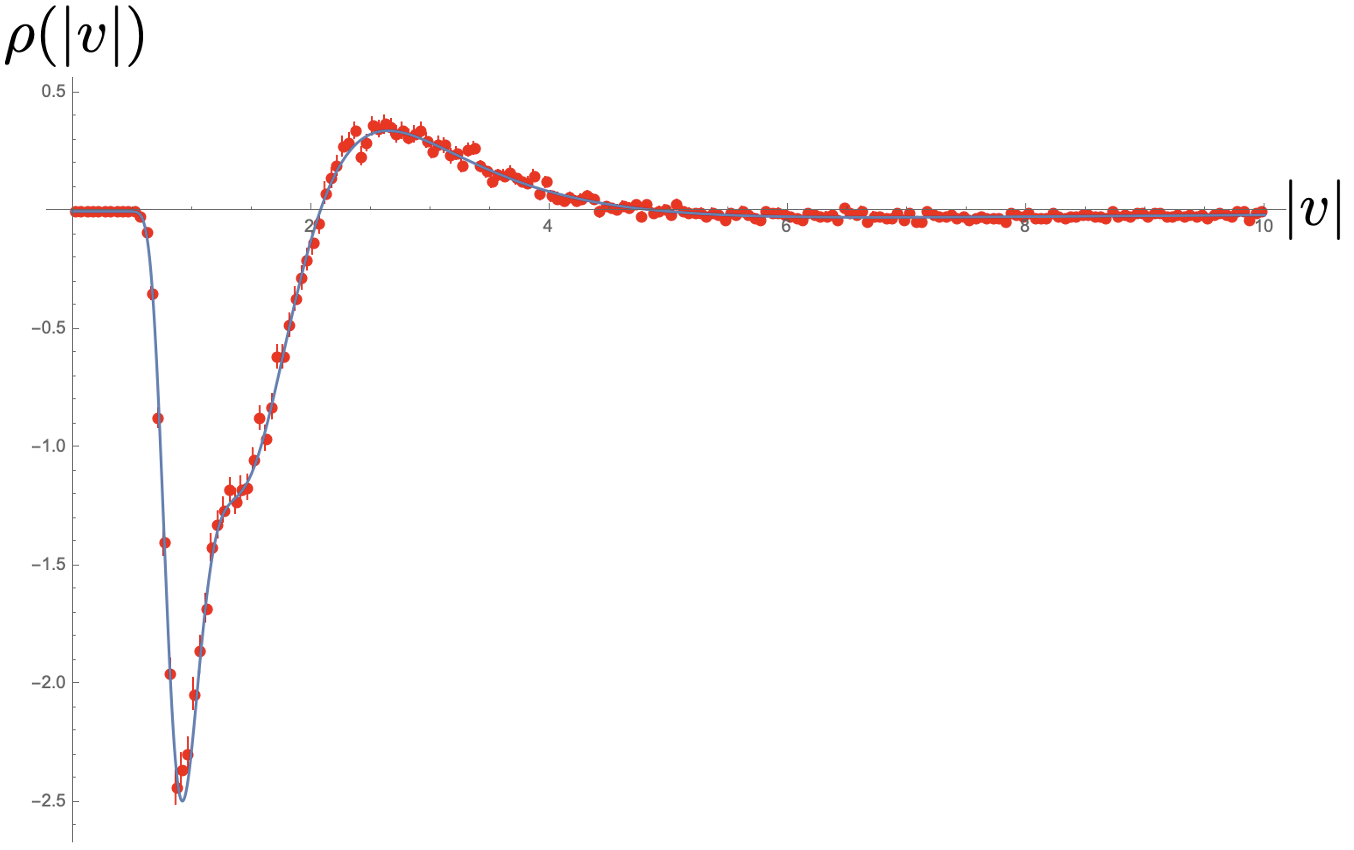}
        \caption{N=6}
    \end{subfigure}%
    ~ 
    \begin{subfigure}[b]{0.5\textwidth}
        \centering
        \includegraphics[height=1.8in]{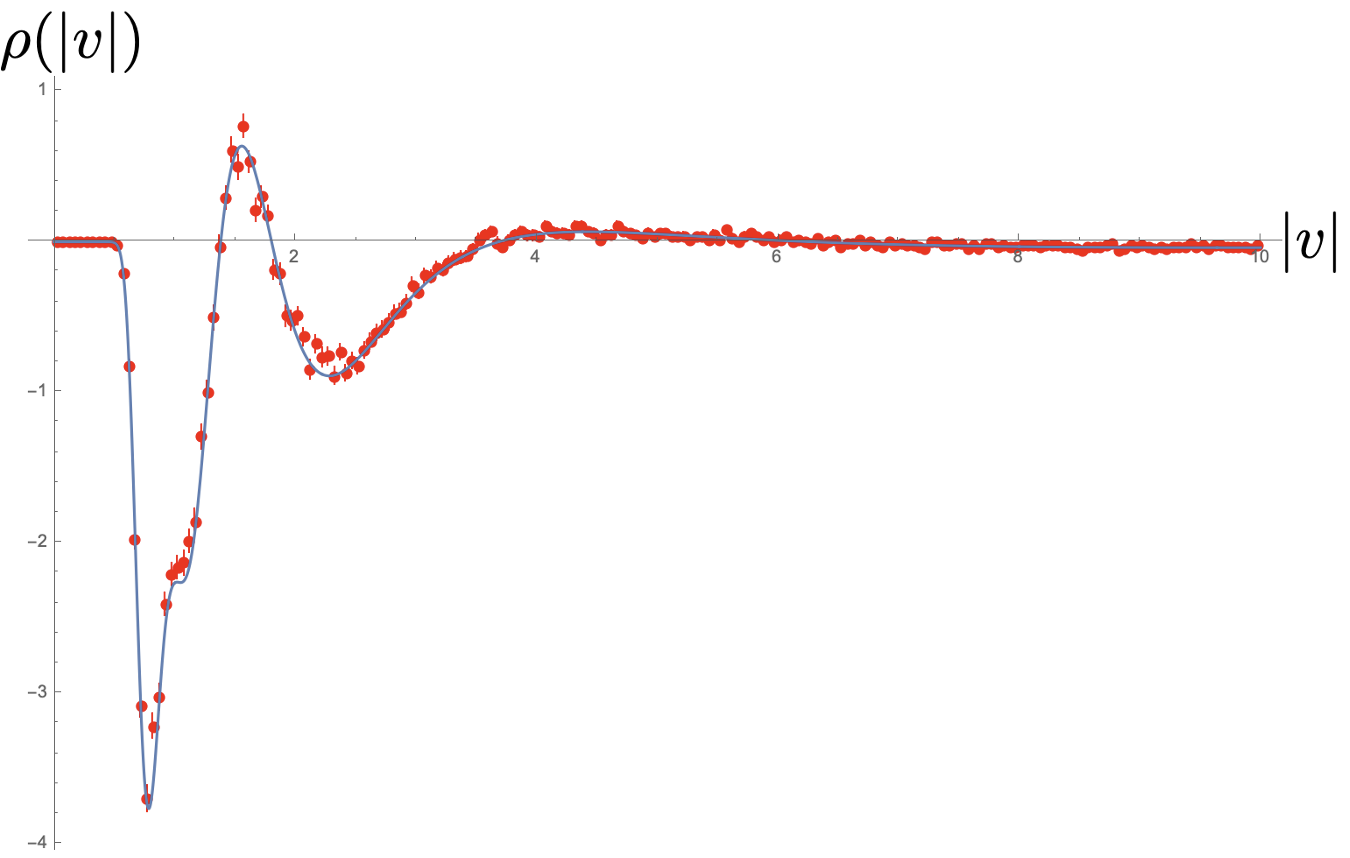}
        \caption{N=7}
    \end{subfigure}
    \caption{Comparison of the analytic expression \eqref{eq:analSignedRho} (in continuous blue) of the signed eigenvalue distributions for $N=4,5,6,7$, with the numerical simulation (in dotted red, with associated numerical errors, done for 10000 iterations)}
\label{fig:signedEigenvaluesp3}    
\end{figure}

\begin{figure}[t!]
    \centering
    \begin{subfigure}[b]{0.5\textwidth}
        \centering
        \includegraphics[height=1.8in]{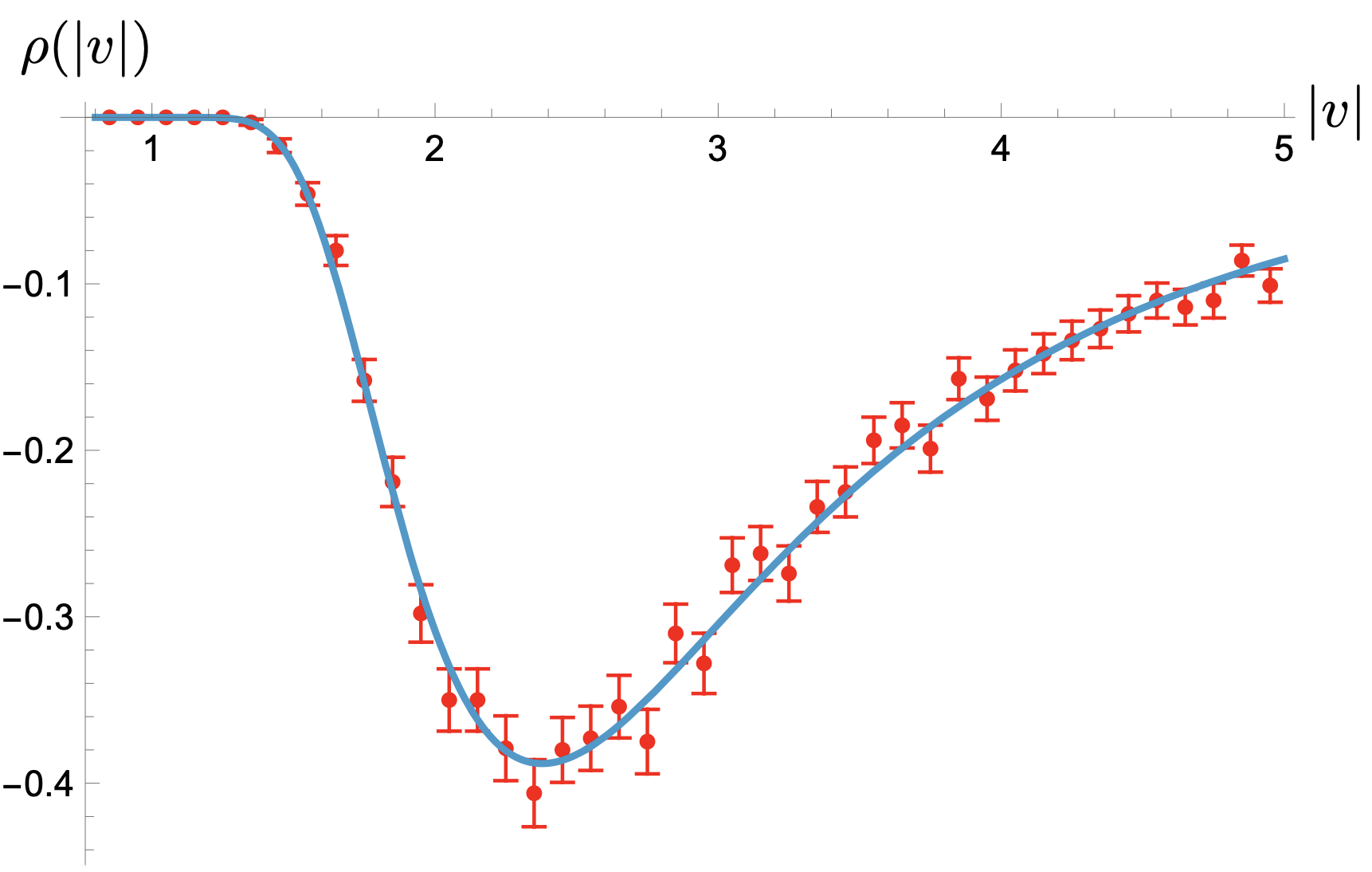}
        \caption{N=4}
    \end{subfigure}%
    ~ 
    \begin{subfigure}[b]{0.5\textwidth}
        \centering
        \includegraphics[height=1.8in]{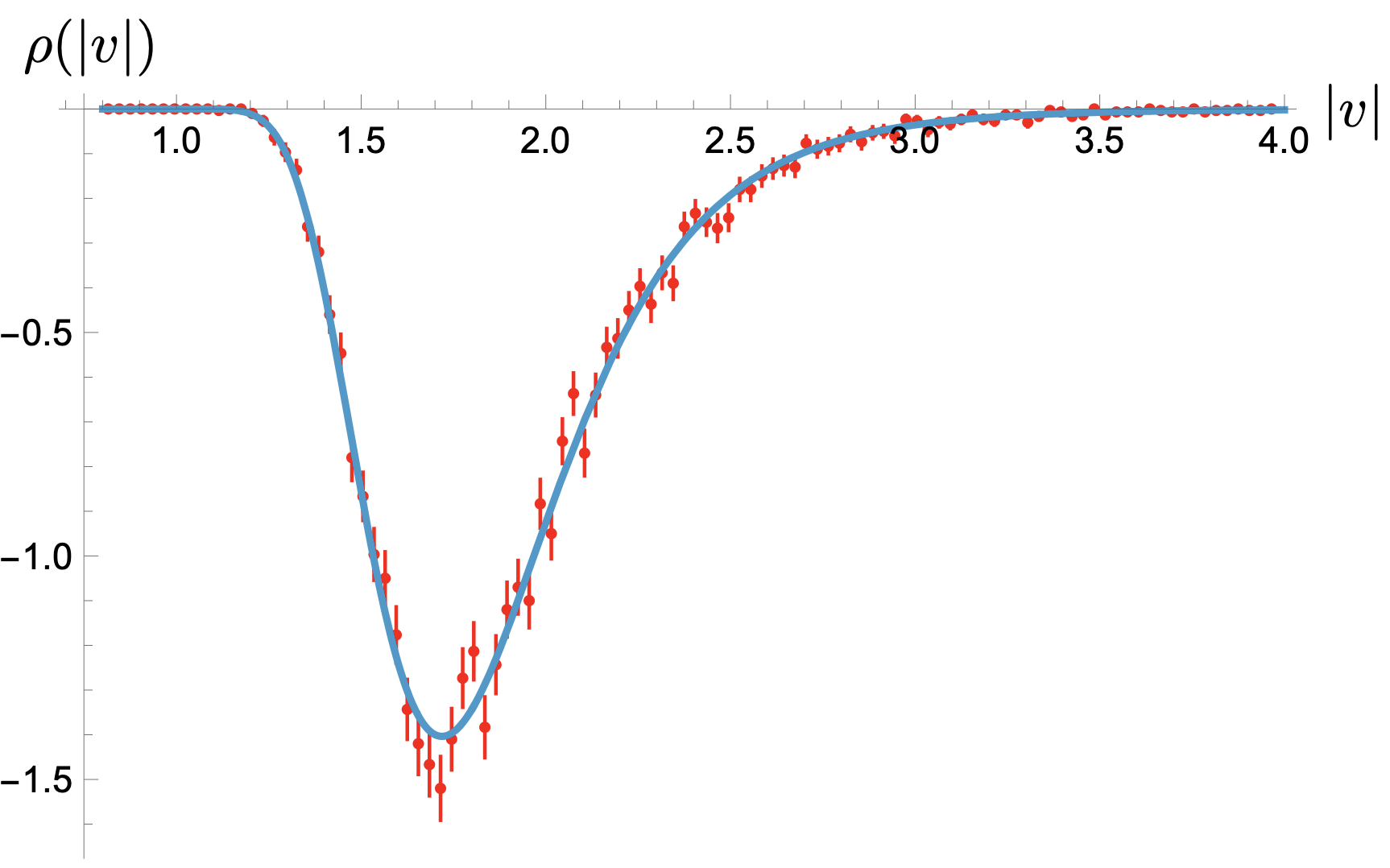}
        \caption{N=5}
    \end{subfigure}
        \centering
    \begin{subfigure}[b]{0.5\textwidth}
        \centering
        \includegraphics[height=1.8in]{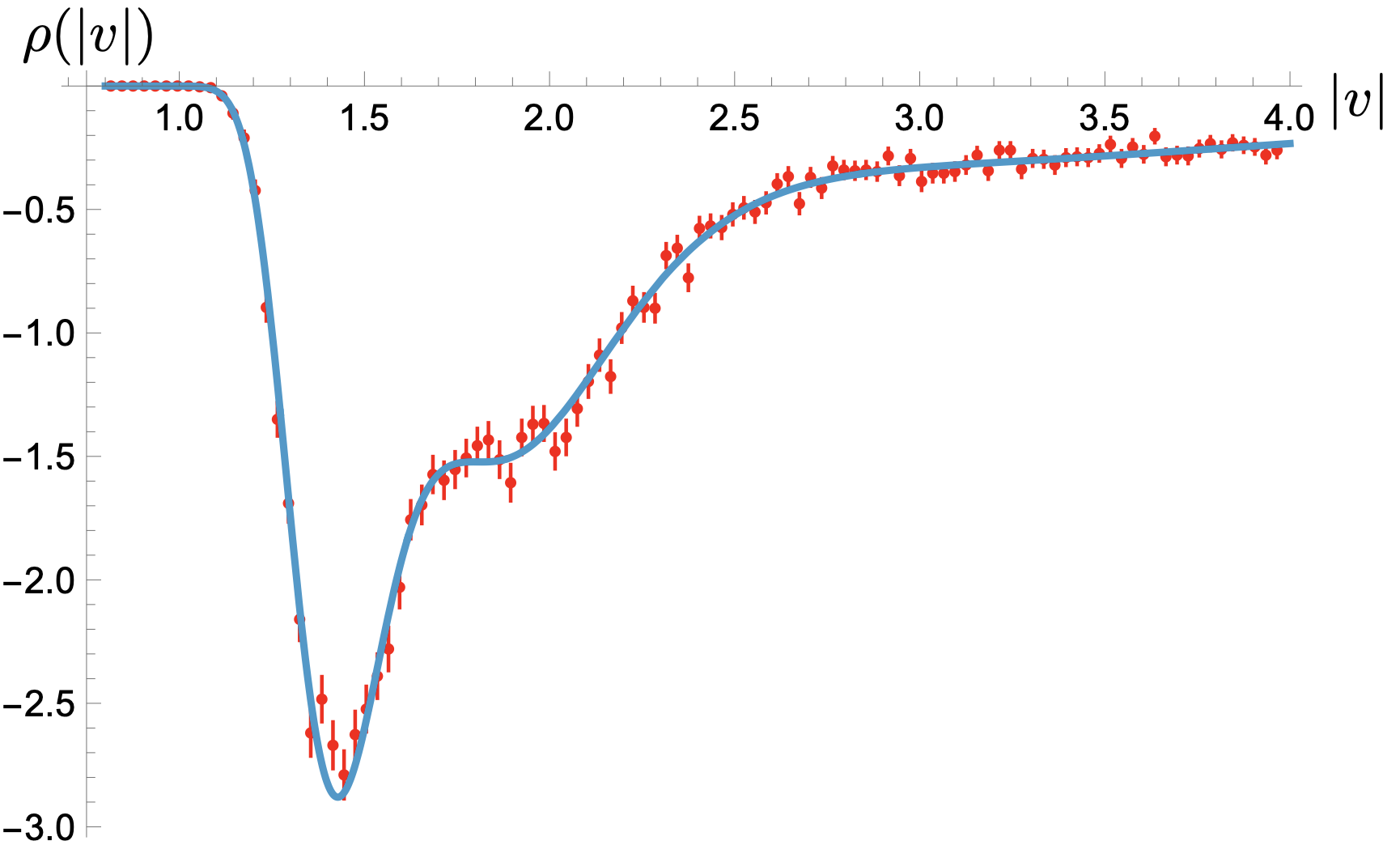}
        \caption{N=6}
    \end{subfigure}%
    ~ 
    \begin{subfigure}[b]{0.5\textwidth}
        \centering
        \includegraphics[height=1.8in]{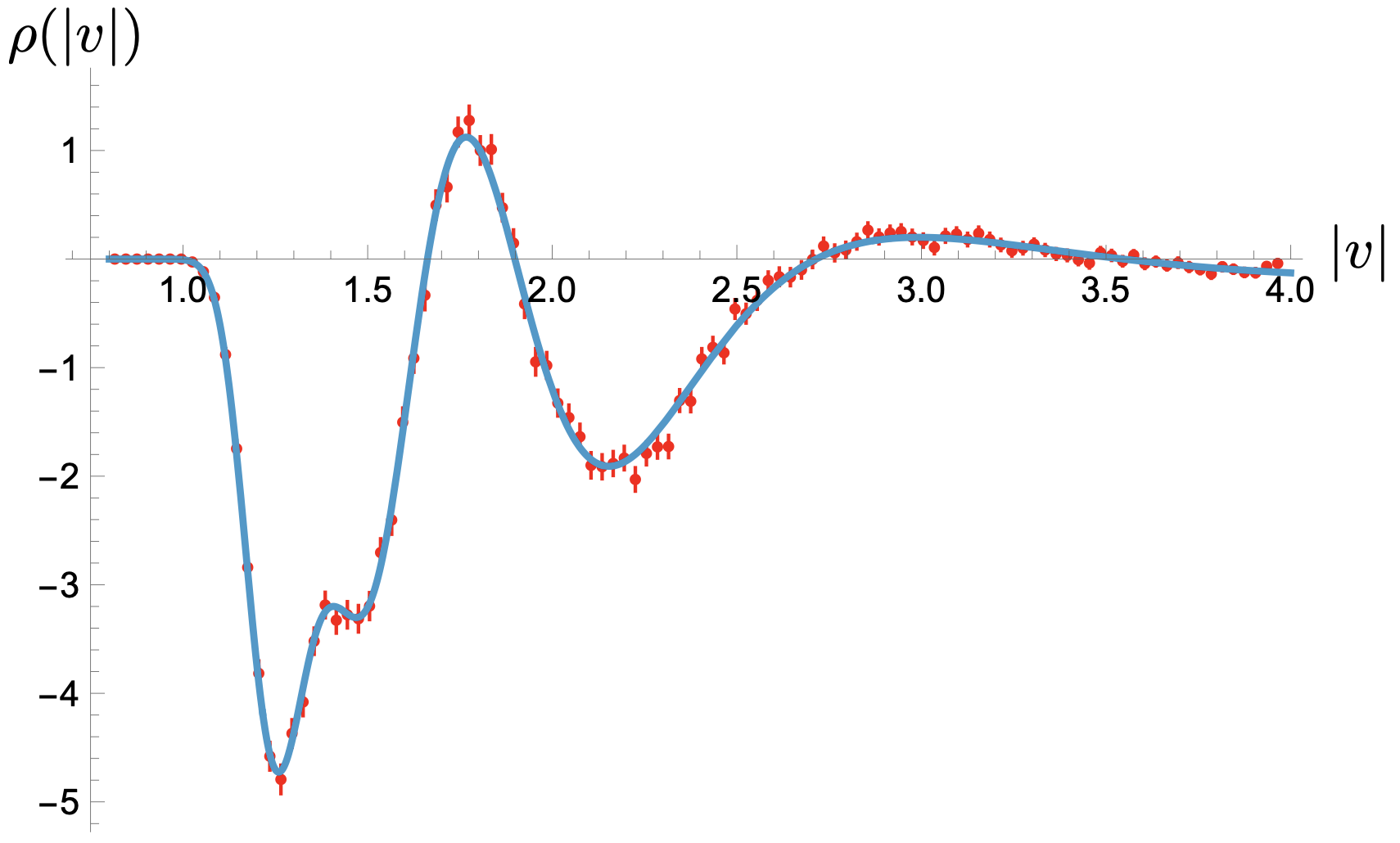}
        \caption{N=7}
    \end{subfigure}
    \caption{Comparison of the analytic expression \eqref{eq:analSignedRho} (in continuous blue) of the signed eigenvalue distributions for $N=4,5,6,7$, and $p=4$ with the numerical simulation (in dotted red, with associated numerical errors, done for 10000 iterations)}
\label{fig:signedEigenvaluesp4}
\end{figure}

We now compare this result with numerical simulations, see Fig.~\ref{fig:signedEigenvaluesp3} for $p=3$ and Fig.~\ref{fig:signedEigenvaluesp4} for $p=4$, and the values $N=4,5,6,7$.  We take $\alpha=1$. We observe that there is essentially perfect agreement with the analytic curves. We have gathered further details about these simulations in the Appendix~\ref{app:numsim}. We make two observations on the shape of the curves. The first is that the curves seem to develop a kink in their first oscillation. The second is that each curve contains a region where the signed distribution quickly drops to zero as $\nu \to 0$. We interpret this behaviour as representing the largest eigenvalue obtained in the limit of $N \to \infty$, as discussed for example in \cite{Kloos:2024hvy}.

\section{The Large $N$ Limit}
\label{sec:signedlargeN}
We proceed to take the large limit $N \to \infty$ of $\tilde{\rho}(\nu)$. Firstly, we introduce our chosen scaling for the vector norm:
\begin{equation}
    \nu = \frac{\tilde{\nu}}{N^{\frac{1}{2(p-2)}}}\,.
\end{equation}
With this scaling, we focus on the regime where $\tilde{\nu}$ is $O(1)$ with respect to $N$. We aim to use the saddle point approximation and to evaluate the limit, therefore we proceed to rewrite all $N$-dependent quantities in terms of exponentials. For the exponentiated constants:
\begin{gather}
    S_{N-i} = \exp\left(\log 2 + \frac{1}{2} (N-i+1) \log \pi - \log \Gamma\Big(\frac{N-i+1}{2}\Big)\right)\,, \\
    \Big(\frac{\alpha p!}{\pi}\Big)^{\frac{1}{2}(p(N-p)+1)} = \exp \left(\frac{pN}{2} \log\Big(\frac{\alpha p!}{\pi}\Big) - \frac{1}{2}(p+1)(p-1) \log\Big(\frac{\alpha p!}{\pi}\Big) \right) \,.
\end{gather}
For the $\nu$-dependent terms:
\begin{gather}
\begin{split}
    \nu&^{-p(p-1)(N-p) - (p+1)(p-1) +p(N-1)} \\
    &= \exp\Big(\frac{1}{2(p-2)}(-p(p-1)(N-p) - (p+1)(p-1) +p(N-1))(\log \tilde{v}^{2(p-2)} - \log N) \Big)\,, 
\end{split}
    \\
    \exp\left(-\frac{\alpha p!}{\nu^{2(p-2)}}\right) = \exp\left(-\frac{\alpha p!}{\tilde{\nu}^{2(p-2)}}N\right) \,.
\end{gather}
Concentrating on the terms involving a $\Gamma$-function, we can apply the logarithm to Stirling's formula to obtain a simplification in the large-$N$ regime\footnote{
We use the common notation $o(a)$ to represent a quantity subleading to $a$ in an asymptotic limit. }:
\begin{equation} \label{eq:gamma-large-n}
    \log \Gamma\Big(\frac{N-i+1}{2}\Big) =\frac{N-i}{2}\log\Big(\frac{N-i+1}{2}\Big) - \frac{N-i+1}{2} + \frac{1}{2}\log 2\pi+o(1) \,.
\end{equation}
By keeping all terms in the expansion of \eqref{eq:analSignedRho} that are of leading order as $N \to \infty$, we obtain the large-$N$ action:
\begin{equation} \label{eq:large-N-constants-contrib}
    \mathcal{S}^{N\gg 1}_{\setminus \psi} = \frac{pN}{2} \log \left(\frac{\alpha p!}{\tilde{\nu}^{2(p-2)}}\right) - \frac{\alpha p!}{\tilde{\nu}^{2(p-2)}}N + \frac{pN}{2} \log2 + \frac{pN}{2} + o(N)\,.
\end{equation}
We note that contributions from the $N \log N$ naive leading order, coming from $\nu$ and $\Gamma$ terms, have exactly cancelled out, leaving the leading order to be linear in $N$.

We now move to the transverse four-fermi contribution, which for the finite-$N$ case we originally chose to compute by means of the exponential derivative operator acting on a quadratic action (equivalently computed by means of Wick contractions), as in \eqref{eq:exponential-derivative-full-formulation}. For large $N$, we instead drop the quadratic action viewpoint in favor of using a Schwinger-Dyson method (see Appendix A in \cite{Sasakura:2022iqd} and \cite{Sasakura:2022axo}) on the action of \eqref{eq:spsi-final-fourfermi} directly. We begin by making some considerations on the symmetries that leave this action invariant. First, we notice that all terms admit the symmetry given by the following scalings by any arbitrary complex parameter:
\begin{equation} \label{eq:unsigned-fermion-gl1c}
    \psi^{\perp i}_{a} \to s \psi^{\perp i}_{a}, \quad \bar{\psi}^{\perp i}_{a} \to s^{-1} \bar{\psi}^{\perp i}_{a}, \quad s \in \text{GL}(1,\mathbb{C}) \,.
\end{equation}
This clearly happens in the quadratic term, but it also happens in the quartic fermion terms because each of them includes two conjugated and two un-conjugated fermions. The complex scaling transformation by the arbitrary $s \in \text{GL}(1, \mathbb{C})$ can be seen as a $\text{U}(1)$ rotation, corresponding to the phase of $c$, representing the compact part of the group transformation, together with the non-compact scaling by a real parameter, corresponding to the norm of $s$. Secondly, the dot-products in \eqref{eq:spsi-final-fourfermi} display an $\text{SO}(N-p)$ symmetry that rotates the transverse fermions within the transverse subspace. Finally, the contractions over $i,j$ indices also display invariance of the action under $\text{SO}(p)$ transformations in this flavour index space. From these symmetries, we propose the following field expectation values to be used in the Schwinger-Dyson method:
\begin{gather}
    \langle \bar{\psi}^{\perp i}_{a} \bar{\psi}^{\perp j}_{b} \rangle = \langle \psi^{\perp i}_{a} \psi^{\perp j}_{b} \rangle = 0 \label{eq:sd-expect-zero}\\
    \langle \bar{\psi}^{\perp i}_{a} \psi^{\perp j}_{b} \rangle = Q\, \delta^{ij} I^{\perp}_{ab} \,, \label{eq:sd-expect-psibarpsi}
\end{gather}
where  the expectation is that of the four-fermi theory 
\eqref{eq:psi-trans-integral-def}, and  
the variable $Q$ is a function of $\tilde{\nu}$ to be determined a posteriori through our application of the saddle point method. 
We want to compute the stationary points of the following effective action, which gives the approximation 
$\mathcal{Z}_{\perp}
=\int dQ \, e^{\widetilde{\mathcal{S}}^{\text{eff}}_{\psi^{\perp}}+o(N)}$ in the large-$N$ asymptotic limit (more explanations on the method can be found in Section~\ref{sec:compoftransgenuine}, where it is used in a more general setting):
\begin{equation}
    \widetilde{\mathcal{S}}^{\text{eff}}_{\psi^{\perp}} := \langle \widetilde{\mathcal{S}}^{\;\setminus\{T,\lambda, \psi^{\parallel}\}}_{\psi^{\perp}} \rangle - (N-p)p \log Q -p(N-p)
\end{equation}
Therefore, using \eqref{eq:sd-expect-zero}, \eqref{eq:sd-expect-psibarpsi} and $\tr I^{\perp} = N - p$, we deal with the large-$N$ effective action of the transverse fermion sector given by:
\begin{align}
    \widetilde{\mathcal{S}}^{\text{eff}}_{\psi^{\perp}} &= (N-p)p Q - \frac{\tilde{v}^{2(p-2)}}{4\alpha p!N} (N-p)^2 p(p-1)Q^2 - (N - p) p \log Q - p(N-p) \nonumber \\
    &= N \big(p Q - \frac{\tilde{v}^{2(p-2)}}{4\alpha p!} p(p-1)Q^2 -  p \log Q - p\big) + o(N) \label{eq:seff-of-q-matrix}.
\end{align}
The saddle point equation $\left. \frac{\partial \mathcal{S}^{\text{eff}}_{\psi^{\perp}}}{\partial Q} \right | _{Q=q}= 0$ implies 
\begin{equation}
    q - \xi  q^2 - 1 = 0, \quad \xi := \frac{\tilde{\nu}^{2(p-2)}}{2\alpha p(p-2)!} \,,
    \label{eq:defofxifirst}
\end{equation}
yielding the solutions:
\begin{equation}
    q = \frac{1 \mp \sqrt{1 - 4\xi}}{2\xi} \,.
\end{equation}
In this form, we can see that the values of $q$ respect two different regimes depending on a critical value of $\xi$, namely $\xi_{c} := 1/4$. Firstly, when $\xi \le \xi_{c}$ we have two real solutions to the saddle point equation. A requirement of a consistent
solution is that for $\xi=0$ the four-fermi theory is a free theory, and hence $q=1$. This is satisfied only by the solution with a negative sign:
\begin{equation} \label{eq:subcritical-signed-saddle}
    \xi \le \xi_{c}: \quad q= \frac{1 - \sqrt{1 - 4\xi}}{2\xi} \,.
\end{equation}
We notice that the solution $q$ can be expressed as the generating function of Catalan numbers with respect to the 
variable $\xi(\tilde{\nu})$.
On the other hand, when $\xi > 1/4$, we have two solutions to the saddle point equation that are complex conjugates of each other, and both must be taken into account for the reality of $\tilde \rho(\nu)$ (or $\mathcal{Z}_{\perp}$), as we will see below:
\begin{equation} \label{eq:supercritial-signed-conjugate-saddlepoints}
    \xi > \xi_{c}: \quad q_{1} = \frac{1 + \mathbf{i}\sqrt{4\xi - 1}}{2\xi}, \quad q_{2} = \frac{1 - \mathbf{i}\sqrt{4\xi - 1}}{2\xi}\,.
\end{equation}
Now the leading-$N$ expression of $\mathcal{Z}_{\perp}$ is given by reinserting the saddle point solutions into \eqref{eq:seff-of-q-matrix}:
\begin{equation} \label{eq:four-fermi-saddlepoint-qstar-contrib}
    \mathcal{Z}_{\perp} =\sum_{q(\tilde{\nu})}\exp\Big( Np \Big(-\frac{1}{2}\xi q (\tilde{\nu})^2 + q (\tilde{\nu}) - \log q (\tilde{\nu}) - 1\Big) + o(N)\Big)\,.
\end{equation}
From the subcritical real solution \eqref{eq:subcritical-signed-saddle} and the supercritical conjugated solutions of \eqref{eq:supercritial-signed-conjugate-saddlepoints}, we see that the large $N$ expression of the signed distribution can be compactly expressed as:
\begin{equation}
    \tilde{\rho}(\nu) = \text{Re}[e^{Np\, h_{p}(x)+o(N)}] \,,
\end{equation}
where we define $h_{p}(x)$ by collecting \eqref{eq:four-fermi-saddlepoint-qstar-contrib} together with the previous \eqref{eq:large-N-constants-contrib}:
\begin{gather}
    h_{p}(x) := \frac{1}{2} \log (p-1) + \frac{1}{x}\left(-1 + \frac{2}{p} - \sqrt{1-x}\right) + \frac{1}{2} \log x - \log(1-\sqrt{1-x}),
    \label{eq:nhb}
    \\ \quad x := \frac{\xi}{\xi_{c}} = 4\xi \,.
\end{gather}
More discussions of the consequences will be given about this result in Section \ref{sec:universality}.

\section{The Genuine Distribution}
\label{sec:genuine}
\def\CR{\nonumber \\} 
\def\eq#1{(\ref{#1})}
\def\s[#1\s]{\begin{align}\begin{split}#1\end{split}\end{align}}
\def\[#1\]{\begin{align}#1\end{align}}
\def\bpsi{{\bar \psi}}
\def\sig{\sigma}
\def\nmS{{\mathcal S}}
\def\ns{{{\mathcal S}_{\rm BF}}} 
\def\nstot{{{\mathcal S}_{\rm tot}}}
\def\ntf{{\tilde f}}
\def\ntJ{{\tilde J}}
\def\nG{\Gamma}
\def\nGs{{\Gamma^{\rm sym}}}
\def\ntv{{|\tilde v|}}
\def\nseff{{{\mathcal S}^{\rm eff}}}
\def\nseffq{{{\mathcal S}^{\rm eff}(Q)}}
\def\nqb{Q^B}
\def\nqf{Q^F}
\def\nhp{{h_p}}
\def\nb{x}
\def\nbp{\nb^{\rm edge}_p}
\def\dels{\delta_{\rm susy}^\kappa}
\def\delps{\tilde\delta_{\rm susy}}
\def\nDelta{\Delta}
\def\ev{z}

\subsection{Setup}
In this section we will compute the genuine distribution \eqref{eq:genuine-distr}. Due to a quartic fermionic and bosonic theory, we will not be able to give a closed expression of the distribution for finite $N$, but we will still be able compute its large $N$ limit. The strategy is basically 
the same as that was first taken in \cite{Sasakura:2022iqd,Sasakura:2022axo} 
and shares much of the procedure with the signed distribution
of the previous sections.
Again, the order of the tensor $T$ is taken to be a general value $p \geq 3$. The eigenvector equation is 
given by $f^{i}_{a}=0$ with \eq{eq:fia-def}.
The gauge-fixing terms can be introduced in the same manner as in the signed case, replacing $f$ with $\tilde f$ in \eq{gauge-f-tildes}. The Jacobian matrix is given by \eq{eq:new-explicit-dftilde}, which is (introducing more convenient notation):
\[
\tilde J_{i a\, j b} :=\frac{\partial \tilde f^{i}_{a} }{\partial v^{j}_{b}} = J_{i a\, j b}+v^{i'}_{a} \frac{\partial G^{i,i'}}{\partial v^{j}_{b}}
\,,
\label{eq:njacobi}
\]
with $G^{i,j}$ being the gauge-fixing conditions of Section \ref{sec:gauge-fixing-procedure} taking non-zero forms only for $i<j$, and:
\[
J_{i a\, j b}=\delta^{i j} \delta_{a b}-\frac{1}{(p-2)!} 
\epsilon^{i j i_{3} \dots i_p} T_{a a_2\cdots a_p } v^{i_3}_{a_3 }v^{i_4}_{a_4}\cdots v^{i_p}_{a_p }\,.
\label{eq:njacobifree}
\]

The major difference from the computations of the signed distribution is that we 
take the absolute value of the determinant of $\tilde J$ as:
\[
\left| \det \ntJ \right|=\lim_{\epsilon, \tilde \epsilon \rightarrow +0}
 \frac{\det \left( \ntJ{}^2+\tilde \epsilon \mathds{1} \right)}{\sqrt{\det \left(\ntJ^T \ntJ +\epsilon \mathds{1}
\right)}}=\lim_{\epsilon,\tilde \epsilon \rightarrow +0} 
\frac{\det \left( \ntJ+\mathbf{i} \sqrt{\tilde \epsilon} \,  \mathds{1}  \right)\det \left( \ntJ-\mathbf{i} \sqrt{\tilde \epsilon}  \,  \mathds{1}  \right)}{\sqrt{\det \left(\ntJ^T \ntJ +\epsilon 
\mathds{1}
\right)}},
\label{eq:detabs}
\]
where  $\mathds{1}$ denotes an identity matrix, 
the superscript $T$ the matrix transpose, and $\epsilon,\tilde \epsilon$ denote positive regularisation parameters.
Note that, as shown in \eq{eq:njacobi}, $\ntJ$ and $\ntJ^T$ are different because of the gauge fixing-term.
Here $\epsilon$ avoids the divergence which occurs when $\tilde J^T  \tilde J$ have zero eigenvalues. 
As for $\tilde \epsilon$, we will later find that $\tilde \epsilon$ is needed to uniquely determine the solution of the Schwinger-Dyson equations in 
the strong coupling regime of the quantum field theory
in the large $N$ limit\footnote{We will find a phase transition point, which separates the weak and strong coupling 
regimes of the quantum field theory in the large $N$ limit, as was found in Section~\ref{sec:signedlargeN}. }.

The rightmost side of \eq{eq:detabs} can be expressed in a field theoretical manner that is more complicated than the simple fermionic determinants of \eqref{eq:signed-unfixed-action} and \eqref{eq:rho-stilde-rel}.
The numerator can be expressed by 
introducing two fermion pairs $\bar{\psi}^{\kappa i}_{a}, \psi^{\kappa i}_{a}$ for $\kappa = 1,2$ and the denominator by two bosons
$\phi^{i}_{a}$ and $\sigma^{i}_{a}$, with $i \in \{1, \dots, p\}, a \in \{1, \dots, N\}$, 
as (see \ref{app:convfieldsints} for the integration measures):
\[
\left | \det \ntJ \right |= \int \mathcal{D}\bpsi \mathcal{D}\psi \mathcal{D} \phi \mathcal{D}\sigma e^\ns
\,, 
\label{eq:njtilde}
\]
where:
\[
\ns=-\phi^2 +2 \mathbf{i} \phi \ntJ \sig -\epsilon \sigma^2+\bpsi^1 \ntJ \psi^1+\bpsi^2 \ntJ \psi^2
+ \mathbf{i} \sqrt{\tilde \epsilon}\, \bpsi^1 \psi^1- \mathbf{i} \sqrt{\tilde \epsilon}\, \bpsi^2 \psi^2
\,,
\label{eq:ns}
\]
with $\phi,\sigma$ standing for the bosons and $\bpsi^1,\psi^1,\bpsi^2,\psi^2$
for the two fermion pairs. 
Here the vector and flavour indices are suppressed for simplicity,
being properly contracted in this expression, 
and they will be recovered, whenever necessary in later discussions.
The limits $\epsilon, \tilde \epsilon \rightarrow +0$ have also been suppressed for notational simplicity, but they are assumed to be taken in the end. 
Note that \eq{eq:njtilde} with \eq{eq:ns} is invariant under the rescaling transformations,
\[
\bpsi^\kappa \rightarrow s_\kappa \bpsi^\kappa,\psi^\kappa \rightarrow s_\kappa^{-1} \psi^\kappa, \quad s_{\kappa} \in \text{GL}(1,\mathbb{C}), \
\ (\kappa=1,2)
\label{eq:noncomp}
\]
with arbitrary complex numbers $s_\kappa\ (\kappa=1,2)$.  
 $\ns$ with $\epsilon,\tilde \epsilon=0$ is also invariant under the following two supersymmetries:
 \[
\dels \sig=\frac{\mathbf{i}}{2} \psi^\kappa, \ \dels \bpsi^\kappa=\phi,\ \dels(\hbox{others})=0,
\label{eq:susy}
 \]
where $\kappa=1,2$ is not being summed over.
 
The eigenvector equations and the averaging over the random tensor can be implemented exactly
in the same way as in the signed case of \eqref{eq:signed-unfixed-part-funct}, 
and we obtain:
\[
\rho(\{v\})
=\int \mathcal{D}T\, \mathcal{D}\lambda\, \mathcal{D}\Psi\,
e^\nstot,
\label{eq:nrhov}
\]
where:
\begin{align}
\nstot
&
:= -\alpha T^2 + \mathbf{i} \lambda \ntf + \ns
\nonumber \\
&=-\alpha T^2 + \mathbf{i} \lambda \ntf -\phi^2+ \nG_{\alpha \beta} \Psi^\alpha \ntJ \Psi^\beta
-\epsilon \sig^2 + \mathbf{i} \sqrt{\tilde \epsilon} \, \bpsi^1 \psi^1- \mathbf{i} \sqrt{\tilde \epsilon} \, \bpsi^2 \psi^2.
\label{eq:nstot}
\end{align}
Here we have collected all the fields into one field 
$\Psi:=(\phi,\sig,\bpsi^1,\psi^1,\bpsi^2,\psi^2)$ 
indexed by $\alpha \in \{1, \dots, 6\}$ that mixes fermions and bosons for convenience, 
and $\nG$ is the following block diagonal matrix:
\[
\nG:=
\left( 
\begin{matrix}
\nG_B &0&0\\0&\nG_F&0 \\ 0&0&\nG_F
\end{matrix}
\right),
\label{eq:ngamma}
\]
where:
\[
\nG_B:=\left( \begin{matrix} 0 &2 i  \\ 0 & 0 \end{matrix} \right),\ \ 
\nG_F:=
\left( \begin{matrix}0 & 1 \\ 0 & 0\end{matrix} \right).
\label{eq:ngammabf}
\]

\subsection{Integration Over $T$}
Let us first look at the coupling terms of $T$ with the fields.
They are contained in $\Psi^\alpha J \Psi^\beta$, as shown in \eq{eq:njacobifree}.
To compute $\Psi^\alpha J \Psi^\beta$, let us first decompose 
$\Psi$ into its parallel and transverse parts against $v$ as was previously performed in Section~\ref{sec:split}:
\begin{equation}
    \Psi^\alpha = I^{\parallel} \Psi^{\alpha} + I^{\perp} \Psi^{\alpha} =: \Psi^{\parallel \alpha}+\Psi^{\perp \alpha}
    \,.
\end{equation}
With the flavour and vector indices being explicit, the transverse part satisfies
$\Psi_{a}^{\perp \alpha i} v^j_{a}=0\ \forall i,j$, and the parallel part can be
expanded in terms of $v^i_a$, namely, $\Psi^{\parallel \alpha i}_{a} =\Psi^{\alpha ii'} v^{i'}_a$, 
by introducing $\Psi^{\alpha ii'}$. Then, as shown in Appendix~\ref{app:derpjp}, 
we find:
\s[
\Psi^\alpha J \Psi^\beta=& \nu^2 \Psi^{\parallel \alpha i i'} \left(
\delta^{ij}\delta^{i'j'}-\delta^{ii'}\delta^{jj'}+\delta^{i j'}\delta^{i'j} \right) \Psi^{\parallel \beta j j'} \\
&+ \Psi_{a}^{\perp \alpha i} \Psi_{a}^{\perp \beta i}  -
\frac{1}{(p-2)!} \epsilon^{i_1 i_2\cdots i_p} T_{a_1 a_2\cdots a_p}
\Psi_{a_1}^{\perp \alpha i_1}\Psi_{a_2}^{\perp \beta i_2}v_{a_3}^{i_3} v_{a_4}^{i_4}
\cdots v_{a_p}^{i_p}.
\label{eq:pjp} 
\s]
There are two important properties: the tensor $T$ couples only to the transverse part, and 
the parallel and the transverse parts decouple from each other.
Then from \eq{eq:nstot} and \eq{eq:pjp}, all the terms containing $T$ can be collected as:
\[
\nmS_T&:=-\alpha T^2 \CR
&\ -T_{a_1 \cdots a_p} \left \llbracket \frac{\mathbf{i}}{(p-1)!} \epsilon^{i_1 i_2\cdots i_p} \lambda_{a_1}^{i_1} v_{a_2}^{i_2} \cdots v_{a_p}^{i_p}+\frac{1}{(p-2)!} \epsilon^{i_1 i_2 \cdots i_p} \nG_{\alpha \beta} \Psi^{\perp \alpha i_1}_{a_1} \Psi_{a_2}^{\perp \beta i_2} v^{i_3}_{a_3}\cdots v_{a_p}^{i_p} \right\rrbracket
\,,
\label{eq:nst}
\]
which is a generalisation of \eq{eq:tensoraction}.
Here:
\begin{equation}
    {\rho}(\{v\}) 
    = \int \mathcal{D} \lambda\, \mathcal{D}\Psi\,
    \underbrace{e^{\nstot - \nmS_{T}}}_{T\text{-indep.}} \int \mathcal{D}T\, e^{\nmS_{T}}
    \,.
    \label{eq:rhovwitht}
\end{equation}
Then the Gaussian integral over $T$  
adds the square of the term 
$\llbracket \cdots \rrbracket$ of \eq{eq:nst} to the action. 
One important fact is that, in the square, the cross terms vanish because 
of $\Psi_{a}^{\perp \alpha i} v_a^{j}=0$, in exactly the same fashion as in \eqref{eq:tensor-averaging-with-par-perp}. After a short computation, we obtain the result of 
the integration over $T$ as:
\[
\nmS^{\backslash \{T\}}
&=\nstot - \nmS_{T}-\frac{1}{4 \alpha \left( (p-1)! \right)^2}\left \llbracket
\epsilon^{i_1 i_2 \cdots i_p}\lambda_{a_1}^{i_1} v_{a_2}^{i_2} \cdots v_{a_p}^{i_p} \right \rrbracket^2 \CR
&\ \ +\frac{\nu^{2 (p-2)}}{2 \alpha p!} \nGs_{\alpha \beta} \nGs_{\alpha'\beta'} (-1)^{[\alpha'][\beta]}
\delta_{i' j'}^{ij} (\Psi^{\perp \alpha i} \cdot \Psi^{\perp \alpha' i'}) (\Psi^{\perp \beta j} \cdot 
\Psi^{\perp \beta' j'})
\,,
\label{eq:nsnot}
\]
where $[\alpha]$ denotes the grade of the field $\Psi^\alpha$, namely, $[\alpha]=0$ for 
$\alpha=1,2$ and $[\alpha]=1$ for $3\leq \alpha\leq 6$. We defined:
\[
\nGs_{\alpha \beta}:=\frac{1}{2} \left( \nG_{\alpha \beta}+(-1)^{[\alpha][\beta]}\nG_{\beta \alpha}\right)\,,
\]
and we recall the notations $\Psi^{\perp \alpha i} \cdot 
\Psi^{\perp \beta j}=\Psi_{a}^{\perp \alpha i}
\Psi_{a} ^{\perp \beta j}$, $\delta_{j j'}^{ii'} =\delta^{ij}\delta^{i' j'}-\delta^{ij'}\delta^{i' j}$.
The genuine density writes now as follows:
\begin{equation}
    {\rho}(\{v\}) 
    = 
    \int \mathcal{D} \lambda\, \mathcal{D}\Psi\,
    e^{\nmS^{\backslash \{T\}}}
    \,.
\end{equation}

\subsection{Integration Over $\lambda$}
From \eq{eq:nstot} and \eq{eq:nsnot} the terms containing $\lambda$ are collected as:
\[
\nmS_\lambda=-\frac{1}{4 \alpha \left( (p-1)! \right)^2}\left \llbracket
\epsilon^{i_1 i_2 \cdots i_p}\lambda_{a_1}^{i_1} v_{a_2}^{i_2} \cdots v_{a_p}^{i_p} \right \rrbracket^2 
+\mathbf{i} \lambda^{i}_{a} \left( f_{a}^i+ v^{i'}_a G^{ii'} \right)
\,,
\label{eq:nslam}
\]
with:
\begin{equation}
    {\rho}(\{v\}) 
    = 
    \int \, \mathcal{D}\Psi\, \underbrace{e^{\nmS^{\backslash \{T\}} - \nmS_\lambda}}_{T, \,\lambda\text{-indep.}} 
    \int \mathcal{D} \lambda
    \, e^{\nmS_\lambda}
    \,.
    \label{eq:rhovnot}
\end{equation}
Since $\nmS_\lambda$ does not depend on the fields, the integration over $\lambda$ is exactly 
the same as what has been done for the signed case in Section~\ref{sec:lamint}. It is necessary to perform a careful analysis 
taking into account the gauge-fixing terms. The result of Section~\ref{sec:lamint} reads:
\s[
\int \mathcal{D} \lambda \, e^{\nmS_\lambda}= 
& \left( \frac{\alpha\, p!}{\pi}\right)^{\frac{p(N-p)+1}{2}} \nu ^{-p(p-1)(N-p)} e^{-\frac{\alpha p!}{\nu^{2(p-2)}}}\\
&\ \ \ \ \ \times \prod_{i<j} \delta \left( v^i \cdot v^j\right)\prod_{i=2}^{p} \delta \left( |v^{i}|^2 -|v^{i-1}|^2 \right) \prod_{i<j} \delta \left( G^{i,j} \right). 
\label{eq:intlamslam}
\s]

\subsection{Computing the Quantum Field Theory}
\label{sec:nft}

\subsubsection{Computation of the Parallel Part}
The integration over $\Psi$ proceeds by first splitting the parallel and transverse components, as was performed in  the signed case
in Section~\ref{sec:split}. The parallel components couple at most quadratically, and the computation proceeds in parallel with that 
of the signed case, including the decoupling of the transverse part in the gauge-fixing sector, as was performed in Section~\ref{sec:psiintegrals}.
Therefore the computation of the parallel part reduces to the computation of some determinants from both the fermions and the bosons. 
The simplicity is that the vanishing limit of the regularisation parameter, $\epsilon,\tilde \epsilon \rightarrow +0$, is regular
in the computation, and the final result is the same as \eq{eq:psi-par-integrated} except for the overall positivity, namely, 
\[
\int \mathcal{D}\Psi^{\parallel}\,  e^{\nmS^{\backslash \{ T,\lambda\}}_{\Psi^\parallel}}=2^{p-1} (p-2) \nDelta,
\label{eq:intpsiparafinal}
\] 
where:
\[
\nDelta:=\det_{i<j,i'<j'} \left( \widetilde{G}^{ij\, i'j'}-\widetilde{G}^{ij\, j'i'} \right)\,,
\]
and:
\[\widetilde{G}^{ij\, i'j'}=\frac{\partial G^{ij}}{\partial v^{i'}_a } v^{j'}_a
\,.
\]
This identical result with the signed case (except for the overall sign) can be attributed to the fact that the extra contributions coming 
from the extra fermions and bosons,
compared with the signed case, indeed cancel with each other because of the presence of the supersymmetry \eq{eq:susy}
in the case that the limit $\epsilon,\tilde \epsilon\rightarrow +0$ is smooth. 

From \eq{eq:rhovnot}, \eq{eq:intlamslam} and \eq{eq:intpsiparafinal}, we obtain in the same style as  \eqref{eq:rhotildebare}:
\s[
\rho(\{v\})=&(p-2) \nDelta \left( \frac{\alpha p!}{\pi}\right)^{\frac{p(N-p)+1}{2}} \nu^{-p(p-1)(N-p)-p^2+1}
e^{-\frac{\alpha p!}{\nu^{2(p-2)}}} \\
&\ \ \times Z_{\perp} \prod_{i<j} \delta\left( \hat v^i \cdot  \hat v^j \right) \prod_{i=1}^{p-1} \delta\left( |v^i|-|v^{i+1}| \right) \prod_{i<j} \delta\left(G^{i,j}\right)
\,,
\label{eq:nrhovbare}
\s]
where $Z_{\perp}$ is similar to \eqref{eq:psi-trans-integral-def} and is defined by:
\[
Z_{\perp}=\int \mathcal{D}\Psi_\perp e^{\nmS_\perp}
\,,
\label{eq:nzperp}
\]
with:
\s[
\nmS_{\perp}&= -(\phi^{\perp})^2+\nG_{\alpha \beta} \Psi^{\perp \alpha i} \cdot \Psi^{\perp \beta i} -\epsilon (\sigma^{\perp})^2 
+ \mathbf{i} \sqrt{\tilde \epsilon} \, \bpsi^{\perp 1} \psi^{\perp 1}- \mathbf{i} \sqrt{\tilde \epsilon} \, \bpsi^{\perp 2} \psi^{\perp 2} \\
&\ \ \ \ \ \ 
+\frac{\nu^{2 (p-2)}}{2 \alpha p!} \nGs_{\alpha \beta} \nGs_{\alpha'\beta'} (-1)^{[\alpha'][\beta]}
\delta_{i' j'}^{ij} (\Psi^{\perp \alpha i} \cdot \Psi^{\perp \alpha' i'}) (\Psi^{\perp \beta j} \cdot 
\Psi^{\perp \beta' j'}).
\label{eq:nsperp}
\s]
For later convenience, $Z_\perp$ in \eq{eq:nzperp} has been normalised for the free field theory:
\[
Z_\perp=1 \hbox{ for } \nu=0.
\label{eq:nnorm}
\]

The smearing in the gauge direction and the integration over $v$ can be performed in the same 
manner as in the signed case described in Section~\ref{sec:gaugeinv}. We obtain:
\[
\rho(\nu)=\frac{\prod_{i=1}^p S_{N-i} }{|{\rm SO}(p)|_\theta }(p-2)  \left( \frac{\alpha p!}{\pi} \right)^\frac{p(N-p)+1}{2} \nu^{-p(p-1)(N-p)-p^2+1 +p(N-1)} 
e^{-\frac{\alpha p!}{\nu^{2(p-2)}}} Z_\perp
\,,
\label{eq:nrhovzperp}
\]
which is identical with \eqref{eq:analSignedRho} except for the missing sign from equation \eqref{eq:intpsiparafinal} and the quartic theory in both fermions and bosons.

\subsubsection{Computation of the Transverse Part in the Large $N$ Limit}
\label{sec:compoftransgenuine}

So far, the computation of the genuine distribution is essentially identical with that of the signed case. This can be attributed to the fact that 
the $\epsilon,\tilde \epsilon\rightarrow +0$ limit is smooth for the parallel part, and the supersymmetry \eq{eq:susy} holds in this limit.
In other words, one can compute the parallel part by setting $\epsilon,\tilde \epsilon$ to zero from the beginning, in which the supersymmetry \eq{eq:susy} exactly holds and the contributions of the additional bosons and fermions cancel with each other.  

However, the computation of the transverse partition function is more difficult and subtle. In the signed case, the quantum field theory describes fermions coupled quartic couplings, and one can simply expand the fermion four-couplings in series. In the genuine case, the integration over 
the bosons with quartic couplings is difficult to compute. It is not generally possible to obtain the integration in a closed form. In addition, 
computing $Z_\perp$ in perturbations of the four couplings would lead to a wrong result. This can 
been seen in the explicit computation performed by a similar field theoretical method for the eigenvalue/vector distribution of 
the real symmetric random tensor (order-three) in \cite{Sasakura:2022axo}. In that computation, the regularisation parameter was essentially important to avoid the divergence of the integrations until the final result. Moreover, the final result in \cite{Sasakura:2022axo} shows that the partition function contains not only 
the polynomials of $\nu^{2(p-2)}$, but also terms with the non-perturbative factors $e^{-{\rm const.}/\nu^{2(p-2)}}$, that cannot be obtained by a series expansion of \eq{eq:nsperp}. 

Because of the above technical difficulties, we will not try to exactly compute $Z_\perp$, but will 
rather determine its leading asymptotic form for large $N$. 
We will apply the method first used in \cite{Sasakura:2022iqd}, 
which was based on the Schwinger-Dyson equation. 
To avoid unnecessary complications of notations/computations, the difference between $N-p$ and $N$ 
will be ignored for large $N$. In particular, 
the dimension will be pretended to be $N$ and the subscript $\perp$ of the fields will be neglected
in this section. 
We start with some assumptions about the 
two-field expectation values, $\langle {\cal O} \rangle := Z_\perp^{-1}\int \mathcal{D} \Psi {\cal O} e^{\nmS_\perp}$:
\s[
&\langle \phi^{i_1}_{a_1} \phi^{i_2}_{a_2}\rangle=\nqb_1 \delta^{i_1 i_2} \delta_{a_1 a_2}, \\
&\langle \phi^{i_1}_{a_1} \sig_{a_2}^{i_2}\rangle=\langle \sig^{i_1}_{a_1} \phi_{a_2}^{i_2}\rangle=
\nqb_2 \delta^{i_1i_2} \delta_{a_1 a_2},\\
&\langle \sig^{i_1}_{a_1} \sig_{a_2}^{i_2}\rangle=\nqb_3 \delta^{i_1i_2} \delta_{a_1 a_2}, 
\label{eq:nassbos}
\s]
for bosons, and:
\s[
&\langle \bpsi^{\kappa_1 i_1}_{a_1} \psi_{a_2}^{\kappa_2 i_2}\rangle=\nqf_{\kappa_1} \delta^{\kappa_1 \kappa_2}\delta^{i_1 i_2} \delta_{a_1 a_2},\\
&\hbox{Others}=0,
\label{eq:nassfer}
\s]
for fermions, respectively, where the superscript $\kappa_i=1,2$ for the fermions serves to label 
the two pairs of fermions introduced in \eq{eq:ns}, and $\nqb_i,\nqf_i$ are functions of $\nu$
to be determined below. 

These forms have been uniquely fixed by requiring some symmetries on these expectation values. 
Since the SO$(N)$ and SO$(p)$
symmetries, in the vector and flavour index spaces respectively,
are the intrinsic symmetries of the eigenvector problem,
it will be reasonable to require them. 
The starting action $\ns$ in \eq{eq:ns} 
has the symmetry \eq{eq:noncomp}, which contains non-compact  directions. 
The breakdown of the symmetry would lead to a non-compact continuous degeneracy of the solutions, 
that would make later analysis involved. 
We assume this does not happen: the symmetry \eq{eq:noncomp} holds in \eq{eq:nassfer}.

In the Schwinger-Dyson method, the values of $Q$ are determined by the stationary points of an effective action which is 
the sum of the expectation value of $\nmS_\perp$ and the logarithmic terms which originate from
the transition of the variables from bosons and fermions to the expectation values (for instance, see Appendix A of \cite{Sasakura:2022iqd}
for the derivation):
\[
\nseff&=\langle \nmS_\perp \rangle + \frac{1}{2} \log \det M^B -\log \det M^F,
\]
where $M^B$ and $M^F$ are the matrices with components: 
\s[
&M^B_{\kappa i a \ \kappa' i' a'}
= \langle X^{\kappa i}_a X^{\kappa' i'}_ {a'} \rangle\ \ (X^{1 i }_a \equiv 
\phi^i_a,\ X^{2 i }_a \equiv \sig^i_a),  \\
&M^F_{\kappa i a \ \kappa' i' a'}
= \langle \bpsi^{\kappa i}_a \psi^{\kappa' i'}_ {a'} \rangle.
\label{eq:nmbmf}
\s]
Using the explicit forms of $\nmS_\perp$ given in Appendix~\ref{app:nsperpexp} and the assumptions \eq{eq:nassbos}
and \eq{eq:nassfer}, we obtain an effective action in terms of $Q$ as:
\s[
\nseffq&= Np\Bigg [ \sum_{\kappa=1}^2 \left(
-\frac{\xi}{2} (\nqf_\kappa)^2 + \nqf_\kappa - \log \nqf_\kappa 
+ \mathbf{i} s_{\kappa} \sqrt{\tilde \epsilon}\, Q_\kappa^F
\right)\\
&\ \ \ \ 
-2 \xi \left(\nqb_1 \nqb_3 +(\nqb_2)^2 \right)- \nqb_1 + 2 \mathbf{i}  \nqb_2-\epsilon \nqb_3 
+\frac{1}{2} \log \left( \nqb_1 \nqb_3 -(\nqb_2)^2 \right)
\Bigg ]
\label{eq:nseffq}
\s]
with $s_1=1, s_2=-1$, and:
\[
\xi=\frac{N \nu^{2(p-2)}}{2 \alpha p (p-2)!},
\label{eq:nbeta}
\]
which is identical to the definition in \eqref{eq:defofxifirst}.
Some details of the derivation of \eq{eq:nseffq} are given in 
Appendix~\ref{app:nsperpexp}.
We assume that the large-$N$ scaling is taken so that $\xi$ is kept finite, namely,
$\nu^{2(p-2)}$ scales like $1/N$.
One reason to take this scaling is that the interesting structure of the distribution, namely its edge, exists in this order. 

An important property of \eq{eq:nseffq} is that the bosonic and fermionic parts do not interact with each other.
Therefore we can solve the stationary point equations by considering them independently. We will find multiple 
solutions for these equations, 
and the proper one can be selected by their values at $\xi=0$ or equivalently $\nu=0$.
Since the case with $\xi=0$ is a free field theory (see \eq{eq:nsperp}), the proper solution should satisfy:
\s[
&\nqb_1=0,\ \nqb_2=\frac{\mathbf{i}}{2},\ \nqb_3=\frac{1}{2}, \\
&\nqf_\kappa=1,
\label{eq:qatzero}
\s]
for $\xi, \epsilon,\tilde \epsilon \rightarrow +0$.

Once we have obtained the solution, the partition function in the leading order of $N$ is determined to be:
\[
Z_\perp=e^{\nseff(Q)-\nmS^{\rm eff}_0+o(N)},
\label{eq:nzlarge}
\]
where $\nmS^{\rm eff}_0$ is the value of $\nseff(Q)$ at $\xi=0$ with \eq{eq:qatzero}, and the subtraction is chosen to be consistent with the normalisation
condition \eq{eq:nnorm}.

\subsubsection{Solving for $\nqf$ (Fermionic Sector)}
The saddle point equation for $Q^F$ is given by:
\[
\frac{\partial \nseffq}{\partial \nqf_\kappa}=Np \left(-\xi\nqf_\kappa +1-\frac{1}{\nqf_\kappa}+  \mathbf{i} s_{\kappa} \sqrt{\tilde \epsilon }\right)=0.
\]
Between the two solutions $\Big(1+\mathbf{i} s_{\kappa} \sqrt{\tilde \epsilon }\pm \sqrt{(1+\mathbf{i} s_{\kappa} \sqrt{\tilde \epsilon })^2-4 \xi}\Big)/(2 \xi)$, 
we pick the one with the $-$ sign that satisfies the condition \eq{eq:qatzero}\footnote{We take the standard branch convention for the square root, i.e. the cut runs 
along the negative real axis.}. Taking $\tilde \epsilon \rightarrow +0$, we obtain:
\[
\nqf_\kappa=\frac{1- \sqrt{1-4 \xi}}{2 \xi}\hbox{ for }\xi \leq \frac{1}{4},
\label{eq:QFbelow}
\]
and:
\[
\nqf_{1}=\frac{1+\mathbf{i} \sqrt{4 \xi-1}}{2 \xi},\ \nqf_{2}=\frac{1-\mathbf{i} \sqrt{4 \xi-1}}{2 \xi} \hbox{ for }\xi\geq 1/4.
\label{eq:QFabove}
\]
An important matter here is that, for $\xi\geq 1/4$,
the  regularization parameter $\tilde \epsilon$ has uniquely required $Q^F_\kappa$ 
to take the solutions with the opposite signs
of the square root to each other. 
If there were no $\tilde \epsilon$, 
the choice would become ambiguous, namely, they could take the solutions with the same sign.
This uniqueness leads to that of the value of 
the saddle point action.

\subsubsection{Solving for $\nqb$ (Bosonic Sector)}
The saddle point equations are given by:
\s[
&\frac{\partial \nseffq}{\partial \nqb_1}=-2 \xi \nqb_3-1+\frac{1}{2}\frac{\nqb_3}{\nqb_1 \nqb_3-(\nqb_2)^2}=0, \\
&\frac{\partial \nseffq}{\partial \nqb_2}=-4 \xi \nqb_2+ 2 i -\frac{\nqb_2}{\nqb_1 \nqb_3-(\nqb_2)^2}=0,\\
&\frac{\partial \nseffq}{\partial \nqb_3}=-2 \xi \nqb_1-\epsilon +\frac{1}{2} \frac{\nqb_1}{\nqb_1 \nqb_3-(\nqb_2)^2}=0.
\s]
The last equation suggests $\nqb_1=0$ in the limit $\epsilon \rightarrow +0$. Assuming this, one can consistently obtain
a solution:
\s[
&\nqb_1=0, \\
&\nqb_2=\frac{\mathbf{i}(1-\sqrt{1-4 \xi})}{4 \xi}, \\
&\nqb_3=\frac{1}{1-4 \xi+\sqrt{1-4 \xi}},
\label{eq:QBbelow}
\s]
which satisfies \eq{eq:qatzero}, for $\xi\leq 1/4$. The solution cannot be extended to $\xi>1/4$, because then the action becomes 
complex, that contradicts the positivity of the denominator of \eq{eq:detabs}. 

The correct solution can be obtained by carefully taking the $\epsilon \rightarrow +0$ limit of the 
solution which satisfies \eq{eq:qatzero}. For $\xi>1/4$ we obtain:
\s[
&\nqb_1=\frac{{\sqrt{\epsilon}}\sqrt{4 \xi-1}}{4 \xi}, \\
&\nqb_2=\frac{\mathbf{i}}{4 \xi},\\
&\nqb_3=\frac{1}{\sqrt{\epsilon}}\frac{\sqrt{4\xi-1}}{4 \xi},
\label{eq:QBabove}
\s]
in the leading order of $\epsilon$.
Here note that it is necessary to keep the $\sqrt{\epsilon}$ order of $\nqb_1$, because
\eq{eq:nseffq} contains $\nqb_1 \nqb_3$, and $\nqb_3$ diverges 
like $1/\sqrt{\epsilon}$.
Though the values of $\nqb$ have no well-defined limits as $\epsilon \rightarrow +0$, the action \eq{eq:nseffq}
does, which is our only concern.

\subsubsection{The Result}
By explicitly putting the above solutions 
\eq{eq:QFbelow}, \eq{eq:QFabove}, \eq{eq:QBbelow}, and~\eq{eq:QBabove}
to \eq{eq:nseffq} and using \eq{eq:nzlarge},
we find a rather compact expression valid for both ranges of $\xi$ as:
\[
Z_\perp= e^{Np\,{\rm Re}[ s^{\rm eff}]+o(N)},
\label{eq:nzperpexp}
\]
where $\rm Re$ denotes the real part, and:
\[
s^{\rm eff}=-\frac{\xi}{2} (Q^F)^2 + Q^F - \log Q^F-1
\label{eq:nh}
\]
with:
\[
Q^F=\frac{1-\sqrt{1-4\xi}}{2 \xi}.
\]

Two comments are in order about this final result. One is that the expression \eq{eq:nh} looks like half 
of the fermionic part of \eq{eq:nseffq}. This observation is correct: the bosonic part exactly cancels half of the fermionic part.
This cancellation could be attributed to the (half) supersymmetry \eq{eq:susy}, but the 
things seem more subtle. The subtlety comes from the fact that we cannot set $\epsilon,\tilde \epsilon$ to zero from the beginning, 
as the above computations show, and therefore the computations in the middle stage are violating the supersymmetry.
We would need a more detailed analysis to understand this cancellation. 
The other comment is that, though the solutions look different between the two ranges of $\xi$, the final
form has the compact expression above for both ranges by taking the real part for $\xi>1/4$. This would imply that 
the partition function has a well-behaved analyticity.

\subsection{Large-$N$ Asymptotic Form of $\rho$ and Location of Edge}
\label{sec:genuinerhovandedge}
It is now straightforward to compute the large-$N$ asymptotic form of $\rho(\nu)$. 
By assuming the same scaling $\nu^{2 (p-2)}$ with $1/N$ as in Section~\ref{sec:signedlargeN}, and 
applying the Stirling's approximation to the sphere volume $S_{N-1}= 2\pi^{N/2}/\Gamma(N/2)$, 
we obtain from \eq{eq:nrhovzperp}:
\[
\rho(\nu)= \left(\frac{2 e\alpha  p! }{\nu^{2(p-2)} N} \right)^{\frac{Np}2} 
\exp
(-\frac{\alpha p! }{\nu^{2 (p-2)}}+o(N))\, Z_\perp
\]
in the leading order of $N$. 
Putting \eq{eq:nzperpexp} into this expression, we finally obtain:
\[
\rho(\nu)= e^{Np\, {\rm Re}[\nhp(\nb)]+o(N)},
\label{eq:nrhofinal}
\]
where $\nhp(\nb)$ is 
defined in \eqref{eq:nhb}.
Here $\nb=\xi/\xi_c$ with $\xi_c=1/4$, or equivalently:
\[
\nb=\frac{\nu^{2 (p-2)}}{\nu_c^{2 (p-2)}}=\frac{\ev_c^2}{\ev^2}
\label{eq:ndefb}
\]
with:
\[
\nu_{c}^{2(p-2)}=\frac{\alpha p (p-2)!}{2 N},\ \ \ev_c=\sqrt{\frac{2N}{\alpha p (p-2)!}},
\label{eq:nvcvalue}
\]
where we have also expressed the distribution in terms of the eigenvalues $\ev$ by using the relation \eq{eq:lamvecrel}\footnote{More rigorously, the distribution of the eigenvalues $\ev$ must be given by $\rho(\nu) |d \nu/d \ev|$ because of the change of 
the measure between $\ev$ and $\nu$. However, the factor is irrelevant in the large-$N$
asymptotic form.}.

The location of the edge of the distribution is determined by the solution to ${\rm Re} [\nhp(\nbp)]=0$.
$\nhp(\nb)$ is a real monotonically increasing function with negative and positive regions in $x<1$, 
while ${\rm Re}[\nhp(\nb)]>0$ for $x\geq 1$. Therefore the solution uniquely exists in the region $x<1$,
and actually satisfies $\nhp(\nbp)=0$. This implies that the signed and the genuine distributions share
the same edge location.
For instance, for $p=3$ we numerically obtain $\nb^{\rm edge}_{p=3}\approx 0.971236$, or:
\[
\nu_{{\rm edge}\,p=3}^{2}\approx  0.971236 \cdot \frac{3 \alpha}{2 N}
\]
from \eq{eq:ndefb} and \eq{eq:nvcvalue}.

The location of the edge of the distribution gives a bound on the injective norm of the random tensor for large-$N$.
The injective norm of a tensor is an operator norm, which for instance has an application to quantum information theory, more specifically, 
defines the geometric measure of entanglement of a multipartite state 
\cite{https://doi.org/10.1111/j.1749-6632.1995.tb39008.x,H_Barnum_2001,PhysRevA.68.042307}.  
The injective norm of an antisymmetric real tensor may be defined by:
\[
\|T\|_{\rm inj}=\max_{w^i \in \mathbb{R}^N \atop |w^i|=1}  \frac{1}{p!} \epsilon^{i_1 \cdots i_p} T_{a_1\cdots a_p} w^{i_1}_{a_1} \cdots w^{i_p}_{a_p}.
\]
The stationary condition of $\epsilon^{i_1 \cdots i_p} T_{a_1\cdots a_p} w^{i_1}_{a_1} \cdots w^{i_p}_{a_p}$ 
with respect to $w^i$ with Lagrange multipliers for their norms leads to the eigenpair equations, 
which lead to the eigenvector equation \eq{eq:eigenproblem-condensed-form} after rescaling. Then we obtain:
\[
\| T \|_{\rm inj}=\frac{1}{\nu^{p-2}_{\rm min}} \ (=z_{\rm max}),
\]
where $\nu_{\rm min}$ denotes the norm of the smallest eigenvector of $T$ (and $z_{\rm max}$ the largest eigenvalue). 
For the random tensor in the large-$N$ limit, it is believed that $\nu_{\rm min}$ 
converges to the edge of the distribution, and therefore the bound is believed to be 
tight\footnote{To our knowledge, this is rigorously proven only for the real symmetric random 
tensor \cite{auffinger2013random,subag2017complexity}. Therefore the rigorous statement 
in our case is the following upper bound,  
\[
\| T_{\rm random} \|_{\rm inj} \leq  \frac{1}{\nu^{p-2}_{\rm edge}},
\]
since the probability of finding an eigenvector of norm $\nu<\nu_{\rm edge}$ 
is suppressed exponentially in $N$ as
$\exp(N\, p\, h_p(x)+o(N))$ with $h_p(x)<0$ for $x<\nbp$.}.
Assuming this and from \eq{eq:ndefb} and \eq{eq:nvcvalue}, we obtain an asymptotic form of the injective norm as:
\[
\| T_{\rm random} \|_{\rm inj} = 2 \left( \frac{p-1}{\nbp N^{p-1}}\right)^{\frac{1}{2}}+o\left(
N^{-\frac{p-2}{2}} \right)\hbox{ for } N\rightarrow \infty,
\]
where the random tensor is normalised as $\langle \sum_{a_1\cdots a_p=1}^p T_{a_1\cdots a_p}^2 \rangle_T=1$ by putting $\alpha=\binom{N}{p}/2$.

\section{Universality}
\label{sec:universality}
The tensor eigenvalue/vector problems have various versions. 
The distributions of 
some of them for random tensors depend only on a common norm of eigenvectors, 
as in our current case\footnote{In our current case, there are $p$ eigenvectors, which have the same norm $\nu$. The distribution 
essentially depends only on $\nu$, because of the rotational symmetry.
A complex case which will be illustrated in Section~\ref{sec:compind} has also a common norm of multiple eigenvectors.}.
In this section we will show that the large-$N$ leading order distributions of 
this kind of the eigenvalue/vector problems, which have been computed 
in the current and previous works, have the same universal form. More precisely, we will
show that, in all the known results of this kind of problems, the genuine and the signed distributions have the form:
\[
&\rho (\nu)= e^{N B\, \Re[\nhp(\nb)]+o(N)}, \label{eq:nuniversal} \\
&\rho_{\rm signed} (\nu)=  \Re [e^{N B\, \nhp(\nb)+o(N)}],
\label{eq:nuniversalsigned}
\]
respectively, in the leading order of $N$, where $\nhp(\nb)$ and $\nb$ were defined in \eq{eq:nhb} and \eq{eq:ndefb}, respectively. 
Here $B$ is a number and $\nu_c$ is a critical value 
of $\nu$ similar to the one encountered in Section~\ref{sec:nft}, where it appeared as a phase transition point of the quantum field theory \eq{eq:nzperp}, 
separating the weak and strong coupling regimes. In \cite{auffinger2013random} 
the transition point was shown to be the energy threshold in the $p$-spin spherical
model, where the stationary points of the energy are all local minima on one side but are dominated by unstable ones on the other side. 
While $\nhp(\nb)$ is universal, $B$ and $\nu_c$ depend on each tensor eigenvalue/vector problem.
However, note that they are common between the genuine and signed distributions of the same problem.

An important consequence of this universality is that the locations of the edges of this kind of the distributions have also a universal form:
\[
\nu_{\rm edge}^{2(p-2)}=\nbp \nu_c^{2 (p-2)},
\label{eq:vedgeuni}
\]   
where $\nbp$ is the unique solution to $\nhp(\nbp)=0$. As for $\nu_c$, there does not seem to exist a 
general formula valid for any problem. However, in many eigenvalue/vector problems, $\nu_c$ has analytical expressions.
In such cases, \eq{eq:vedgeuni} gives a useful formula for the location of the edge, which corresponds to the largest eigenvalue 
being useful in various applications \cite{10.5555/3240740}.
As mentioned above, referring to \cite{auffinger2013random}, the region $\nu<\nu_c$ corresponds to that of the dominance of local minima, while unstable ones dominate in the region $\nu>\nu_c$. 

As for the value of $B$, we can make an interesting observation from the known results. 
In the following cases, we will find:
\[
NB= \hbox{total dimensions of eigenvectors}.
\]
In fact, as shown in \eq{eq:nrhofinal}, $NB=Np$ for the current case, and there are 
$p$ $N$-dimensional real eigenvectors; $NB=N$ for the problem in Section~\ref{sec:rrtens}, and there is one real $N$-dimensional eigenvector;
$NB=2Np$ for that in Section~\ref{sec:compind}, and there are $p$ complex $N$-dimensional eigenvectors; $NB=2N$ for that in Section~\ref{sec:compsym}, and there is one complex $N$-dimensional eigenvector.

\subsection{Real Eigenvalues of Real Symmetric Random Tensors} 
\label{sec:rrtens}
The eigenvector equation of this case is given by:
\[
v_{a_1}=T_{a_1a_2\cdots a_p} v_{a_2}v_{a_3}\cdots v_{a_p},
\]
where $T$ is a symmetric real tensor and $v$ a real eigenvector of the tensor. By 
normalising the eigenvector, the corresponding eigenvalue $z$ is given by $z=1/\nu^{p-2}$.

This is the most classic case of the eigenvalue/vector distribution of random tensors,
and has been computed in \cite{auffinger2013random} in the context of the 
$p$-spin spherical model for spin glasses \cite{crisanti1992spherical}.
Following \cite{auffinger2013random}, the leading-$N$ distribution 
is given by:
\[
\rho_{\rm \, real}(u)=e^{N {\rm Re}[\Theta_p(u)]+o(N)},
\]
where:
\[
\Theta_p(u)=\frac{1}{2} \log(p-1) -\frac{p-2}{4 (p-1)} u^2 -I_1 (u), 
\label{eq:ntheta}
\]
with:
\[
I_1(u)=-\frac{u}{E_\infty^2} \sqrt{u^2-E_\infty^2}-\log (-u+\sqrt{u^2-E_\infty^2})+\log E_\infty,
\label{eq:nione}
\]
and $E_\infty=2 \sqrt{(p-1)/p}$. $u=-E_\infty$ corresponds to the critical point, and 
$u$ can be shown to be related to the eigenvalue $z$ of $T$ by $-\sqrt{N} u=z$ from the definition of $u$ as the energy level of the $p$-spin
spherical model for spin glasses \cite{crisanti1992spherical}. 
Because of $z=1/\nu^{p-2}$, $u$ and $\nb$ in \eq{eq:ndefb} are related by $u=-E_\infty/\sqrt{\nb}$.
Putting this into \eq{eq:ntheta} and \eq{eq:nione}, we indeed obtain:
\[
\Theta_p(u)=h_p(\nb).
\]
Therefore the distribution of this case is given by \eq{eq:nuniversal} with $B=1$ 
and $\nu_c^{2(p-2)}=\alpha p / (2 N (p-1))$ (this can be derived from 
$\sqrt{N} E_\infty= 1/\nu_c^{p-2}$)\footnote{\cite{auffinger2013random} takes the standard normal distribution for $T$, that corresponds to $\alpha=1/2$.}.

The signed distribution of this case has been computed in \cite{Sasakura:2022zwc}, 
and it agrees with the form in \eq{eq:nuniversalsigned} with the same $B$ and $\nu_c$.

\subsection{Complex Random Tensors with Independent Indices}
\label{sec:compind}
The eigenvalue/vector problem of this case has some similarities with the current case. 
It is described by a system of eigenvector equations:
\s[
&v^{(1)}_{a_1}=T_{a_1a_2\cdots a_p} (v^{(2)}_{a_2})^* \cdots (v^{(p)}_{a_p})^*,\\
&v^{(2)}_{a_2}=T_{a_1a_2\cdots a_p} (v^{(1)}_{a_1})^* (v^{(3)}_{a_3})^* \cdots  (v^{(p)}_{a_p})^*,\\
&\ \ \ \ \ \  \vdots \\
&v^{(p)}_{a_p}=T_{a_1a_2\cdots a_p} (v^{(1)}_{a_1})^* \cdots  (v^{(p-1)}_{a_{p-1}})^*,
\s]
where $T$ is a complex tensor with independent indices, whose dimensions may be different from each other,
and $v^{(i)}$ denotes the complex eigenvector in the $i$-th index vector space. From the system of the equations one can readily show that all the 
eigenvectors have a common norm $\nu=|v^{(i)}|\equiv \sqrt{v^{(i)*}_a v^{(i)}_a}\ (\hbox{no sum for }i)$. 
Moreover the distribution is invariant under the rotational symmetry in each index space.
Therefore the distribution can essentially be represented by a function of $\nu$.  
The signed distribution has been computed in \cite{Sasakura:2024awt} 
for $p=3$ by doing a gauge-fixing procedure as in the current case. 
By extending the result to general $p$ \cite{workinprogress} and assuming that all the indices have the same dimension $N$,  the signed distribution 
in the leading order is given by:
\[
\rho_{\rm signed\, cp} = \Re [e^{Np\, h_{\rm cp}+o(N)}],
\]
where:
\[
h_{\rm cp}=-\log g -\frac{1}{g} +\log p -1+2 Q-g \left(1-\frac{1}{p} \right) Q^2 -2 \log Q
\]
with $g=Np \nu^{2(p-2)}/\alpha$:
\[
Q=p \cdot \frac{\sqrt{1+\frac{4g}{p}y^2}-1}{2 g y},
\]
and $y$ being a solution to the equation:
\[
p-2+2 y-p \sqrt{1+\frac{4gy^2}{p}}=0.
\]
By taking the appropriate branch of the solution of $y$, the critical point can be shown to exist at $g=g_c=p/(4 (p-1))$. 
Then by putting $g=g_c\,  \nb$ to the above equations, we obtain:
\[
h_{\rm cp}=2 \nhp(\nb).
\]  
Therefore the signed distribution of this case is given by \eq{eq:nuniversalsigned} with $B=2 N p$ and 
$\nu_c^{2(p-2)}=\alpha/(4 (p-1) N)$.

\subsection{Complex Symmetric Random Tensors}
\label{sec:compsym}
In \cite{Majumder:2024ntn} three different kinds of 
complex eigenvalue/vector problems of symmetric random tensors for $p=3$, which have different characterisations by their symmetries, 
have been studied. 
Among them that with $U(N,\mathbb{C})$ symmetry depends essentially only on the norm $\nu$, 
corresponding to the symmetric tensor analogue of the problem in Section~\ref{sec:compind}. The eigenvector equation of this case is given by
(only the $p=3$ case was considered in \cite{Majumder:2024ntn}):
\[
v_a=T_{abc}v_b^*v_c^*,
\]
where $T$ is a complex symmetric tensor and $v$ is a complex eigenvector. The signed distribution of this case in the leading order
of $N$ is given by:
\[
\rho_{\rm signed\, sym\, cp}(\nu)= \Re [e^{N h_{\rm sym \, cp}+o(N)}]
\]
with:
\[
h_{\rm sym\, cp}=\log 2 -\frac{2}{3 g} -\log l +\frac{l}{g(1+l)},
\]
where $g=2 N \nu^2 /(3 \alpha)$, and $l$ is obtained by taking the appropriate branch of the solution to the 
stationary condition $\partial h_{\rm sym\, cp}/\partial l=0$ as:
\[
l=\frac{1-2 g -\sqrt{1-4 g}}{2 g}.
\] 
The critical point is at $g=1/4$, and therefore putting $g=\nb/4$ into the above equations, we obtain:
\[
h_{\rm sym\, cp}=2 h_{p=3}(\nb).
\] 
Therefore in this case we have obtained \eq{eq:nuniversalsigned} with $B=2 N$ and $\nu_c^2 = 3 \alpha/(8N)$.

\section{Conclusions}
\label{sec:concl}

In this work, we defined real eigenpairs of a real antisymmetric tensor of 
order $p$ and dimension $N$ as pairs of a real eigenvalue and $p$ orthonormal 
$N$-dimensional real eigenvectors, and computed the signed and the genuine 
distributions of such eigenvalues for Gaussian random real antisymmetric tensors by 
the quantum field theoretic method, which had been successfully applied in 
the previous studies \cite{Sasakura:2022zwc,Sasakura:2022iqd,Sasakura:2022axo,Kloos:2024hvy,Sasakura:2024awt,Delporte:2024izt,Majumder:2024ntn}.
The asymptotic distributions for large $N$ were computed 
by using the Schwinger-Dyson equations.
We found a large-$N$ universality which holds across the  
complex \cite{Sasakura:2024awt}, complex symmetric \cite{Majumder:2024ntn}, 
real symmetric \cite{auffinger2013random}, and real antisymmetric tensors:
their large-$N$ asymptotic forms of the distributions
of the eigenvalues $\ev$ are all expressed by $e^{N \,B \, h_p(\ev_c^2/\ev^2)+o(N)}$,
where the function $h_p(\cdot)$ depends only on $p$, while $B$ and $\ev_c$ differ for each 
case, $NB$ being the total dimension of the eigenvectors and $\ev_c$ being
determined by the phase transition point of the quantum field theory.
We conjecture that the same universality holds for all cases in which 
the eigenvector distribution for random tensor
essentially depends only on a common norm of eigenvectors. 

As in the previous cases \cite{Sasakura:2022zwc,Kloos:2024hvy,Sasakura:2024awt,Delporte:2024izt,Majumder:2024ntn},
the functional behavior of the signed distribution for large-$N$ in \eq{eq:nuniversalsigned} changes qualitatively at the phase transition point. 
For $x<1$ it is monotonic, while at $x>1$ it is oscillatory,
because $h_p(x)$ is real for $x < 1$, while it is complex for $x>1$. 
This change of behavior is consistent with the picture proven 
in \cite{auffinger2013random} that the signs of $\det \tilde J$ of the 
eigenvectors are the same for $x<1$, while they take both at $x>1$.

We obtained an analytic expression of the signed distribution for finite $N$, 
but we did not compute the finite $N$ analytic expression for the genuine one in this work. 
As shown in \eq{eq:analSignedRho} and \eq{eq:nrhovzperp},
the only difference between the two is the transverse quantum field theories. 
The quantum field theory of the signed distribution is purely fermionic, and therefore
the partition function can in principle be exactly computed 
by expanding it in terms of the four-fermi interaction terms. 
In other words, it is intrinsically
perturbative. On the other hand, 
that of the genuine distribution contains also bosons, and 
it is not clear whether the four-boson interaction terms can be treated by
doing expansions. 
In fact, the previous exact computation for another case \cite{Sasakura:2024awt} 
shows that it should not be expanded, because the transverse partition function
contains non-perturbative terms proportional to $\exp(-{\rm const.}/\nu^{2(p-2)})$.
We have to develop a non-perturbative method to get an exact expression for the 
genuine distribution for finite $N$. 

The exact computability of the signed distribution for finite $N$ makes the 
signed distribution valuable \cite{Kloos:2024hvy}, 
though it is not genuine as a distribution. 
Note that \eq{eq:analSignedRho} and \eq{eq:nrhovzperp} are the same except for 
the transverse part, and also that, if we assume the large-$N$ universality, 
the function $h_p(x)$ can be read from \eq{eq:nuniversalsigned},
that means that we essentially know the large-$N$ genuine distribution  
\eq{eq:nuniversal} as well. 
Knowing $h_p(x)$, the phase transition point and the largest eigenvalue
can be read from the signed distribution. 
It also enables us to crosscheck our results with the Monte-Carlo simulations for finite $N$, as was shown in the text. We believe that, similarly to the indices of quantum field theories, the signed distribution can provide valuable information about the contents of the landscape of saddle points of tensor theories.

For the genuine distribution, the partition function displays supersymmetry, and we expect that, in order to obtain the exact and finite $N$ result,
resurgence theory may be useful in identifying the transseries and the instanton-like
terms mentioned above. 
We also expect that due to supersymmetry, the transseries (and also perturbation series) will truncate as observed in  \cite{Dunne:2016jsr} known as Cheshire Cat resurgence \cite{Kozcaz:2016wvy}.
This will be an interesting future work.

\section*{Acknowledgements}
We are grateful for the help and support provided by the Scientific Computing and Data Analysis section of Core Facilities at OIST with which part of the simulations was done. NS is supported in part by JSPS KAKENHI Grant No.~25K07153.
NS is thankful to S.~Nishigaki for brief correspondence.

\newpage
\appendix
\section{Conventions} \label{sec:appendix-conventions}
\subsection{Fields and Integrals}
\label{app:convfieldsints}
The $\lambda$ and $\bar{\psi}, \psi$ fields of the partition function initially appear from expressing delta functions and determinants within the field theoretic framework. We pick the following convention for $n$-dimensional delta function integrals, obtained in the context of Fourier transforms that avoid having a rescaling factor in the exponential:
\begin{equation} \label{eq:delta-function-convention}
    \delta(x) = \frac{1}{(2\pi)^{n}} \int \prod_{i} d\mu_{i}\, e^{\mathbf{i} \mu \cdot x} \,.
\end{equation}
With this convention, we define the bosonic integration measure $\mathcal{D}\lambda$:
\begin{equation}
    \mathcal{D}\lambda\, := \frac{1}{(2\pi)^{n}} \prod_{i,a} d\lambda^{i}_{a}
    \,.
\end{equation}
We take the fermionic integration measure $\mathcal{D}\bar{\psi} \mathcal{D}\psi$ to have the following definition:
\begin{equation} \label{eq:fermi-measure-convention}
    \mathcal{D}\bar{\psi}\, \mathcal{D}\psi\, := \prod_{i,a} (d\psi^{i}_{a}\, d\bar{\psi}^{i}_{a})
    \quad \Longrightarrow \quad \int \mathcal{D}\bar{\psi}\, \mathcal{D}\psi\, e^{\bar{\psi} M \psi} = \det M \,,
\end{equation}
which is automatically normalised when $M = \mathds{1}$ and does not pick up a sign in parallel-transverse transformation of the measure. Similarly, the joint bosonic field integration measure of Section \ref{sec:genuine} is defined by:
\begin{equation} \label{eq:bosonmeasure}
   \mathcal{D}\phi\mathcal{D}\sigma := \frac{1}{\pi^n} d\phi\, d\sigma = \frac{1}{\pi^{n}} \prod_{i,a} d\phi^{i}_{a}\, d\sigma^{i}_{a}
   \,,
\end{equation}
so that $\int \mathcal{D}\phi\mathcal{D}\sigma e^{-\phi^2-\sigma^2}=1$.

Finally, we have the result for the surface integral of an $n$-dimensional hypersphere of unit radius:
\begin{equation} \label{eq:hypersphere-area}
    S_{n-1} = \frac{2\pi^{n/2}}{\Gamma\big(\frac{n}{2}\big)}
\end{equation}

\subsection{Matrices and Contractions}
Many matrix-to-$\epsilon$ contractions appear in our calculations, so we collect the following simple manipulations involving a general matrix $M$ and its determinant:
\begin{align}
    \epsilon^{i_{1} \dots i_{p}} \det M &= \epsilon^{i'_{1} \dots i'_{p}} M^{i_{1} i'_{1}} \dots M^{i_{p} i'_{p}} \\
    \Rightarrow (M^{-1})^{i''_{1} i_{1}} \epsilon^{i_{1} i_{2} \dots i_{p}} \det M &= \epsilon^{i''_{1} i'_{2} \dots i'_{p}} M^{i_{2} i'_{2}} \dots M^{i_{p} i'_{p}} \label{eq:eps-det-2m}\\
    \Rightarrow (M^{-1})^{i''_{1} i_{1}} (M^{-1})^{i''_{2} i_{2}} \epsilon^{i_{1} i_{2} i_{3} \dots i_{p}} \det M &= \epsilon^{i''_{1} i''_{2} i'_{3} \dots i'_{p}} M^{i_{3} i'_{3}} \dots M^{i_{p} i'_{p}}
\end{align}
Recall that in the particular case where $M \in \text{O}(p)$, this simplifies further using $M^{-1} = M^{T}$ and if $M \in \text{SO}(p)$ then also $\det M = 1$.
\\\\
Manipulation of matrices and indexed objects sometimes includes symmetrisations and antisymmetrisations. Our conventions
are taken to be:
\begin{align}
    M^{(i_1\dots i_p)}&:=\frac{1}{p!}\sum_{\sigma}M^{\sigma(i_1)\dots\sigma(i_p)}\\
    M^{[i_1\dots i_p]}&:=\frac{1}{p!}\sum_{\sigma}\text{Sign}(\sigma)M^{\sigma(i_1)\dots\sigma(i_p)}\,.
\end{align}
For objects with lower indices taking values on $\{1, \dots, N\}$ we also use the double bracket notation, e.g.:
\begin{equation}
    \llbracket P_{a_{1} \dots a_{p}} \rrbracket := P_{[a_{1} \dots a_{p}]}
\end{equation}
We also take the following convention for the generalised Kronecker delta:
\begin{equation}
 \delta^{i_{1} \dots i_{p}}_{i'_{1} \dots i'_{p}} =\sum_{\sigma} {\rm Sign} (\sigma)\, \delta^{i_1}_{\sigma(i'_1)} \delta^{i_2}_{\sigma(i'_2)}\cdots 
 \delta^{i_p}_{\sigma(i'_p)}\,,
\end{equation}
with $\sigma(i'_1),\sigma(i'_2),\cdots,\sigma(i'_p)$ being permutations of $i'_1,i'_2,\cdots,i'_p$ and ${\rm Sign}(\sigma)=\pm 1$ being the signature of $\sigma$.

\subsection{Parallel and Transverse Subspace Measures} \label{sec:appendix-conventions-partrans}
Finally, we introduce notation for integration measures on parallel and transverse subspaces with respect to the eigenvector tuple. We consider an integral over a $\lambda$-field that splits in the following way as an example:
\begin{equation} \label{eq:gram-par-trans-split}
    \lambda^{i}_{a} = \lambda^{\parallel ij} v^{j}_{a} + \lambda^{\perp ij} \bar{v}^{j}_{a}
\end{equation}
Where the $v^{i}$ span the parallel subspace of $\mathbb{R}^{p}$ and $\bar{v}^{i}$ span the transverse subspace $\mathbb{R}^{N-p}$. It follows that integration over the $p \times N$ variables $\lambda^{i}_{a}$ precisely corresponds to integrating over the $p \times p$ coefficients $\lambda^{\parallel ij}$ and the $p \times (N-p)$ coefficients $\lambda^{\perp ij}$. Let $\mathcal{D}[\lambda^{\parallel, \perp}]$ represent the ``matrix-like" integration over the parallel or alternatively the transverse subspace coefficients. Then we can derive the change of measure through the following:
\begin{align}
    \mathcal{D}\lambda &= \Big|\det \left(\frac{\partial \lambda^{i}_{a}}{\partial\{\lambda^{\parallel jk}, \lambda^{\perp jk}\}}\right)\Big|\, \mathcal{D}[\lambda^{\parallel}]\, \mathcal{D}[\lambda^{\perp}] \nonumber\\
    &= \sqrt{\Big|\det \left(\frac{\partial \lambda}{\partial\{\lambda^{\parallel}, \lambda^{\perp}\}}\right)^{T}\left(\frac{\partial \lambda}{\partial\{\lambda^{\parallel}, \lambda^{\perp}\}}\right)\Big|}\, \mathcal{D}[\lambda^{\parallel}]\, \mathcal{D}[\lambda^{\perp}] \nonumber\\
    &= \sqrt{\Big|\det \left(\frac{\partial \lambda}{\partial \lambda^{\parallel}}\right)^{T}\left(\frac{\partial \lambda}{\partial \lambda^{\parallel}}\right)\Big|} \, \mathcal{D}[\lambda^{\parallel}] \times \sqrt{\Big|\det \left(\frac{\partial \lambda}{\partial \lambda^{\perp}}\right)^{T}\left(\frac{\partial \lambda}{\partial \lambda^{\perp}}\right)\Big|}\, \mathcal{D}[\lambda^{\perp}] \nonumber\\
    &=: \Big|\det \Big(\frac{\partial \lambda}{\partial \lambda^{\parallel}}\Big)\Big| \, \mathcal{D}[\lambda^{\parallel}] \times \Big|\det \Big(\frac{\partial \lambda}{\partial \lambda^{\perp}}\Big)\Big| \, \mathcal{D}[\lambda^{\perp}] \label{eq:gramian-change-meas}
\end{align}
In the first two lines, we use $\partial/\partial\{\lambda^{\parallel jk}, \lambda^{\perp jk}\}$ to schematically represent taking the Jacobian of the transformation of $\lambda$ with respect to the new separate variables (or, more precisely, coefficients) $\lambda^{\parallel}$ and $\lambda^{\perp}$. In the third line, squaring the Jacobian matrix separates the contributions of the $\partial/\partial \lambda^{\parallel}$ and $\partial/\partial \lambda^{\perp}$ derivatives within the Jacobian's block form, allowing us to split the two determinants. In this way, we interpret the determinant notation of the last line in \eqref{eq:gramian-change-meas} in the Gramian sense. We take the following to be equivalent notations for the measures integrating an action:
\begin{align} \label{eq:par-trans-meas-notation}
    \int \mathcal{D}\lambda^{\parallel}\, \mathcal{D}\lambda^{\perp} e^{f(\lambda^{\parallel},\, \lambda^{\perp})} &= \Big|\det \Big(\frac{\partial \lambda}{\partial \lambda^{\parallel}}\Big)\Big| \int \mathcal{D} [\lambda^{\parallel}]\, \mathcal{D}\lambda^{\perp}\, e^{f(\lambda^{\parallel ij}, \, \lambda^{\perp})} \nonumber\\
    &=\Big|\det \Big(\frac{\partial \lambda}{\partial \lambda^{\parallel}}\Big)\Big| \Big|\det \Big(\frac{\partial \lambda}{\partial \lambda^{\perp}}\Big)\Big| \int \mathcal{D} [\lambda^{\parallel}]\, \mathcal{D}[\lambda^{\perp}]\, e^{f(\lambda^{\parallel ij}, \, \lambda^{\perp ij})}
\end{align}
Where, again, the two determinants of rectangular matrices should be interpreted in the Gramian sense of \eqref{eq:gramian-change-meas}. It is worth noting that in the parallel transverse split of \eqref{eq:gram-par-trans-split}, the parallel determinant just yields the metric of \eqref{eq:partrans-metric}:
\begin{gather}
    \Big|\det \Big(\frac{\partial \lambda}{\partial \lambda^{\parallel}}\Big)\Big| = \sqrt{\Big|\det \left(\frac{\partial \lambda}{\partial \lambda^{\parallel}}\right)^{T}\left(\frac{\partial \lambda}{\partial \lambda^{\parallel}}\right)\Big|} = \sqrt{|\det (v\cdot v)|^{p}} = |\det g|^{\frac{p}{2}} \\
    \Longrightarrow \mathcal{D}\lambda^{\parallel} = |\det g|^{\frac{p}{2}} \mathcal{D}[\lambda^{\parallel}] \label{eq:par-meas-with-metric}
\end{gather}
The analogous result holds for the metric defined on the transverse subspace basis vectors (although it is never required in calculations):
\begin{gather}
    \bar{g}^{ij} = \bar{v}^{i} \cdot \bar{v}^{j} \quad \Longrightarrow \quad \Big|\det \Big(\frac{\partial \lambda}{\partial \lambda^{\parallel}}\Big)\Big| = |\det \bar{g}|^{\frac{p}{2}} \\
    \Longrightarrow \mathcal{D}\lambda^{\perp} = |\det \bar{g}|^{\frac{p}{2}} \mathcal{D}[\lambda^{\perp}] \label{eq:trans-meas-with-metric}
\end{gather}
Also, repeating the calculation with Grassmann variables $\psi$ and $\bar{\psi}$ yields the same Jacobian factors but with inverse power:
\begin{gather}
    \mathcal{D}\psi^{\parallel} = |\det g|^{-\frac{p}{2}} \mathcal{D}[\psi^{\parallel}] \\
    \mathcal{D}\bar{\psi}^{\parallel} = |\det g|^{-\frac{p}{2}} \mathcal{D}[\bar{\psi}^{\parallel}]
\end{gather}
This follows from the Jacobian rule for Grassmann integration.

\section{Alternating Between $\widetilde{f}$ and $f$ Conditions} 
\label{sec:appendix-justify-ftilde-f}
In this appendix we want to show that, starting from the $\delta$-functions of the bosonic sector, in the rest of our partition function we are always able to alternate between taking the gauge-fixed solution variables to be zeroed as $\tilde{f}= 0$ or taking the non-gauge-fixed solution variables and the gauge-fixing variables to be zeroed as $f = 0$, $G = 0$ together. The result that we will derive is:
\begin{equation} \label{eq:delta-ftildes-to-delta-f-g}
    \prod_{i,a} \delta(\tilde{f}^{i}_{a}) = |\det g|^{\frac{p}{2}} \left(\prod_{i} \delta(f^{i}_{a} v^{i}_{a})\right) \left(\prod_{i > j} \delta(g^{ij})\right) \left(\prod_{i < j} \frac{1}{|g^{jj}|} \delta(G^{i,j}) \right) \delta(f^{\perp})
\end{equation}
The second product of $\delta$-functions both sets the vectors $v^{i}$ to span a fully $p$-dimensional subspace of $\mathbb{R}^{N}$, as previously shown in Section \ref{sec:gauge-fixing-procedure}, and, by the equivalence in equation \eqref{eq:fv-off-diag-metric}, also sets the off-diagonal projections $f^{i} \cdot v^{j}, i \neq j$ to be equal to zero. Together with the first product of $\delta$-functions, this forces the parallel coefficients of all $f$ variables to zero and, further multiplying with the perpendicular sector $\delta(f^{\perp})$, this implies all $f$ variables being zeroed. The remaining $\delta$-functions explicitly zero the gauge-fixing variables. Therefore, we interpret \eqref{eq:delta-ftildes-to-delta-f-g} as implementing $\tilde{f}^{i}_{a} = 0 \Leftrightarrow f^{i}_{a}, G^{i,j} = 0$ of Section \ref{sec:gauge-fixing-procedure} by $\delta$-functions while still accounting for the redundancy in the $f^{i}_{a} = 0$ system.

To show \eqref{eq:delta-ftildes-to-delta-f-g}, we begin by the usual split between parallel and transverse components of our solution conditions $\tilde{f}^{i}_{a}$:
\begin{equation}
    \widetilde{f}^{i}_{a} = \widetilde{f}^{\parallel i}_{a} + \widetilde{f}^{\perp i}_{a} = I^{\parallel}_{ab} \widetilde{f}^{i}_{b} + I^{\perp}_{ab} \widetilde{f}^{i}_{b}
\end{equation}
Then, we can split the $\delta$-functions over these coefficients using \eqref{eq:gramian-change-meas}, \eqref{eq:par-meas-with-metric} and \eqref{eq:trans-meas-with-metric}:
\begin{align}
    \prod_{i,a} \delta(\widetilde{f}^{i}_{a}) &= \int \mathcal{D}\lambda\, e^{\mathbf{i} \lambda^{i} \cdot \widetilde{f}^{i}} = \int \mathcal{D}\lambda^{\parallel}\, e^{\mathbf{i} \lambda^{\parallel i} \cdot \widetilde{f}^{\parallel i}} \int \mathcal{D}\lambda^{\perp}\, e^{\mathbf{i} \lambda^{\perp i} \cdot \widetilde{f}^{\perp i}} \nonumber\\
    &= \left(|\det g|^{\frac{p}{2}} \int \mathcal{D}[\lambda^{\parallel}]\, e^{\mathbf{i} \lambda^{\parallel ij} \widetilde{f}^{\parallel ik} g^{jk}} \right) \left(|\det \bar{g}|^{\frac{p}{2}} \int \mathcal{D}[\lambda^{\perp}]\, e^{\mathbf{i} \lambda^{\perp ij} \widetilde{f}^{\perp ik} \bar{g}^{jk}}\right) \nonumber\\
    &=: \delta(\widetilde{f}^{\parallel}) \delta(\widetilde{f}^{\perp}) \label{eq:deltas-ftilde-formal-split}
\end{align}
In the above, we absorb the Jacobian factors into the definitions of $\delta(\widetilde{f}^{\parallel})$ and $\delta(\widetilde{f}^{\perp})$. Concentrating on the $\delta$-functions for the parallel solution variables, we split the summation between diagonal, upper-triangular and lower-triangular entries in the integration variable index space:
\begin{align}
    &\int \mathcal{D}[\lambda^{\parallel}]\, \exp \left( \mathbf{i}\lambda^{\parallel ij} \tilde{f}^{i}_{a} v^{j}_{a} \right) = \int \mathcal{D}[\lambda^{\parallel}]\, \exp \left( \mathbf{i} \lambda^{\parallel ij} (f^{i}_{a} v^{j}_{a} + G^{ik} g^{kj})\right) \nonumber\\
    = &\frac{1}{(2\pi)^{p^2}}\int (\prod_{i} d\lambda^{\parallel ii})\, (\prod_{i < j} d\lambda^{\parallel ij})\, (\prod_{i > j} d\lambda^{\parallel ij})\, \exp \left( (\sum_{i} + \sum_{i < j} + \sum_{i > j}) \left( \mathbf{i} \lambda^{\parallel ij} (f^{i}_{a} v^{j}_{a} + G^{ik} g^{kj}) \right) \right)
    \,.
\end{align}
Concentrating on the lower triangle:
\begin{align}
    &\frac{1}{(2\pi)^{\frac{1}{2}p(p-1)}} \int (\prod_{i > j} d\lambda^{\parallel ij})\, \exp \left( \sum_{i > j} \mathbf{i}\lambda^{\parallel ij} (f^{i}_{a} v^{j}_{a} + G^{ik} g^{kj}) \right) \nonumber\\
    =& \frac{1}{(2\pi)^{\frac{1}{2}p(p-1)}} \int (\prod_{i > j} d\lambda^{\parallel ij})\, \exp \left( \sum_{j < p} \mathbf{i}\lambda^{\parallel pj} g^{pj} \right) \exp \left( \sum_{j < p-1} \mathbf{i} \lambda^{\parallel p-1\,j} (f^{p-1}_{a} v^{j}_{a} + G^{p-1\,k} g^{kj})\right) \dots \nonumber\\
    \begin{split}
        =& \prod_{j < p} \delta(g^{pj}) \times \frac{1}{(2\pi)^{\frac{1}{2}p(p-1) - (p-1)}} \nonumber\\
        &\hspace{1.7cm}\times \int (\prod_{i < p} \prod_{j < i} d\lambda^{\parallel ij})\, \exp \left( \sum_{j < p-1} \mathbf{i} \lambda^{\parallel p-1\,j} (g^{p-1\,j} + G^{p-1\,k} g^{kj})\right) \dots
    \end{split}\\
    =& \prod_{i>j} \delta \left( g^{ij} \right) \label{eq:app-deltas-off-diag-g}
    \,.
\end{align}
In the second line, we make use of equation \eqref{eq:fv-off-diag-metric} to obtain an explicit expression for $f^{i}_{a} v^{j}_{a}$. Therefore, the lower triangle sets $g$ to be fully diagonal and, by definition of $g$ on non-zero real vectors, invertible on the domain of interest. Using this fact, we expand and compute the upper triangle:
\begin{align}
    &\frac{1}{(2\pi)^{\frac{1}{2}p(p-1)}} \int (\prod_{i < j} d\lambda^{\parallel ij})\, \exp \left( \sum_{i < j} \mathbf{i} \lambda^{\parallel ij} (f^{i}_{a} v^{j}_{a} + G^{ik} g^{kj}) \right) \nonumber\\
    =& \frac{1}{(2\pi)^{\frac{1}{2}p(p-1)}} \int (\prod_{i < j} d\lambda^{\parallel ij})\, \exp \left( \sum_{i < j} \mathbf{i} \lambda^{\parallel ij} (\underbrace{g^{ij}}_{= 0} + G^{ik} g^{kj}) \right) \nonumber\\
    = & \frac{1}{(2\pi)^{\frac{1}{2}p(p-1)}} \int (\prod_{i < j} d\lambda^{\parallel ij})\, \exp \left( \sum_{i < j} \mathbf{i} \lambda^{\parallel ij} G^{ij} g^{jj} \right) \nonumber\\
    =& \prod_{i < j} \delta \left( G^{ij} g^{jj} \right) = \prod_{i < j} \frac{1}{|g^{jj}|} \delta \left( G^{ij} \right) \label{eq:app-deltas-gauge-zero}
    \,.
\end{align}
Finally, we use \eqref{eq:app-deltas-gauge-zero} in the diagonal sector:
\begin{align}
    &\frac{1}{(2\pi)^{p}} \int (\prod_{i} d\lambda^{\parallel ii})\, \exp \left( \sum_{i} \mathbf{i} \lambda^{\parallel ii} (f^{i}_{a} v^{i}_{a} + G^{ik} g^{ki}) \right) \nonumber\\
    =&\frac{1}{(2\pi)^{p}} \int (\prod_{i} d\lambda^{\parallel ii})\, \exp \left( \sum_{i} \mathbf{i} \lambda^{\parallel ii} f^{i}_{a} v^{i}_{a} \right) \nonumber\\
    =& \prod_{i} \delta(f^{i}_{a} v^{i}_{a}) \label{eq:app-deltas-remaining-diag}
    \,,
\end{align}
such that, by combining \eqref{eq:app-deltas-off-diag-g}, \eqref{eq:app-deltas-gauge-zero}, \eqref{eq:app-deltas-remaining-diag} and \eqref{eq:deltas-ftilde-formal-split}, noting that $\tilde{f}^{\perp}|_{G = 0} = f^{\perp}$, we derive the expression \eqref{eq:delta-ftildes-to-delta-f-g}. Additionally, we also note that fixing $g^{ij} = 0, i \neq j$ implies the invertibility of $g$ from $v^{i} \neq 0$ giving both a well defined $p$-dimensional subspace given by the vectors and a well defined transverse subspace, as used in \eqref{eq:trans-meas-with-metric}.

\section{Rewriting the Four-Fermi Theory}
\label{sec:rewritingfourfermi}
In Section \ref{sec:four-fermi}, we rewrite the action with quartic fermionic interactions $S_{\perp}$ as the application of an exponential derivative operator to a theory quadratic in the fermions $S_{\perp, 2}$. First, let $D$ be a differential operator, then we formally define the following using the usual interpretation of the exponential:
\begin{equation}
    e^{D} \phi = \Big( \sum^{\infty}_{n = 0} \frac{1}{n!} D^{n} \Big) \phi = \phi + D\phi+\frac{1}{2}D^2\phi+\dots
\end{equation}
We are interested in the form where $D$ is a quadratic differential operator, given for example by:
\begin{equation}
    D = g_{ij} \frac{\partial}{\partial k_{i}} \frac{\partial}{\partial k'_{j}}
\end{equation}
Where $g$ is a constant and $k, k'$ are two arbitrary vectors of dummy couplings. Then, we consider applying the exponential of this operator on what we can view as an ``action with a lesser degree of interactions":
\begin{align}\label{eq:exponenting-differentiation}
    e^{\frac{\partial}{\partial k} \cdot g \cdot \frac{\partial}{\partial k'}} e^{k \cdot X + k' \cdot Y} &= \sum^{\infty}_{n = 0} \frac{1}{n!} \Big(\frac{\partial}{\partial k} \cdot g \cdot \frac{\partial}{\partial k'} \Big)^{n} e^{k \cdot X + k' \cdot Y} \nonumber\\
    &= \sum^{\infty}_{n = 0} \frac{1}{n!} (X \cdot g \cdot Y)^{n} e^{k \cdot X + k' \cdot Y} = e^{X \cdot g \cdot Y + k \cdot X + k' \cdot Y}
\end{align}

\section{Derivation of \eq{eq:pjp}}
\label{app:derpjp}
There are three types of contributions in $\Psi^\alpha J \Psi^\beta$, namely,
$\Psi^{\parallel\alpha} J \Psi^{\parallel\beta}$, $\Psi^{\perp\alpha} J \Psi^{\parallel\beta}\, (=\Psi^{\parallel\alpha} J \Psi^{\perp\beta})$,
and $\Psi^{\perp\alpha} J \Psi^{\perp\beta}$.

\subsection{$\Psi^{\parallel\alpha} J \Psi^{\parallel\beta}$ Contribution}
Since $\Psi^{\parallel\alpha}$ is expanded in terms of $v^i$, let us compute the contractions of $J$ with them. We obtain:
\[
v_{a_1}^{i_1'} J_{i_1 a_1\ i_2 a_2 } v_{a_2}^{i_2'}&=
v_{a_1}^{i_1'}\left( 
\delta^{i_1 i_2} \delta_{a_1 a_2}-
\frac{1}{(p-2)!} \epsilon^{i_1 i_2 \cdots i_p} T_{a_1 a_2 \cdots a_p}v_{a_3}^{i_3}\cdots v_{a_p}^{i_p}\right)v_{a_2}^{i_2'} \CR
&=\delta^{i_1 i_2} g^{i_1' i_2'} -\frac{1}{(p-2)!} \epsilon^{i_1 i_2\cdots i_p} \epsilon^{j i_2' i_3
i_4\cdots i_p} g^{i_1' j} \CR
&=\nu^2 (\delta^{i_1 i_2}\delta^{i_1' i_2'} -\delta^{i_1 i_1'} \delta^{i_2 i_2'}+\delta^{i_1 i_2'}
\delta^{i_2 i_1'}).
\]
Here from the first line to the second, we have used an identity from the eigenvector equation:
\[
T_{a_1 a_2 \cdots a_p} v_{a_2}^{i_2} v_{a_3}^{i_3}
\cdots v_{a_p}^{i_p}=\epsilon^{i_1 i_2 \cdots i_p} v_{a_1}^{i_1},
\]
which has been given in \eq{eq:partrans-v-epsilon-identity}.
From the second line to the last, we have used $g^{ij}=\nu^2 \delta^{ij}$.

\subsection{$\Psi^{\perp\alpha} J \Psi^{\parallel\beta}$ Contribution}
A similar computation derives:
\[
J_{i_1 a_1\ i_2 a_2 } v_{a_2}^{i_2'} &= \left( 
\delta^{i_1 i_2} \delta_{a_1 a_2}-
\frac{1}{(p-2)!} \epsilon^{i_1 i_2 \cdots i_p} T_{a_1 a_2 \cdots a_p}v_{a_3}^{i_3}\cdots v_{a_p}^{i_p}\right)v_{a_2}^{i_2'} \CR
&=\delta^{i_1 i_2} v^{i_2'}_{a_1} -\frac{1}{(p-2)!} \epsilon^{i_1 i_2\cdots i_p} \epsilon^{j i_2' i_3
i_4\cdots i_p} v_{a_1}^j. 
\]
Since this is a linear combination of $v^i$, the contraction with $\Psi^{\perp\alpha}$ vanishes:$\Psi^{\perp\alpha} J \Psi^{\parallel\beta}=0$.
Similarly, $\Psi^{\parallel\alpha} J \Psi^{\perp\beta}=0$.

\subsection{$\Psi^{\perp\alpha} J \Psi^{\perp\beta}$ Contribution}
We immediately obtain the last two terms of \eq{eq:pjp} by inserting the explicit expression of $J$.

\section{The Explicit Form and the Expectation Value of $\nmS_\perp$}
\label{app:nsperpexp}
Since \eq{eq:nsperp} is compact but not convenient for explicit computations, let us first write down the explicit form.
From \eq{eq:ngamma} and \eq{eq:ngammabf} we obtain:
\[
\nmS_\perp=\nmS^{(2)}_B+\nmS^{(2)}_F+\nmS^{(4)}_{BB}+\nmS^{(4)}_{BF}+\nmS^{(4)}_{FF},
\]
where:
\s[
&\nmS^{(2)}_B=-\phi^i\cdot \phi^i+2 \mathbf{i} \phi^i\cdot \sig^i-\epsilon \sig^i\cdot \sig^i, \\
&\nmS_F^{(2)}=\bpsi^{\kappa i}\cdot \psi^{\kappa i} + \mathbf{i}\sqrt{\tilde \epsilon} \, \bpsi^{1i}\cdot \psi^{1i}- \mathbf{i} \sqrt{\tilde \epsilon} \, \bpsi^{2i} \cdot\psi^{2i}, \\
&\nmS_{BB}^{(4)}=-\frac{\nu^{2(p-2)}}{\alpha p!} \delta_{i'j'}^{ij}\left( \phi^i\cdot \phi^{i'} \sig^{j} \cdot \sig^{j'}+\phi^i\cdot \sig^{i'} \sig^{j} \cdot \phi^{j'}\right), \\
&\nmS^{(4)}_{BF}=\frac{\mathbf{i} \nu^{2(p-2)}}{\alpha p!} \delta_{i'j'}^{ij} \left(
\phi^i\cdot \bpsi^{\kappa i'} \sig^{j}\cdot\psi^{\kappa j'}+\sig^{i} \cdot\bpsi^{\kappa i'}\phi^j\cdot \psi^{\kappa j'}
\right),\\
&\nmS^{(4)}_{FF}=-\frac{\nu^{(2(p-2)}}{4 \alpha p!} \delta^{ij}_{i'j'} \left(
\bpsi^{\kappa i} \cdot \bpsi^{\kappa' i'}\psi^{\kappa j}\cdot \psi^{\kappa' j'}
+\bpsi^{\kappa i}\cdot \psi^{\kappa'i'} \bpsi^{\kappa' j'}\cdot \psi^{\kappa j}
\right),
\s]
where $\delta^{ij}_{i'j'}=\delta^{ii'}\delta^{jj'}-\delta^{ij'}\delta^{ji'}$.

The computation of $\langle \nmS_\perp \rangle$ with \eq{eq:nassbos} and \eq{eq:nassfer} in the leading order 
of $N$ can be described as follows. It is straightforward for the quadratic terms. For instance:
\[
\langle \phi^i_{a} \phi^i_a \rangle=Q^B_1 \delta_{ii} \delta_{aa}=N p Q^B_1.
\]
Similar computations lead to the linear terms in \eq{eq:nseffq}.

As for the interaction terms we ignore the connected expectation values of the four fields and subleading orders. For instance:
\s[
\langle \phi^i\cdot \phi^{i'} \sig^{j} \cdot \sig^{j'}\rangle&=\langle \phi^i_a \phi^{i'}_a \sig^{j}_b  \sig^{j'}_b \rangle\\
&=\langle \phi^i_a \phi^{i'}_a \rangle \langle \sig^{j}_b  \sig^{j'}_b \rangle 
+\langle \phi^i_a \sig^{j}_b\rangle \langle   \sig^{j'}_b \phi^{i'}_a\rangle 
+\langle \phi^i_a \sig^{j'}_b\rangle \langle \phi^{i'}_a \sig^{j}_b   \rangle 
+\langle \phi^i_a \phi^{i'}_a \sig^{j}_b  \sig^{j'}_b \rangle_c \\
& =Q^B_1 \delta^{ii'} \delta_{aa} Q^B_3 \delta^{jj'}\delta_{bb} +Q^B_2 \delta^{ij}\delta_{ab} Q^B_2 \delta^{j'i'}\delta_{ba}+Q^B_2 \delta^{ij'}\delta_{ab} Q^B_2 \delta^{i'j} \delta_{ab} + o(N^2)\\
&= N^2 Q^B_1 Q^B_3 \delta^{ii'}\delta^{jj'}+o(N^2). 
\s]
From the second to the third line, we have ignored the connected four-field correlation function, and from the 
third to the last line, we have only taken the most leading term in $N$.
Similar computations and $\delta^{ij}_{[i'j']} \delta^{ii'} \delta^{jj'}=p^2-p$ 
lead to the quadratic terms of \eq{eq:nseffq}.

Since \eq{eq:nassbos} and \eq{eq:nassfer} have diagonal forms in the vector and the flavour indices, 
we readily obtain $\det M^B=(Q^B_1Q^B_3-(Q^B_2)^2)^{Np}$ and $M^F=(Q^F_1 Q^F_2)^{Np}$. These lead
to the logarithmic terms of \eq{eq:nseffq}.

\section{Numerical Simulations}
\label{app:numsim}

We have gathered our numerical computations in a
Mathematica notebook at the link: \url{https://github.com/dlprtn/AntisymmetricTensorEigenvalues}.

Firstly, it contains the numerical evaluation of the transverse action by acting with the exponential of the quadratic derivative operator on the pfaffian of the quadratic transverse action \eqref{eq:pf-wick-contractions}. 

Secondly, it details how to obtain the generalised eigenvector equations for the totally antisymmetric Gaussian tensors. We relied on the Mathematica numerical solver to exhaustively search for all solutions to the eigenvector equations. Equation \eqref{eq:analSignedRho} presented the signed distribution for a fixed $p$ and fixed finite $N$. By choosing each, we compared the distribution with numerical simulations and verified that theoretical and empirical results match. The chosen parameters for numerical comparison are $p=3$, $p=4$ for $N=4,5,6,7$. This means that we considered the eigenvalues of order-three and order-four tensors. The averaging was done by sampling over $10000$ tensors, each having totally-antisymmetric entries that are randomly generated according to the Gaussian distribution in \eqref{eq:tensor-gaussian-measure}. As explained in Sec.~\ref{sec:gaugefreedom}, while the distribution is independent of gauge, we chose to fix the conditions~\eqref{eq:chosen-gauge-conditions}, as well as the discrete symmetry by $v_i^i>0$, $1\leq i\leq p-1$, in the numerical solver of the eigenvalue equations to get unique solutions among gauge-equivalent solutions. 

In the simulation of the signed distribution,
each solution must be weighted by the sign of the Jacobian $\det \frac{\partial \tilde f}{\partial v}$. The sign actually depends on 
the gauge conditions (including the discrete gauge choice), as is explicit in \eqref{eq:mtilde-det-ma}. In the numerical simulation, however, we rather computed 
the sign by taking the product of the non-zero eigenvalues of $\frac{\partial f} {\partial v}$, ignoring the zero eigenvalues. 
This procedure is independent of the gauge choice, 
and should be equivalent to removing the global sign factor which is explained below \eq{eq:pmdeltaamb}.

\let\oldbibliography\thebibliography 
\renewcommand{\thebibliography}[1]{\oldbibliography{#1}
\setlength{\itemsep}{-1pt}}
\bibliographystyle{JHEP}
\bibliography{TensorEigenvalues.bib}

\end{document}